\documentclass[12pt]{article}
\usepackage{color}
\usepackage{amsmath}
\usepackage{amsthm}
\usepackage{amsfonts}
\usepackage{amssymb}
\usepackage{graphicx}
\usepackage{float}
\usepackage{geometry}
\usepackage[utf8]{inputenc}

\usepackage[all]{xy}
\usepackage{amscd}
\usepackage{longtable}
\usepackage{mathrsfs}
\usepackage{enumitem}
\usepackage[table,dvipsnames]{xcolor}

\usepackage{tabularray}
\usepackage{booktabs}

\usepackage{soul}

\usepackage{listings}
\usepackage{color}

\definecolor{dkgreen}{rgb}{0,0.6,0}
\definecolor{gray}{rgb}{0.5,0.5,0.5}
\definecolor{mauve}{rgb}{0.58,0,0.82}

\usepackage{caption}
\usepackage[utf8]{inputenc}
\usepackage[english]{babel}

\usepackage[thinlines]{easytable}

\usepackage{tikz}
\usetikzlibrary{arrows, shapes.arrows, shapes.geometric, shapes.multipart, decorations.pathmorphing, positioning, swigs}

\usepackage{pdfpages}

\usepackage{listings} %for code in the paper 
\usepackage{xcolor}

\usepackage{natbib}
\usepackage{comment}
\usepackage{color}
\usepackage[backref=page]{hyperref}
\hypersetup{
 colorlinks=true,
 linkcolor=blue,
 filecolor=magenta,
 citecolor=darkRed,
 urlcolor=cyan,
 pdftitle={
 Overcoming an extreme positivity violation to distinguish the causal effects of surgery and anesthesia via a separable effects model
 },
 }

\definecolor{codegreen}{rgb}{0,0.6,0}
\definecolor{codegray}{rgb}{0.5,0.5,0.5}
\definecolor{codepurple}{rgb}{0.58,0,0.82}
\definecolor{backcolour}{rgb}{0.95,0.95,0.92}
\definecolor{darkRed}{rgb}{0.60,.03,.03}

\lstdefinestyle{mystyle}{
    backgroundcolor=\color{backcolour},   
    commentstyle=\color{codegreen},
    keywordstyle=\color{magenta},
    numberstyle=\tiny\color{codegray},
    stringstyle=\color{codepurple},
    basicstyle=\ttfamily\footnotesize,
    breakatwhitespace=false,         
    breaklines=true,                 
    captionpos=b,                    
    keepspaces=true,                 
    numbersep=5pt,                  
    showspaces=false,                
    showstringspaces=false,
    showtabs=false,                  
    tabsize=2
}

\theoremstyle{definition}

\newtheorem*{counter example}{Counter Example}

\newtheorem{assumption}{Assumption}

\def\ci{\mbox{\ensuremath{\perp\!\!\!\perp}}}

\allowdisplaybreaks[1]

\begin{document}

\def\spacingset#1{\renewcommand{\baselinestretch}%
{#1}\small\normalsize} \spacingset{1}

\title{ {\Large 
% Past title ideas: 
%A Separable effects approach to identify the impact of prenatal anesthesia exposure on childhood behavior disorders 
%Distinguishing the effects of prenatal exposure to surgery and anesthesia on development of DIBD using a separable effects model. 
%Using a separable effects model to overcome extreme positivity violation and distinguish the causal effects of surgery and anesthesia.
Addressing an extreme positivity violation to distinguish the causal effects of surgery and anesthesia via separable effects
}}
\author{ {Amy J. Pitts, Ling Guo, Caleb Ing, Caleb H. Miles\thanks{Amy J.~Pitts is a doctoral student, Department of Biostatistics, Columbia
 University, New York, NY 10032, USA; Ling Guo is Biostatistician, Department of Anesthesiology, Columbia
 University, New York, NY 10032, USA; Caleb
 Ing is Associate Professor, Department of Anesthesiology, Columbia
 University, New York, NY 10032, USA; and Caleb
 H. Miles is Assistant Professor, Department of Biostatistics, Columbia
 University, New York, NY 10032, USA (email: cm3825@cumc.columbia.edu).}\hspace{.2cm}}}
\date{}

%\clearpage 
\maketitle
%\thispagestyle{empty}

%\vspace*{5cm}
\bigskip

\begin{abstract}

\noindent The U.S. Food and Drug Administration has cautioned that prenatal exposure to anesthetic drugs during the third trimester may have neurotoxic effects; however, there is limited clinical evidence available to substantiate this recommendation. One major scientific question of interest is whether such neurotoxic effects might be due to surgery, anesthesia, or both. Isolating the effects of these two exposures is challenging because they are observationally equivalent, thereby inducing an extreme positivity violation. To address this, we adopt the separable effects framework of Robins and Richardson (2010) to identify the effect of anesthesia (alone) by blocking effects through variables that are assumed to completely mediate the causal pathway from surgery to the outcome. We apply this approach to data from the nationwide Medicaid Analytic eXtract (MAX) from 1999 through 2013, which linked 16,778,281 deliveries to mothers enrolled in Medicaid during pregnancy. Furthermore, we assess the sensitivity of our results to violations of our key identification assumptions.
\end{abstract}	
\noindent%
{\it Keywords:} Anesthesiology, Causal inference, Claims data, Overlap violation, Positivity violation, Separable effects \vfill

%\doublespacing

\newpage 
\setcounter{page}{1}
\spacingset{1.5}

%\tableofcontents

%-------------------------------------------------------
\section{Introduction} 
\label{section:intro}

In 2016, the U.S. Food and Drug Administration issued a cautionary statement regarding the potential impact of prolonged or repeated exposure to anesthetic drugs during the third trimester of pregnancy or in children under the age of three on their brain development \citep{FDA}. This warning was issued after both preclinical and clinical studies raised concerns about the safety of general anesthesia on the developing brain, citing potential neurotoxic effects %\citep{jevtovic2003early, loepke2008assessment, vutskits2016lasting, flick2011cognitive, ing2012long, wilder2009early, doberschuetz2016follow, davidson2016neurodevelopmental}. 
\citep{jevtovic2003early, loepke2008assessment, vutskits2016lasting, ing2021prospectively, reighard2022anesthetic}.
As summarized in \cite{ing2022anesthesia} and \cite{singh2024anesthetic}, while preclinical studies are convincing, human studies have had limited sample sizes and yielded mixed findings. The fundamental question of the safety of commonly used anesthetics continues to be a topic of debate. % \citep{singh2024anesthetic, ing2022anesthesia}. 

A recent study using data from the nationwide Medicaid Analytic eXtract (MAX) from 1999 through 2013 revealed that in pregnant mothers requiring appendectomy or cholecystectomy, a 31\% elevated risk of a disruptive or internalizing behavioral disorder (DIBD) diagnosis was observed in their offspring after matching on many potential confounders \citep{ing2024behavioural}. However, whether this increased risk of DIBD is caused by the anesthetic medications, or other factors associated with surgery such as perioperative inflammation, remains unclear \citep{ende2024behavioural, vutskits2024developmental}. The relationship between anesthesia and negative neurotoxic outcomes has only been examined through the association of surgery and anesthesia jointly with these outcomes. This is because in existing studies, there are no instances in which a mother receives anesthesia without subsequent surgery, nor the converse, thereby creating an extreme positivity (also known as overlap) violation. This challenge inspires our current work, where we propose an approach that, under strong assumptions, allows us to address this extreme positivity violation to isolate the causal effect of prenatal exposure to general anesthesia required for surgery on the development of DIBD. Given the strength of these assumptions, we do not claim to have fully overcome this challenge in practice, but in a similar spirit to the practice of applying causal methods in an attempt to mitigate the challenge of confounding that is ever-present in observational studies \citep{hernan2018c}, if our assumptions are not too strongly violated, we are able to at least partially address the positivity challenge and likewise obtain an estimate that is closer to the causal estimand of interest, namely the causal effect of anesthesia in the absence of surgery.

The typical positivity assumption requires that all individuals in the study have a non-zero conditional probability of receiving one or both treatments given their covariates. The possibility of observing treatment levels of interest in each patient subgroup is essential for using data from one treatment arm (in which the counterfactual of interest is observed) to learn about the distribution of the corresponding counterfactual under another treatment arm (in which it is not). However, positivity assumptions play an important role in other domains beyond causal inference in which data sampled from one distribution is used to learn about features of another distribution. Prominent examples include missing data \citep{robins1994estimation}, supervised learning under covariate shift \citep{%shimodaira2000improving,
zadrozny2004learning}, %sugiyama2007covariate}, 
nonprobability survey sampling \citep{chen2020doubly}, and data fusion \citep{li2023efficient}. Typical approaches for overcoming positivity violations include trimming patients with poor overlap \citep{cochran1973controlling,kurth2006results, crump2009dealing},  stabilized weights \citep{robins2000marginal}, overlap weights \citep{li2018balancing}, incremental propensity score interventions \citep{kennedy2019nonparametric}, and simply relying on extrapolation. Under extreme positivity violations as in our case, typical approaches do not apply as they would either lead to the exclusion of all observations, or require at least one instance where anesthesia is administered without surgery, which is never observed in our data set. Instead, we will use a separable effects approach to address the positivity violation and achieve nonparametric identification of our causal estimands.

%Robins and Richardson (2010)
\cite{robins_2010} first introduced the idea of separable effects as an alternative way of interpreting mediated effects as ``manipulable'' parameters, i.e., parameters that could be verified under some form of randomized experiment. Since then, the separable effects framework has been extended to address topics such as spillover effects \citep{shpitser2017modeling}, mediation in survival analysis and longitudinal settings \citep{ didelez2019defining,aalen2020time,di2024longitudinal}, intercurrent events \citep{Young2020causal,stensrud2021generalized, strensrud_2022, stensrud2023conditional,  maltzahn2024separable}, and noncompliance \citep{wanis2024separable}. In the typical application of a separable effects model, one begins with a clearly-defined post-exposure (mediating) variable, and then proceeds by positing the existence of separable components of the exposure that justify an extended version of the original causal directed acyclic graph (DAG). In contrast, we offer an alternative perspective, where we begin with clearly-defined separable components of the exposure that are themselves of substantive interest \emph{a priori}, and then we proceed by identifying post-exposure (mediating) variables that justify our extended DAG as a means of addressing the extreme positivity violation between the separable components with which we began. We further contribute to this literature by developing sensitivity analyses with causally-interpretable sensitivity parameters, making connections to back-door and front-door adjustment, and discussing possible extensions to address milder, more commonplace forms of positivity violations.

The remainder of the manuscript is organized as follows. First, we give an overview of the data in Section \ref{section:data}. Section \ref{section:formulating} describes our causal estimands and assumptions, which serve as the foundation for the identification results given in Section \ref{section:identification_est}. In the latter section, we also discuss the relationship between the identification formula and the pure direct effect as described in the mediation literature. 
Section \ref{section:connections} explores relationships to the standard back-door and front-door identification results, as well as extensions to more standard positivity violations. Section \ref{section:estimation} describes how we carry out estimation in the MAX data. Section \ref{section:sensitivity} describes a sensitivity analysis to examine the implications of potential violations of one of our key identification assumptions. Section \ref{section:simulation} demonstrates the performance of our proposed approach in a simulation study. We discuss results from the MAX data application in Section \ref{section:real_data}. Finally, we conclude with a discussion in Section \ref{section:discussion}.

%-------------------------------------------------------
\section{Medicaid Analytic eXtract Data} \label{section:data}

We analyze data from the Medicaid Analytic eXtract for 47 states and Washington D.C., covering the period from 1999 to 2013, resulting in 16,778,231 mother-child pairs \citep{max}. We excluded Maine, Montana, and Connecticut due to the inability to link mother and infant data. Our primary exposure of interest, $A$, represents the decision for a pregnant mother to undergo either an appendectomy or cholecystectomy during pregnancy. These specific procedures were chosen as they are the most common surgeries performed during pregnancy \citep{sachs2017risk}. This exposure consists of two components. The first is the component we are interested in isolating the effect of, i.e., exposure to general anesthetic medications, which we denote $O$. More precisely, we are interested in neurotoxic effects of these anesthetic medications, %on neurotoxicity, 
so other perioperative complications or effects that may arise due to anesthetic management were not included, such as pulmonary aspiration during intubation. The second component, which we denote $N$, is all of the other factors that do not involve neurotoxic effects of anesthetic medications, which mainly consists of the surgery itself, but also other perioperative complications or effects. For ease of discussion, we will simply refer to each of these as anesthesia and surgery, with the understanding that there is some additional nuance to these components. The primary outcome, $Y$, is the age at which the child first receives an ICD-9-CM diagnosis of a DIBD, which we follow the definition from \cite{ing2024behavioural} as a composite outcome consisting of any diagnosis for a disruptive behavioral disorder (attention-deficit/hyperactivity disorder (ADHD) or conduct, impulse control, or oppositional defiant disorders) or an internalizing behavioral disorder (bipolar disorder, depression, or anxiety). The evaluation period begins at the child's date of birth and extends until censoring, which was defined as either December 31st, 2013 or any period during which a child experienced three consecutive months without Medicaid eligibility. We define the outcome in terms of time to diagnosis due to the high rate of non-administrative censoring.

We consider the same set of potential confounders $\bf C$ as those that were matched on in \cite{ing2024behavioural}. These were defined as observations up to the time of exposure, including sociodemographic factors, year of childbirth, healthcare utilization before pregnancy, maternal comorbidities, psychotropic medication prescriptions, and Medicaid eligibility and coverage type. A full list of the variables in ${\bf C}$ is available in Table \ref{tab:tab1} in the Appendix \ref{a:baseline_table}. %Due to the time of anesthetic exposure being absent in the unexposed mothers by definition, we created a pseudo-exposure time for them, as we describe further in Section \ref{section:real_data}.

The vector ${\bf M}$ consists of post-exposure variables that are considered to potentially be on the causal pathway from surgery to the behavioral deficit outcome $Y$. There is also a clear temporal ordering between ${\bf M}$ and $Y$, as all variables in ${\bf M}$ occur prior to or at birth, i.e., time zero for $Y$. As we elaborate on in the following section, our key identifying assumption is that ${\bf M}$ completely mediates the effect of surgery on DIBD, and is also unaffected by anesthesia. 
Since we suspect that any neurotoxic effect of anesthesia on the fetus should be unrelated to post-operative maternal or obstetric complications, we consider the following peri-operative events as variables in ${\bf M}$: any adverse events or complications resulting from the surgical procedure including post-operative infection, hemorrhage, wound dehiscence, injury from abdominopelvic accidental puncture/laceration, blood transfusion, presence of a foreign body,  intra-operative or peri-operative aspiration, or post-operative intubation. Prolonged hospital length of stay of over seven days after the surgical procedure is also included, which is used as a proxy for other unmeasured perioperative complications. 
Further post-operative events within 30 days following the procedure were also included: significant post-operative pain defined as new opioid prescriptions filled and then refilled with the total prescription covering at least seven days; any emergency room visits or inpatient hospitalization not solely attributable to prenatal care; or any need for further related procedures such as endoscopic retrograde cholangiopancreatography, hernia repair, and foreign body removal. All variables listed thus far only occurred among mothers who received surgery and anesthesia, creating a monotonic relationship between these variables and $A$, which we account for in our statistical model. Additional variables in ${\bf M}$ for which this monotonic relationship does not hold include preterm birth, low birth weight (defined as weight less than 2,500g), C-section, and thromboembolism.
%These $M$ variables are crucial in isolating the effect of just anesthetic neurotoxicity on the development outcome of the child.  

%-------------------------------------------------------
\section{Defining the Causal Estimands and Model} \label{section:formulating}

\subsection{Counterfactuals and causal estimands}
\label{section:estimands}

We begin by assuming the existence of counterfactual (or potential) outcomes \citep{neyman1923application,rubin1974estimating}. Given the presence of multiple exposures, there are multiple interventions to consider, each yielding a distinct interpretation. First, $Y(a)$ denotes the time at which a child is diagnosed with a DIBD had they been assigned (possibly contrary to fact) to receive status $a$ of appendectomy or cholecystectomy and accompanying anesthesia. Existing studies, whether explicitly or implicitly, have targeted distributions of counterfactuals under interventions such as this one under an assumption of no unmeasured confounding between the decision to undergo surgery and the outcome interest. Since we are specifically interested in targeting the effect of anesthesia in isolation, we wish to consider estimands defined in terms of interventions on anesthesia independently of surgery. To this end, we introduce the counterfactual $Y(n,o)$, which is the time to DIBD diagnosis had the child been prenatally exposed to anesthesia status $o$ and surgery status $n$, where $o$ and $n$ need not agree. In particular, we focus on the counterfactuals $Y(n=0,o=0)$, $Y(n=1,o=1)$, and $Y(n=0,o=1)$. The counterfactual $Y(n=1,o=0)$ is of lesser interest for our purposes, both due to its impractical interpretation and our inability to identify its distribution from our observed data distribution, as we discuss further in Section \ref{section:estimation}. On the former point, it may be less clear how to interpret the effect of anesthesia in the presence of surgery, since withholding anesthesia may have additional harmful effects that are not of interest since these are generally avoided in modern medicine. We assume the equivalences $Y(n=0,o=0)=Y(a=0)$ and $Y(n=1,o=1)=Y(a=1)$, which are observed among unexposed and exposed mothers, respectively. These equivalences are essentially tautological in our case given our definitions of $A$, $N$, and $O$. However, unlike these counterfactuals, we do not observe the counterfactual $Y(n=0,o=1)$ for any subject in the data, since no mother receives anesthesia without also undergoing surgery. Despite this, we will discuss how its distribution can in theory be identified under the separable effects model.

One may also be interested in estimands defined in terms of the counterfactual $Y(o)$, i.e., the time to DIBD diagnosis had a child been assigned to receive anesthesia status $o$. If we take the anesthesia intervention to be one such that it does not affect surgery (and assume there are no non-neurotoxic effects of anesthesia captured in $N$), then by composition $Y(o)$ is equal to $Y(N,o)$, where $N$ is the random variable representing the observed surgery status, or equivalently, $(1-N)Y(n=0,o)+NY(n=1,o)$. Given the identification and interpretational challenges of the counterfactual $Y(n=1,o=0)$ mentioned above, we will not focus on contrasts of the counterfactuals $Y(o=1)$ and $Y(o=0)$, but rather of $Y(n=0,o=0)$, $Y(n=0,o=1)$ and $Y(n=1,o=1)$.

% might remove 
%{\color{blue} To bridge the gap between causal diagrams and the counterfactual world, we can use a single world intervention graph (SWIG), seen in Figure \ref{fig:dags}.c. Using Single World intervention graphs (SWIG) can help encode the counterfactual independencies associated with specific hypothetical interventions \citep{richardson2013single, robins2022interventionist}. The nodes that have been split into a random component on the left and a fixed component on the right are the separate components of the exposure. The child nodes represent corresponding counterfactual random variables $M(n,o)$, $Y(n,o)$. Analyzing Figure \ref{fig:dags}.c we see that $Y(n,o)$ is d-separated from $n$ given $M(n,o)$. This implies that given $M(n,o)$ that $Y(n,o)$ does not depend on $n$ under the SWIG global Markov property \citep{richardson2013single}. In addition, $M(n,o)$ is d-separated from $o$, and thus $M(n,o)$ is independent of $o$. It should be noted that we chose to label $M$ and $Y$ with $(n,o)$  rather than just $n$ or $o$. This was intentionally done to highlight that any SWIG is a population causal graph.  The SWIG also helps motivate the story of controllable interventions that could be used to create a hypothetical 4-arm trial to test our assumption. However, this will never be implemented in practice due to ethical violations. }

As mentioned, in previous studies, the effect of anesthesia on DIBD has inherently been examined through approximations of the contrasts of distributions of $Y(a=1)$ and $Y(a=0)$ \citep[e.g.]{bartels2009anesthesia, wilder2009early, ing2012long, ing2024behavioural}. One such estimand is $\psi^{\textrm{joint}} \equiv \mathrm{Pr}\{Y(a=1)\leq t \}/\mathrm{Pr}\{Y(a=0)\leq t \}$,
which captures the effect of surgery and anesthesia jointly on development of DIBD by time $t$, rather than the effect of anesthesia alone. Since $A$ completely determines both $O$ and $N$ and we assume the effect of $A$ on $Y$ to be entirely through $O$ and $N$, we can decompose the joint effect of surgery and anesthesia as follows:
\begin{align}
\begin{split}
\label{eq:decomp}
    \frac{\mathrm{Pr}\{Y(a=1)\leq t \}}{\mathrm{Pr}\{Y(a=0)\leq t \}} &= \frac{\mathrm{Pr}\{Y(n=1,o=1)\leq t \}}{\mathrm{Pr}\{Y(n=0,o=0)\leq t \}} \\
    &= \underbrace{\frac{\mathrm{Pr}\{Y(n=1,o=1)\leq t \}}{\mathrm{Pr}\{Y(n=0,o=1)\leq t \}}}_{\textrm{effect of surgery}} \times \underbrace{\frac{\mathrm{Pr}\{Y(n=0,o=1)\leq t \}}{\mathrm{Pr}\{Y(n=0,o=0)\leq t \}}}_{\textrm{effect of anesthesia}}.
\end{split}
\end{align} 
Our effect of primary interest will be the second term capturing the effect of anesthesia alone, i.e., the relative risk of a child developing DIBD by time $t$ if their mother were to be assigned to receive anesthesia, but to not undergo surgery compared with that if their mother were to be  assigned to receive neither surgery nor anesthesia.
The other term in the decomposition captures the effect of surgery in the presence of anesthesia, and has an analogous interpretation, but contrasting levels of surgery while assigning the mother to receive anesthesia. Though not our primary effect of interest, this estimand provides complementary information about how the effect of anesthesia relates to the joint effect of both anesthesia and surgery. 

%These causal contrasts elude to the existence of a four-arm $(N, O)$ randomized trial. That is patients are randomized to all the combinations of surgery and anesthesia. However, this 4-arm trial is unethical and will not be conducted in the real world. 

\subsection{Separable effects causal model}
\label{section:assumptions}

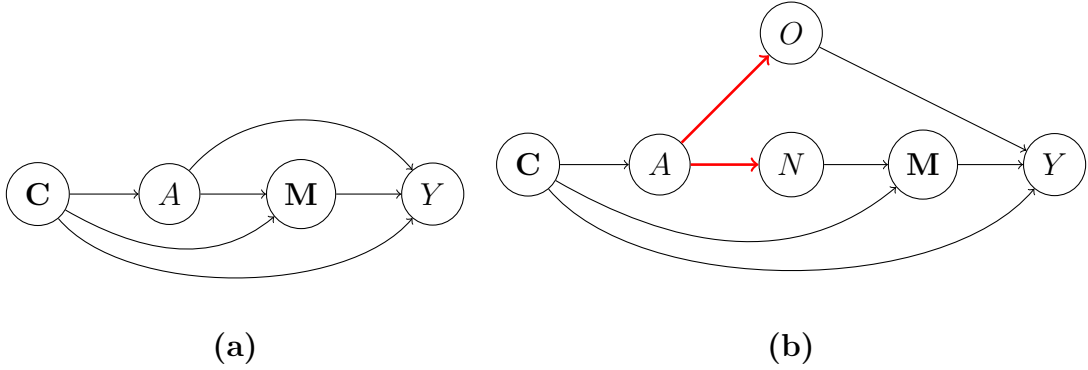
\begin{figure}[h]
    \centering
    \begin{tabular}{cc}
      
         \begin{tikzpicture}
        \begin{scope}[on grid,node distance=1.75cm]
        \node[circle, draw=black] (c) {${\bf C}$};
        \node[circle, draw=black, right= of c] (a) {$A$};
        \node[circle, draw=black, right= of a] (m) {${\bf M}$};
        \node[circle, draw=black, right= of m] (y) {$Y$};
        \end{scope}
        
        \draw[->] (c) -- (a);
        \draw[->] (a) -- (m);
        \draw[->] (m) -- (y);
        \draw[->] (a) to  [out=50,in=130, looseness=1] (y);
        \draw[->] (c) to  [out=330,in=220, looseness=1] (m);
        \draw[->] (c) to  [out=310,in=230, looseness=0.75] (y);
    \end{tikzpicture}  &
    \begin{tikzpicture}
        \begin{scope}[on grid,node distance=1.75cm]
        \node[circle, draw=black] (c) {${\bf C}$};
        \node[circle, draw=black, right= of c] (a) {$A$};
        \node[circle, draw=black, right= of a] (n) {$N$};
        \node[circle, draw=black, above= of n] (o) {$O$};
        \node[circle, draw=black, right= of n] (m) {${\bf M}$};
        \node[circle, draw=black, right= of m] (y) {$Y$};
        \end{scope}
        
        \draw[->] (c) -- (a);
        \draw[->] (n) -- (m);
        \draw[->] (m) -- (y);
        \draw[->, red, line width= 1] (a) -- (o);
        \draw[->, red, line width=1] (a) -- (n);
        \draw[->] (o) -- (y);
        \draw[->] (c) to  [out=330,in=220, looseness=1] (m);
        \draw[->] (c) to  [out=310,in=230, looseness=0.75] (y); 
    \end{tikzpicture} \\
     \textbf{(a)}  & \textbf{(b)}
    \end{tabular}
    \caption{The directed acyclic graph displays the supposed data-generating mechanisms for our scenario of interest. In (a) the observable data structure is displayed.  Here ${\bf C}$ represents the baseline covariates, $A$ condition requiring surgery with anesthesia, ${\bf M}$ harmful effect specific to surgery, and $Y$ time to diagnosis of a disruptive or internalizing behavioral disorder (DIBD) in the child.  In (b) the extended data structure is displayed highlighting the separable effects. The bold red arrows denote a deterministic function such that $A$ fully determines both surgery $N$ and anesthesia $O$.  }
    \label{fig:dags}
\end{figure}

To introduce the separable effects model and contrast our application with how it is commonly framed, we first recount the motivating thought experiment posed by \cite{pearl_2009} to provide a substantive public health motivation for mediation. Pearl imagined that a procedure is developed to remove nicotine from cigarettes, and further supposed that the entire effect of nicotine on myocardial infarction (MI) ($Y$) is through its effect on hypertension (${\bf M}$), and the effect of the remaining components of smoking have no effect on hypertension. The DAG underlying this story is shown in Figure \ref{fig:dags}.a, where the exposure $A$ is cigarette smoking, and ${\bf C}$ is taken to be the empty set for simplicity. 
He then argued that a cross-world nested counterfactual of the form $Y\{a = 1,{\bf M}(a = 0)\}$ can then be interpreted as the risk of MI if one were assigned to smoke nicotine-free cigarettes. Furthermore, he argued that the expectation of this counterfactual is identified under his structural causal model, which involves so-called ``cross-world'' counterfactual independence assumptions. \cite{robins_2010} considered an extended version of this DAG, shown in Figure \ref{fig:dags}.b, that captures the additional elements in Pearl's story. In particular, the extended graph captures the nicotine ($N$) and non-nicotine ($O$) components of cigarettes, as well as their deterministic relationship to smoking, illustrated by bold red arrows, and their relationships to hypertension and MI. Under this extended graph,
they showed that the mediation parameter $E[Y\{a = 1,{\bf M}(a = 0)\}]$ is equivalent to the parameter $E\{Y(n=0,o=1)\}$, which involves separate interventions on the nicotine and non-nicotine components of the cigarette. They further showed that this parameter is also identified by the g-formula under their cross-world assumption-free model known as the Finest Fully Randomized Causally Interpretable Structured Tree Graph (FFRCISTG). Thus, this causal parameter is ``manipulable'', meaning it could hypothetically be estimated by randomly assigning individuals to smoke nicotine-free cigarettes. Importantly, this identification is shown to hold in spite of the fact that exposure to non-nicotine components and lack of exposure to nicotine components are never observed in the same person, since they are observationally equivalent (i.e, $N=O=A$) in the (hypothetical) observed data. 

Returning to our anesthesia setting, the DAGs in Figure \ref{fig:dags} represent our model of the causal and deterministic relationships among baseline covariates (${\bf C}$), the decision to undergo surgery ($A$), the surgery itself ($N$), anesthesia ($O$), time to DIBD diagnosis ($Y$), and our collection of post-exposure variables (${\bf M}$). As in the smoking example, we have observational equivalence among $A$, $O$, and $N$, i.e., $N=O=A$. That is, the decision to undergo surgery completely determines whether the patient receives anesthesia and the surgical procedure, which we indeed observe to be the case in our data. As before, this equivalence is represented in the extended DAG in Figure \ref{fig:dags}.b by the bold red arrows from $A$ to $O$ and $A$ to $N$. However, it is still natural to conceive of intervening on anesthesia and surgery separately, at least hypothetically, such that the counterfactual $Y(n=0,o=1)$ is well-defined. For our purposes, we do not need to conceive of an intervention assigning surgery without anesthesia, though it is certainly conceptually possible. 

%the extended DAG in Figure \ref{fig:dags}.b encodes additional key assumptions about the relationships between variables. %Below are formal description of those assumptions. 
\begin{comment}
\begin{assumption}[Equivalence among $A$, $O$ and $N$] 
\label{assn:deterministic}
$N=O=A$.
    %\begin{equation} \label{assn:bold_red} \tag{A4}
    %     N(a)=O(a)=a. %\ \ w.p.1.
    %\end{equation}
\end{assumption}
\end{comment}
%Additionally, we make the following assumption that intervening to set $A$ to a certain level is equivalent to intervening to set $N$ and $O$ to that same level.
%\begin{assumption}[Equivalence of component-wise interventions]
%\label{assn:equivalence} 
%    $Y(a=k)=Y(n=k,o=k)$ for each $k\in\{0,1\}$. 
%\end{assumption}
%This assumption is essentially tautological in our case given our definitions of $A$, $N$, and $O$.
%In our case, the red arrows now indicate observational equivalence between the decision to undergo surgery, surgery itself, and anesthesia. The exchangeability assumption aligns with the finest fully randomized causally interpretable structural tree graph (FFRCISTG) counterfactual model and the extended DAG. 
%This causal model can be defined with respect to the DAG in Figure \ref{fig:dags}.a. 

In addition to the counterfactuals $Y(a)$ and $Y(n,o)$, we will also assume the existence of counterfactuals $Y(a,{\bf m})$ and $Y(n,o,{\bf m})$ that result from interventions on ${\bf M}$ in addition to $A$, $N$, and $O$, and that $Y(n,o,{\bf m})=Y(a,{\bf m})$ when $n=o=a$. 
We assume that the distribution of the observed variables and their corresponding counterfactuals follow a FFRCISTG corresponding to the DAG in Figure \ref{fig:dags}.a. In particular, the following consistency and exchangeability assumptions are implied by this model. %and suffice for nonparametric identification of our causal estimands. %FFRCISTG corresponding to the DAG in Figure \ref{fig:dags}.a.

\begin{assumption}[Consistency]
\label{assn:consistency} 
If $A=a$, then ${\bf M}(a)={\bf m}$; %If $N=n$, then $M(n)=M$. 
if $A=a$ and ${\bf M}={\bf m}$, then $Y(a,{\bf m})=Y$. %If $O=o$ and $N=n$, then $Y(n,o)=Y$. If $O=o$ and $M=m$, then $Y(o,m)=Y$.
\end{assumption}
These equations imply that the counterfactuals ${\bf M}(a)$ and $Y(a,{\bf m})$ are well defined. In particular, this requires that different surgery types or durations and modes of anesthesia delivery do not lead to different developmental outcomes or values of the intermediate variables ${\bf M}$, and likewise that different versions within the same levels of ${\bf M}$ have no effect on the developmental outcome. In our case, these may not hold precisely, but may be reasonable approximations. Alternatively, if this assumption is violated, then our estimands could be interpreted as averaging over multiple versions of the surgery and anesthesia according to their distribution in our observed data, as discussed in \cite{vanderweele2013causal}. To address potential consistency violations with respect to appendectomy vs.~cholecystectomy, we also analyze effects of these surgeries separately in Section \ref{section:real_data}. % and the Supplementary Materials. 
\begin{assumption}[Exchangeability]
\label{assn:exchangeability} 
\begin{equation*} \label{eq:exchange1} %\tag{A3.1}
    Y(a,{\bf m}) \ci A \mid {\bf C} \textrm{ for all }a \textrm{ and } {\bf m}. 
\end{equation*}
\begin{equation*} \label{eq:exchange3} %\tag{A3.3}
    {\bf M}(a)  \ci A \mid {\bf C} \textrm{ for all }a.
\end{equation*}
\begin{equation*} \label{eq:exchange2} %\tag{A3.2}
    Y(a,{\bf m}) \ci {\bf M}(a) \mid A=a, {\bf C} \textrm{ for all }a \textrm{ and } {\bf m}.
\end{equation*}
\end{assumption}
%\ref{eq:exchange1} Setting the surgery outcomes $M$ to the level $m$ there is no unobserved confounding of the effect of the mother's surgery status $A$ on the child's diagnosis of DIBD $Y$ given covariates $C$. 
%\ref{eq:exchange2} Setting the surgery outcomes $M$ to the level $m$ there is no unobserved confounding of the effect of the mother's negative surgery outcome $M$ on the child's diagnosis of DIBD $Y$ given the covariates $C$ and the mother's surgery status $A$.  
%\ref{eq:exchange3}  There is no unobserved confounding of the effect of the mother's surgery status $A$ on the outcomes related to surgery $M$ given the covariates $C$. 

%To satisfy the exchangeability assumption, we must be able to adjust for any confounders of the relationships between (a) the surgical procedure with anesthesia exposure during pregnancy and DIBD, (b) the surgical procedure with anesthesia exposure during pregnancy and the mediating variables, and (c) the mediating variables and DIBD. 
These ensure that the mother-child pairs exposed to surgery and anesthesia during pregnancy are comparable to those who did not within strata of ${\bf C}$, and that mother-child pairs are comparable across different values of the mediating variables within strata of ${\bf C}$ and exposure status. %Assumption \ref{eq:exchange1} states that when setting the surgery outcomes $M$ to a specific value $m$ there is no unobserved confounding of the effect of the mother's treatment status $A$ on the child's diagnosis of DIBD $Y$ given covariates $C$.  Assumption \ref{eq:exchange2} assumes that when setting the surgery outcomes $M$ to a specific value $m$ there is no unobserved confounding of the effect of the mother's negative surgery outcome $M$ on the child's diagnosis of DIBD $Y$ given the covariates $C$ and the mother's treatment status $A$. Finally, assumption \ref{eq:exchange3} asserts no unobserved confounding of the effect of the mother's treatment status $A$ on the outcomes related to surgery $M$ given the covariates $C$. 
These are standard no-unmeasured-confounding assumptions, and are commonly used for identification of mediated effects. However, unlike the standard mediation setting, we do not require any cross-world counterfactual independencies.

\begin{assumption}{(No direct effect of $N$ on $Y$)}
\label{assn:ny}
$Y(n, o, {\bf m}) = Y(n^*, o,{\bf m}) = Y(o, {\bf m})$ for all $n$, $n^*$, $o$, and ${\bf m}$.
\end{assumption}
\begin{assumption}{(No effect of $O$ on ${\bf M}$)}
\label{assn:om}
${\bf M}(n,o) = {\bf M}(n,o^*) = {\bf M}(n)$ for all $n$, $o$, and $o^*$.
\end{assumption}
Assumption \ref{assn:ny} reflects the absence of an arrow directly from $N$ to $Y$, and requires that there is no direct effect of surgery on the development of DIBD with respect to the intermediate variables ${\bf M}$. In other words, the entire effect of surgery is through the measured surgery-related outcomes in ${\bf M}$. Assumption \ref{assn:om} reflects the absence of an arrow from $O$ to ${\bf M}$, and requires that there is no direct effect of anesthesia on the measured surgery-related outcomes in ${\bf M}$. These are analogous to other exclusion restrictions that commonly arise in causal inference, such as in the instrumental variable and front-door adjustment settings. %criterion that the effect of an exposure is completely mediated by some intermediate observed variables \citep{pearl1995causal}, as well as the assumption in the instrumental variable setting that there is no direct effect of the instrument on the outcome with respect to the exposure of interest. 
We develop sensitivity analyses for violations of Assumptions \ref{assn:ny} and \ref{assn:om} in Section \ref{section:sensitivity} and Appendix \ref{a:sens_om}, respectively. 
%Assumptions \ref{assn:ny} and \ref{assn:om} can also be seen in the SWIG displayed in Figure \ref{fig:dags}.c where the absence of backdoor paths helps create d-separation implies these independencies. 

%[MOVING THIS HERE SINCE IT TOUCHES ON MULTIPLE ASSUMPTIONS AND SEEMS BETTER TO DISCUSS AFTER INTRODUCING THEM ALL]
One important limitation is that we do not include the underlying inflammation (appendicitis or cholecystitis) necessitating surgery in ${\bf C}$. Therefore, we need to assume that the underlying inflammation does not have a direct effect on DIBD diagnosis with respect to surgery and anesthesia in order for our exchangeability assumption to hold. The reason for this is that controlling for the underlying inflammation in ${\bf C}$ would yield yet another extreme positivity violation, as it is observationally equivalent to surgery and anesthesia. 
%In some sense, this also rules out other variables that are equivalent to the exposure $A$ and have a direct effect on our outcome, as these could additionally be considered as components of treatment. For example, we assume that the underlying inflammation requiring surgery has no direct effect on our outcome. Such a violation arising from a deterministic form of confounding by indication is challenging, and we do not address it here. 
In principle, with the right data, one could consider the underlying inflammation to be part of the component $N$, and augment ${\bf M}$ to include variables that completely mediate the effect of both surgery and the underlying inflammation on $Y$ without being affected by $O$. Alternatively, the underlying inflammation could be considered as a third component of the exposure; \cite{robins_2010} discuss alternative identification results for additional components. In the past few years, acute appendicitis has started to be treated with antibiotics, so it is possible that some mothers with appendicitis may be avoiding surgery, in which case the underlying inflammation would no longer be a third component as its relationship to the surgery and anesthesia exposures would no longer be deterministic. Thus, given access to more recent data, one could potentially evade the problem of unmeasured confounding due to the inflammation by restricting one's analysis to mothers with acute appendicitis and contrasting those treated with surgery and those treated with antibiotics. 
For our purposes, we will assume that no such effect of the underlying inflammation exists, while acknowledging that this may not hold, which could yield some residual confounding bias \citep{ende2024behavioural}. However, we suspect that any residual confounding bias incurred by the underlying inflammation is weaker than that which would arise by ignoring the effect of surgery.

%The positivity assumptions below do not conflict with the extreme positivty violation we are trying to overcome. Here we are describing restrictions based on the observable data $A$, rather than the extended components $O$ and $N$ which create the extreme violation. 
In addition to the above assumptions implied by the FFRCISTG corresponding to the DAGs in Figure \ref{fig:dags}, we also require a positivity assumption. However, as we have discussed, this is not a positivity assumption with respect to the anesthesia exposure, $O$, as we know this is violated when conditioning on surgery, $N$. Instead, we require a positivity assumption with respect to the joint exposure, i.e., the decision to undergo surgery $A$, as well as a positivity assumption on ${\bf M}$ among the exposed.
\begin{assumption}[Positivity]
\label{assn:positivity}
\begin{equation*} \label{eq:pos1} %\tag{A1.1}
      \mathrm{Pr}(A = a \mid {\bf c})>0 \textrm{ for all } a \in \{0,1\} \textrm{ and } {\bf c} \textrm{ such that } \mathrm{Pr}({\bf C}={\bf c}) > 0. 
\end{equation*}
\begin{equation*} \label{eq:pos2} %\tag{A1.2}
      \mathrm{Pr}({\bf M}={\bf m} \mid A = 1, {\bf c}) >0 \textrm{ for all } {\bf m} \textrm{ and } {\bf c} \textrm{ such that } \mathrm{Pr}({\bf C}={\bf c}) > 0. 
\end{equation*}
\end{assumption}
%Two positivity assumptions are required: one for the treatment and covariates and another for the mediator under exposure to anesthesia, surgery and covariates. 
%This assumption pertains to observable data, and does not conflict with the extreme positivity violation we are trying to overcome. 
The first part of the assumption %(\ref{eq:pos1}) 
ensures that every mother in a stratum of measured covariates has a non-zero chance of receiving surgery during pregnancy. The second part %(\ref{eq:pos2}) 
ensures that for any mother who undergoes surgery and anesthesia, any value of the mediator can be observed with a non-zero probability. Importantly, we only require positivity for those exposed to anesthesia and surgery ($A=1$), not for the general case ($A = a \in \{0,1\}$), since the mediator ${\bf M}$ includes outcomes that cannot occur without surgery.  % since the even within the observable data; we only require it for the group exposed to anesthesia and surgery ($A=1$). Requiring positivity for the general case ($A=a\in\{0,1\}$) would violate this assumption because the mediator $M$ includes indicators of negative outcomes related to surgery, which cannot be observed in patients who did not undergo surgery. 

Lastly, since we have a time-to-event outcome subject to right censoring, we do not observe the time to diagnosis $Y$ directly, but rather $\tilde{Y}\equiv\min\{Y,T_{\textrm{censor}}\}$ and $\Delta\equiv I(Y\leq T_{\textrm{censor}})$, where $T_{\textrm{censor}}$ is the censoring time. We will use the following standard non-informative censoring assumption. 
\begin{assumption}[Non-informative censoring] \label{assn:censoring}
        $Y \ci T_{\textrm{censor}} \mid {\bf M}, A, {\bf C}$.
\end{assumption}
Under this assumption, censoring is accounted for in the usual manner in survival analysis. Therefore, in the following, we will regard quantities that are in terms of the distribution of $Y$ as being effectively nonparametrically identified even though this is technically not observed for all subjects.

One critique of the separable effects framework as it is commonly used is that for a given intermediate variable ${\bf M}$, one must posit that there exist some separable components $N$ and $O$ of the exposure $A$ that satisfy the DAG in Figure \ref{fig:dags}.b. Ideally, one would come up with a compelling story that specifies what $N$ and $O$ are that justifies this DAG and makes the estimands readily interpretable. However, this can be difficult to conceive of, and $N$ and $O$ are at times left vaguely-defined or completely unspecified, leading to causal estimands with unclear interpretations. In our case, however, the set-up of the model is inverted. We begin with precisely-defined variables $N$ and $O$, and then come up with a set of variables ${\bf M}$ that justifies the DAG in Figure \ref{fig:dags}.b. While in typical separable effects applications, ${\bf M}$ is usually specified to be a single variable of interest, in our case we take it to be as many variables as are needed to plausibly satisfy the exclusion restrictions in the extended DAG. Furthermore, since our $N$ and $O$ are precisely defined, our causal estimands have clear interpretations. 

\section{Nonparametric Identification} \label{section:identification_est}

%\subsection{Identification results}
%\label{section:identification}
Identification of $\psi^{\textrm{anesthesia}}$ and $\psi^{\textrm{surgery}}$ follows directly from identification of  $\psi_a \equiv \mathrm{Pr}\{ Y(n=a, o=a) \leq t \}$ for $a\in\{0,1\}$ and $\psi_{01} \equiv \mathrm{Pr}\{ Y(n=0, o=1) \leq t \}$. The former is identified by the backdoor adjustment formula/g-formula: 
\begin{align*}
    \psi_a &\equiv \mathrm{Pr}\{Y(n=a, o=a)\leq t\} =\mathrm{Pr}\{Y(a)\leq t\}  \\
    &= \sum_{{\bf m},{\bf c}} \mathrm{Pr}(Y \leq t \mid {\bf M}={\bf m}, A=a, {\bf C}={\bf c} ) f({\bf m} \mid A=a, {\bf C}={\bf c}) f({\bf c}) \equiv \Psi_{a}(P),
\end{align*}
where we present results in terms of discrete ${\bf M}$ and ${\bf C}$ for ease of notation, and throughout $P$ will be taken to be the joint distribution of $\{{\bf C},A,{\bf M},Y\}$. 
%When $a\not=a'$, namely $n=0, o=1$ seen in $\mathrm{Pr}\{ Y(n=0, o=1) \leq t \}$, then 
Identification of $\psi_{01}$ is more challenging due to the events $N=0$ and $O=1$ not being observed simultaneously in our data. %However, identification is still possible. 
Following \cite{robins_2010}, under the assumptions presented in the previous section, which follow from the FFRCISTG corresponding to the extended DAG in Figure \ref{fig:dags}.b, we have
\begin{align*}
    \psi_{a,a^*} &\equiv \mathrm{Pr}\{Y(n=a, o=a^*)\leq t\}\\ 
    &= \sum_{{\bf m},{\bf c}} \mathrm{Pr}(Y\leq t \mid {\bf M}={\bf m}, A=a^*, {\bf C}={\bf c} )f({\bf m}\mid A=a,{\bf C}={\bf c})f({\bf c}) \equiv \Psi_{a,a^*}(P)
\end{align*}
for $\{a,a^*\}\in\{\{0,0\},\{0,1\},\{1,1\}\}$, where $\psi_{a,a}\equiv\psi_a$, and we can see that $\Psi_{a,a}(P)$ agrees with $\Psi_{a}(P)$ above. 
%This shows that $\mathrm{Pr}\{ Y(n=0, o=1) \leq t  \}$ is a well-defined function of the observed data distribution. 
Thus, our causal contrasts are identified by 
$\psi^{\textrm{joint}} = \Psi_{1}(P)/\Psi_0(P)$, $\psi^{\textrm{anesthesia}} = \Psi_{01}(P)/\Psi_0(P)$, and $\psi^{\textrm{surgery}} = \Psi_1(P)/\Psi_{01}(P)$. 
    %The above estimands only requires the observable data which has the structure shown in Figure \ref{fig:dags}.a.  

%\subsection{Relationship to mediation}
%\label{section:mediation} 
%The identification results presented in Section \ref{section:identification} can be compared to established mediation techniques as described by \citep{robins_2010}. 
\cite{robins_2010} also showed that under the extended DAG in Figure \ref{fig:dags}.b, $\psi^{\textrm{joint}}$, $\psi^{\textrm{anesthesia}}$, and $\psi^{\textrm{surgery}}$ are equivalent to the total effect, 
%In the mediation literature, it is common to decompose the total effect of interventions, such as the joint effect of surgery and anesthesia, into two distinct components: the 
pure natural direct effect (PNDE), and the total natural indirect effect (TNIE), respectively (though we focus on relative risks at time $t$ rather than mean differences). Thus, the terms in the decomposition in (\ref{eq:decomp}) are equivalent to the terms in the standard mediation decomposition:
\begin{align*}
    \frac{\mathrm{Pr}\{Y(a=1) \leq t\}}{\mathrm{Pr}\{Y(a=0) \leq t\}} %&= \frac{\mathrm{Pr}\{Y(A=1, M(A=1)) \leq t \}}{\mathrm{Pr}\{Y(A=0, M(A=0)) \leq t \}} \\
    &= \underbrace{\frac{\mathrm{Pr}[Y\{a=1, {\bf M}(a=1)\} \leq t ]  }{\mathrm{Pr}[Y\{a=1, {\bf M}(a=0)\} \leq t ] }}_{\text{TNIE}} \times  \underbrace{\frac{\mathrm{Pr}[Y\{a=1, {\bf M}(a=0)\} \leq t ] }{\mathrm{Pr}[Y\{a=0, {\bf M}(a=0)\} \leq t ] }}_{\text{PNDE}}. 
\end{align*}
In fact, $\Psi(P)$ is often referred to as the ``mediation formula'', and was first shown by \cite{pearl2001direct} to identify $E[Y\{a=1, {\bf M}(a=0)\}]$ under certain assumptions. 
Analogously to the presence of the counterfactual $Y(n=0,o=1)$ in our estimands, the TNIE and PNDE both involve the cross-world counterfactual $Y\{a=1, {\bf M}(a=0)\}$, which is not directly observable in our real data. The primary difference is that the parameter $E\{Y(n=0,o=1)\}$ is still manipulable, meaning that despite $Y(n=0,o=1)$ never being observed in our data, it could conceivably arise from a joint intervention setting $N$ to 0 and $O$ to 1, whereas there is no such single intervention that would (with certainty) yield a realization of the cross-world counterfactual $Y\{a=1, {\bf M}(a=0)\}$. In order to identify the TNIE and the PNDE without resorting to separable effects, an additional, untestable cross-world counterfactual independence assumption is required. %; however, the decomposition of $A$ into $N$ and $O$ and the exclusion restriction Assumptions \ref{assn:ny} and \ref{assn:om} are not needed.
Since we have the equivalences $\psi^{\textrm{anesthesia}}=\mathrm{TNIE}$ and $\psi^{\textrm{surgery}}=\mathrm{PNDE}$ under the separable effects model, they are both identified by the same functionals $\Psi_{01}(P)/\Psi_0(P)$ and $\Psi_{1}(P)/\Psi_{01}(P)$. The result is that despite mediation not being the focus of our study, we may use existing estimation methodology from mediation analysis to estimate $\psi^{\textrm{anesthesia}}$ and $\psi^{\textrm{surgery}}$.

We could have alternatively used an effect decomposition that involved multiplying and dividing by the term $\mathrm{Pr}\{Y(n=1, o=0)\leq t\}$. Under our separable effects model, the terms in such a decomposition would have instead been equivalent to those in the decomposition of the total effect into the \emph{total} natural direct effect and the \emph{pure} natural indirect effect, which involve the term $\mathrm{Pr}[Y\{a=0,{\bf M}(a=1)\}\leq t]$. However, as we have discussed, we are not interested in studying the effect of anesthesia in the presence of surgery, %in part because of positivity violations incurred by the fact that some events in ${\bf M}$ can only occur among the exposed mothers. Thus, 
so we will not consider this alternative decomposition.

\section{Relationships to Standard Identification Results} \label{section:connections}
The basic intuition behind why it is possible to isolate the effects of surgery and anesthesia under a separable effects model despite the extreme positivity violation is that we can use the intermediate variables to ``block'' the effect of surgery without ``blocking'' that of anesthesia. To help formalize this intuition a bit more, in this section we make connections to two familiar identification results: back-door and front-door adjustment. We also discuss potential generalizations to standard, less extreme forms of positivity violation. 

% Moving this section here, as the back-door formula is more familiar
\subsection{Relationship to back-door adjustment}
\label{section:back-door}
%We can frame the idea of separable effect in the context of the back door criterion. The main goal of the back door criterion is to identify a minimally sufficient adjustment to block all backdoor paths. Using the causal structure depicted in Figure \ref{fig:dags}.b the main goal is to understand 
We begin by considering the parameter $\mathrm{Pr}\{Y(o=1) \leq t\}$ capturing the effect of $O$ on $Y$. For simplicity, we will consider ${\bf C}$ to be the empty set; allowing for nonempty ${\bf C}$ simply amounts to conditioning on ${\bf C}$ throughout. The set $\{N\}$ (or likewise, $\{A\}$) is a sufficient adjustment set satisfying the back-door criterion, as seen in Figure  \ref{fig:dags}.b. However, due to the deterministic relationship between $O$ and $N$, we are not actually able to adjust for $N$. Instead, we can consider an alternative sufficient adjustment set, $\{{\bf M}\}$, which also satisfies the back-door criterion. Assuming positivity for $O$ holds with respect to ${\bf M}$, we have $\mathrm{Pr}\{Y(o=1) \leq t\} = E\{ \mathrm{Pr}( Y \leq t \mid O=1, {\bf M}) \}$.

Now suppose that we can conceive of the intervention on $O$ in such a way that it does not affect the value of $N$, i.e., $N(o)=N$. This yields the decomposition
\begin{align*}
    \mathrm{Pr}\{Y(o=1) \leq t\} &= \mathrm{Pr}\{Y(o=1,N) \leq t\} \\
    &= \mathrm{Pr}\{Y(o=1, n=0) \leq t\} \mathrm{Pr}(N=0) + \mathrm{Pr}\{Y(o=1, n=1) \leq t\} \mathrm{Pr}(N=1).
\end{align*}
The counterfactual term in the second summand, $\mathrm{Pr}\{Y(o=1, n=1) \leq t\} = \mathrm{Pr}\{Y(a=1)\leq t\}$ is identified by the back-door formula/g-formula to be $\mathrm{Pr}(Y \leq t \mid A=1)$.  Thus, our desired estimand  $\mathrm{Pr}\{Y(o=1, n=0) \leq t\}$ can be solved for in the above equation in terms of the two quantities that we have just shown are identified by the back-door formula, and the result is readily seen to coincide with the identification formula $\Psi_{01}(P)$ given above. %, as we show in Appendix \ref{a:fdc}.

%The back door criterion helps give intuition behind our approach.  In a sense, the effect of surgery $N$ will be blocked when we adjust for the post-exposure variable $M$, letting us isolate the effect of anesthesia $O$ on developmental outcomes $Y$. Therefore, $M$ becomes a tool to help learn more about $O$. The key is to adequately identify these post-exposure variables that capture the entire effect of $N$, a task that can be challenging.  

\subsection{Relationship to front-door adjustment}
\label{section:front-door}
\begin{figure}[h]
    \centering
    \begin{tikzpicture}
        \begin{scope}[on grid,node distance=1.5cm]
        %\draw[help lines](-1,-1) grid(4,4);%\leq -- comment this line to hide the grid
     %   \node[circle, draw=black] (c) {$C$};
        \node[circle, draw=black] (a) {$A$};
        \node[circle, draw=black, right= of a] (m) {${\bf M}$};
        \node[circle, draw=black, fill=lightgray, above= of m] (o) {$U$};
        \node[circle, draw=black, right= of m] (y) {$Y$};
        \end{scope}
        
        \draw[->] (a) -- (m);
        \draw[->] (m) -- (y);
        \draw[->] (o) -- (y);
        \draw[->] (o) -- (a);
       % \draw[->] (c) -- (a);
       % \draw[->] (c) to  [out=330,in=220, looseness=1] (m);
       % \draw[->] (c) to  [out=310,in=230, looseness=0.75] (y);
    \end{tikzpicture}
    \caption{Pearl's front-door criterion DAG, where $U$ represents an unmeasured confounder.}
    \label{fig:fdc_dag}
\end{figure}
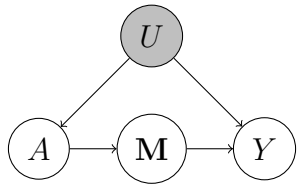

We can also re-derive the identification formula $\Psi_{01}(P)$ via the front-door formula in a similar fashion, but by first beginning with the parameter $\mathrm{Pr}\{Y(n=0) \leq t\}$ capturing the effect of $N$ on $Y$. 
%We can frame the idea of separable effect in the context of the front door criterion. Judea Pearl's
The front-door criterion makes it possible to identify the total effect of an exposure on an outcome in the presence of unmeasured confounding \citep{pearl1995causal}. The typical DAG representing this setting is shown in Figure \ref{fig:fdc_dag}. 
Briefly, the front-door criterion requires that a set of variables ${\bf M}$ completely mediate the effect of the exposure on the outcome, and that there is no unmeasured confounding of the exposure--mediator or mediator--outcome relationships. %\citep{pearl1995causal, pearl_2009}.  Formally the assumptions are: (i) $M$ intercepts all directed paths from $A$ to $Y$, (ii) there is no back-door pathway between $A$ and $M$, and (iii) every back-door path between $M$ and $Y$ is blocked by $A$.  
Under these assumptions, $\mathrm{Pr}\{Y(a)\leq t\}$ is nonparametrically identified by
\begin{equation*} \label{eq:fdc}
    \sum_{{\bf m}}\mathrm{Pr}({\bf M}={\bf m} \mid A=a) \sum_{a'} \mathrm{Pr}( Y\leq t \mid A=a', {\bf M}={\bf m})\mathrm{Pr}(A=a').
\end{equation*}

Now consider modifying the extended DAG in Figure \ref{fig:dags}.b by collapsing $A$ and $O$ into the same node (which in fact yields an equivalent model \citep{robins_2010}). Additionally, we will once again take ${\bf C}$ to be the empty set for simplicity. Then the resulting DAG is isomorphic to the DAG in Figure \ref{fig:fdc_dag}, where $N$ maps to $A$, the combined $A$/$O$ node maps to the unobserved confounding variable $U$, and ${\bf M}$ and $Y$ map to themselves. %, depicted in Figure \ref{fig:dags}.b, 
Although $O$ is observed in our case, we are unable to adjust for it due to the extreme positivity violation, so it plays a similar role to that of an unmeasured confounder. Thus, the front-door criterion is satisfied for purposes of identifying the effect of $N$ on $Y$, and the front-door formula for $\mathrm{Pr}\{Y(n=0)\leq t\}$ is $\sum_{{\bf m}}\mathrm{Pr}({\bf M}={\bf m} \mid N=0) \sum_{n} \mathrm{Pr}( Y\leq t \mid N=n, {\bf M}={\bf m})\mathrm{Pr}(N=n)$. 
Now suppose that we can conceive of an intervention on $N$ in such a way that it has no effect on the value of $O$, i.e., $N(o)=N$. Then we have an analogous decomposition to the one shown above:
\begin{align*} \label{eq:fdc_firstproof}
    \mathrm{Pr}\{Y(n=0) \leq t\} &= \mathrm{Pr}\{Y(n=0,O) \leq t\} \\
    &= \mathrm{Pr}\{Y(n=0, o=0) \leq t\} \mathrm{Pr}(O=0) + \mathrm{Pr}\{Y(n=0, o=1) \leq t\} \mathrm{Pr}(O=1).
\end{align*}
Since the term $\mathrm{Pr}\{Y(n=0, o=0) \leq t\}=\mathrm{Pr}\{Y(a=0) \leq t\}$ is identified by the back-door formula/g-formula, we can solve the above equation for $\mathrm{Pr}\{Y(n=0,o=1)\leq t\}$ in terms of the identified quantities $\mathrm{Pr}\{Y(n=0)<t\}$ and $\mathrm{Pr}\{Y(a=0)<t\}$, using the front-door and back-door formulas, respectively. Once again, the result can easily be seen to coincide with the identification formula $\Psi_{01}(P)$, as we show in Appendix \ref{a:fdc}. 
Interestingly, this means that under a separable effects model, the front-door formula can be re-derived entirely using the back-door formula.

\subsection{Generalizing to less extreme forms of positivity violations}
\label{sec:less-extreme}
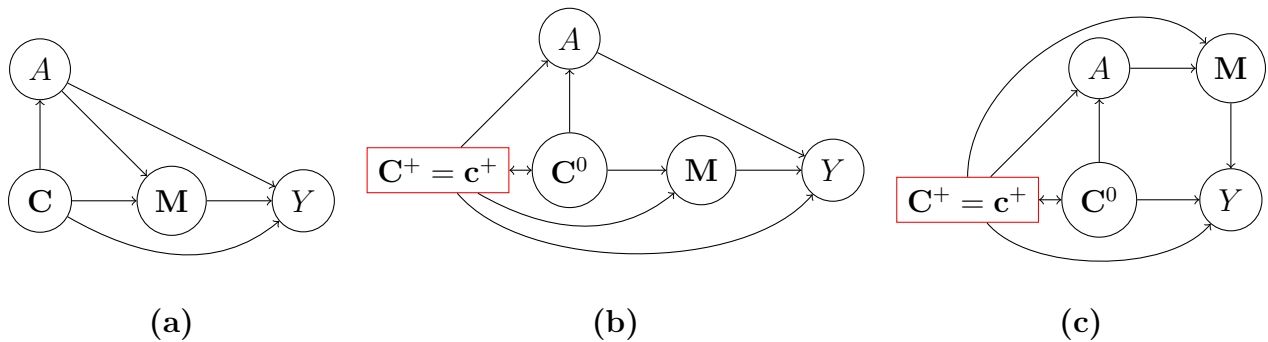
\begin{figure}[h]
    \centering
    \begin{tabular}{ccc}
    \begin{tikzpicture}
        \begin{scope}[on grid,node distance=1.75cm]
        %\node[rectangle, draw=black] (c+) {$C^+=c^+$};
        \node[circle, draw=black] (c) {${\bf C}$};
%        \node[circle, draw=black, right= of a] (n) {$N$};
        \node[circle, draw=black, above= of c] (a) {$A$};
        \node[circle, draw=black, right= of c] (m) {${\bf M}$};
        \node[circle, draw=black, right= of m] (y) {$Y$};
        \end{scope}
        
%        \draw[->] (c+) -- (c0);
        \draw[->] (c) -- (m);
        \draw[->] (m) -- (y);
        \draw[->] (c) -- (a);
%        \draw[->, red, line width=1] (a) -- (n);
        \draw[->] (a) -- (m);
        \draw[->] (a) -- (y);
        \draw[->] (c) to [out=330,in=220, looseness=1] (y);
%        \draw[->] (c+) to  [out=330,in=220, looseness=1] (m);
%        \draw[->] (c+) to  [out=310,in=230, looseness=0.75] (y); 
    \end{tikzpicture}   &
    \begin{tikzpicture}
        \begin{scope}[on grid,node distance=1.75cm]
        \node[rectangle, draw=red] (c+) {${\bf C}^+={\bf c}^+$};
        \node[circle, draw=black, right= of c+] (c0) {${\bf C}^0$};
%        \node[circle, draw=black, right= of a] (n) {$N$};
        \node[circle, draw=black, above= of c0] (a) {$A$};
        \node[circle, draw=black, right= of c0] (m) {${\bf M}$};
        \node[circle, draw=black, right= of m] (y) {$Y$};
        \end{scope}
        
        \draw[<->] (c+) -- (c0);
        \draw[->] (c+) -- (a);
        \draw[->] (c0) -- (m);
        \draw[->] (m) -- (y);
        \draw[->] (c0) -- (a);
%        \draw[->, red, line width=1] (a) -- (n);
        \draw[->] (a) -- (y);
        \draw[->] (c+) to  [out=330,in=220, looseness=1] (m);
        \draw[->] (c+) to  [out=310,in=230, looseness=0.75] (y); 
    \end{tikzpicture}  &
    \begin{tikzpicture}
        \begin{scope}[on grid,node distance=1.75cm]
        \node[rectangle, draw=red] (c+) {${\bf C}^+={\bf c}^+$};
        \node[circle, draw=black, right= of c+] (c0) {${\bf C}^0$};
%        \node[circle, draw=black, right= of a] (n) {$N$};
        \node[circle, draw=black, above= of c0] (a) {$A$};
        \node[circle, draw=black, right= of a] (m) {${\bf M}$};
        \node[circle, draw=black, right= of c0] (y) {$Y$};
        \end{scope}
        
        \draw[<->] (c+) -- (c0);
        \draw[->] (c+) -- (a);
        \draw[->] (c0) -- (y);
        \draw[->] (m) -- (y);
        \draw[->] (c0) -- (a);
%        \draw[->, red, line width=1] (a) -- (n);
        \draw[->] (a) -- (m);
        \draw[->] (c+) to  [out=90,in=135, looseness=1] (m);
        \draw[->] (c+) to  [out=310,in=230, looseness=0.75] (y); 
    \end{tikzpicture} \\
     \textbf{(a)}  & \textbf{(b)}   &   \textbf{(c)}
    \end{tabular}
    \caption{(a) The causal DAG in the general positivity violation case. (b) A causal DAG within the stratum ${\bf C}^+={\bf c}^+$ in which the back-door formula can be used with sufficient adjustment set ${\bf M}$. (c) A causal DAG within the stratum ${\bf C}^+={\bf c}^+$ in which the front-door formula can be used.}
    \label{fig:less-extreme-dag}
    \end{figure}

Thus far, we have only discussed an extreme form of positivity violation. However, it stands to reason that if this approach works for the extreme case, then these concepts should at least in principle be adaptable to less severe cases. Indeed, this generalization is relatively straightforward. Consider the DAG in Figure \ref{fig:less-extreme-dag}.a and a stratum ${\bf C}={\bf c}$ for which positivity fails for exposure level $a$, i.e., $\mathrm{Pr}(A=a\mid {\bf C}={\bf c})=0$ (otherwise, identification can follow as usual from the g-formula assuming no unobserved confounding). 
Then we can partition ${\bf C}$ into a set of variables ${\bf C}^+\subset {\bf C}$ for which $\mathrm{Pr}(A=a\mid {\bf C}^+={\bf c}^+)>0$, where ${\bf c}^+$ is the corresponding subvector of ${\bf c}$, and the remaining variables ${\bf C}^0\equiv {\bf C}\setminus {\bf C}^+$. Conditional on ${\bf C}^+={\bf c}^+$, the relationship between ${\bf C}^0$ and $A$ is at least partially deterministic, since when ${\bf C}={\bf c}$, we must have $A\neq a$ with probability one. Thus, we cannot directly use the back-door formula to identify $E\{Y(a)\}$ using ${\bf C}=\{{\bf C}^+,{\bf C}^0\}$. 

Suppose that conditional on ${\bf C}^+={\bf c}^+$, there is no direct effect of ${\bf C}^0$ on $Y$ with respect to ${\bf M}$, and that there is no effect of $A$ on ${\bf M}$ as shown in the DAG in Figure \ref{fig:less-extreme-dag}.b. These exclusion restrictions need not hold within other strata of ${\bf C}^+$, but must hold conditional on ${\bf C}^+={\bf c}^+$. Then within stratum ${\bf C}^+={\bf c}^+$, ${\bf M}$ is a sufficient adjustment set, and we may use the back-door formula to adjust for ${\bf M}$ and identify $E\{Y(a)\mid {\bf C}^+={\bf c}^+\}$ by $E\{E(Y\mid {\bf M}, A=a, {\bf C}^+={\bf c}^+)\mid {\bf C}^+={\bf c}^+\}$ as in Section \ref{section:back-door}.

Alternatively, suppose that conditional on ${\bf C}^+={\bf c}^+$, there is no direct effect of $A$ on $Y$ with respect to ${\bf M}$, and that there is no effect of ${\bf C}^0$ on ${\bf M}$ as shown in the DAG in Figure \ref{fig:less-extreme-dag}.c. Again, these exclusion restrictions need not hold within other strata of ${\bf C}^+$, but must hold conditional on ${\bf C}^+={\bf c}^+$. Then within stratum ${\bf C}^+={\bf c}^+$, we may use the front-door formula as in Section \ref{section:front-door} to identify $E\{Y(a)\mid {\bf C}^+={\bf c}^+\}$ by \[\sum_{{\bf m}}\mathrm{Pr}({\bf M}={\bf m} \mid A=a, {\bf C}^+={\bf c}^+) \sum_{a'=0}^1 E( Y \mid {\bf M}={\bf m}, A=a', {\bf C}^+={\bf c}^+)\mathrm{Pr}(A=a'\mid {\bf C}^+={\bf c}^+).\]

More closely analogous to our case, if one were also interested in an intervention on ${\bf C}^0$ and preferred to identify the more refined estimand $E\{Y(a,{\bf c}^0)\mid {\bf C}^+={\bf c}^+\}$, then one could potentially use the same approach outlined in the previous subsections where the law of iterated expectations is applied to $E\{Y(a,{\bf C}^0)\mid {\bf C}^+={\bf c}^+\}$ and the result is solved for $E\{Y(a,{\bf c}^0)\mid {\bf C}^+={\bf c}^+\}$, having already identified $E\{Y(a)\mid {\bf C}^+={\bf c}^+\}$ via one of the two approaches we have just described. However, this would further require that $E\{Y(a,{\bf c}')\mid {\bf C}^+={\bf c}^+\}$ is identifiable for all ${\bf c}'\neq {\bf c}^0$, which may not be the case if positivity is also violated for these other values.

The practicality of this approach will, of course, depend on the scenario. As in our case, one would need to be willing to believe the exclusion restrictions encoded in the DAGs in Figures \ref{fig:less-extreme-dag}.b or \ref{fig:less-extreme-dag}.c and to have measured the relevant variables ${\bf M}$. Furthermore, one would need to identify all strata with positivity violations in order to recover identification of the marginal parameter $E\{Y(a)\}$, which may be challenging in practice, particularly when ${\bf C}$ is not low-dimensional. 
%The key is to identify subsets of the sample, based on covariate information, where positivity violations occur. This allows us to create two distinct subsets: one with an extreme positivity violation and another with no violation.
%In the subset without violations, estimation is straightforward. For the subset with an extreme violation, we can apply the approach outlined in this paper. This is only feasible if post-exposure information is available, where the total effect of $N$ goes through $M$ to influence the outcome $Y$. Once estimates have been obtained for both subsets, they can be combined using an averaging method. %If there does not exist variables that satisfiy the conditions of $M$ for the whole stratum but rather just a subset of the stratum then you can redefined your group of interest and preform your analysis within the stratum where $M$ is collected.   

\section{Estimation} \label{section:estimation}
As previously mentioned, since our identification formulas for $\Psi_{a,a^*}(P)$ are the same as those for mediated effects with a survival outcome, we may used methodology previously developed for the latter setting. We adapt a simple substitution estimator proposed by \cite{vanderweele2011causal} for the survival function under a Cox proportional hazards model for the outcome to allow for multiple mediators. 
%We model the conditional distribution of $Y$ given $A$, $M$, and $C$ usin the Cox proportional hazards model, and maximum likelihood estimation for the remainder of the nuisance parameters. 
Specifically, we use the model $\lambda_Y(y \mid a,{\bf m},{\bf c};\theta) = \lambda_0(y)\exp(\theta_1 a + \theta_2'{\bf m} + \theta_3' a{\bf m}+ \theta'_4 {\bf c})$.
For the joint distribution of all elements in ${\bf M}$, we model each term in the factorization
\begin{align*}
    \mathrm{Pr}({\bf m} \mid a , {\bf c}) = \mathrm{Pr}(m_1 \mid a , {\bf c})\prod_{j=2}^k \mathrm{Pr}(m_j \mid a,{\bf c},m_1,\ldots,m_{j-1})
\end{align*}
individually, while encoding the knowledge that many elements of ${\bf M}$ can only occur in the presence of the exposure, i.e., $A=0$ implies $M_j=0$ for all $j=1,\ldots,\ell$. It is this positivity violation that prevents the identification formula from being well defined with $a=1$ and $a^*=0$, and hence why we cannot estimate the effect of anesthesia in the presence of surgery. We order the elements of ${\bf M}$ such that the first $\ell$ elements have the constraint that $A=0$ implies $M_j=0$. Thus, we only need to estimate their conditional distribution given $A=1$; conditional on $A=0$, the estimated probabilities are assigned to be zero. We fit logistic regression models with main effect terms on data from only the exposed. The events corresponding to the remaining elements of ${\bf M}$ can occur with or without the exposure. We use logistic regression models for each of these variables using data from both exposed and unexposed mothers. Due to the constraint among the first $\ell$ variables, there are no instances in which both $A=0$ and $M_j=1$ for $j\leq \ell$, so for all such $j$ we include interaction terms between $A$ and $M_j$, but no main effect terms for $M_j$, so as to avoid models not being identified. 

These model estimates are then substituted into the identifying functional to produce a substitution estimator of the risk of diagnosis by time $t$ for each $\{a,a^*\}\in\{\{0,0\},\{0,1\},\{1,1\}\}$:
%We will focus our analysis on analyzing the survival curve $S_{a^*a} = 1-\hat{\Psi}_{a^*a}$ 
%for each counterfactual at a specific time $t$:
\begin{equation}
    \hat{\Psi}_{a,a^*} \equiv \Psi_{a,a^*}(\hat{P}) = 1-\frac{1}{n}\sum_{i=1}^n\sum_{{\bf m} \in \{0,1\}^k} \exp\left\{ - \hat{\Lambda}_0(t) \exp\left(\hat{\theta}_1a + \hat{\theta}_2'{\bf m} + \hat{\theta}_3' a{\bf m} + \hat{\theta}_4'{\bf C}_i\right) \right\} \widehat{\mathrm{Pr}}({\bf m}\mid a^* , {\bf C}_i),
\end{equation}
where $\hat{\Lambda}_0(t)$ is the Nelson--Aalen estimator of the cumulative baseline hazard at time $t$. 
%To estimate the standard error of the estimate of 95\% confidence intervals the standard practice is to use the bootstrapping method. However, 
We used the Bayesian bootstrap \citep{rubin1981bayesian} for standard error and confidence interval estimation in order to avoid instability due to the relatively rare exposure. %, which uses a Dirichlet weighting scheme, allowing for bootstrap estimates to be smoother functionals of the empirical distribution compared to non-parametric bootstrapping due to the continuous nature of the weighting scheme. In addition, using the Dirichlet weighting scheme prevents observation from getting a zero weight thus helping eliminate corner cases. This is essential in applications where either the outcome or the exposure is rare. 
%We provide sample code for the estimation procedure applied to the lower-dimensional simulation setting in Appendix \ref{a:pseudo_code}. 

\section{Sensitivity Analysis} \label{section:sensitivity} 
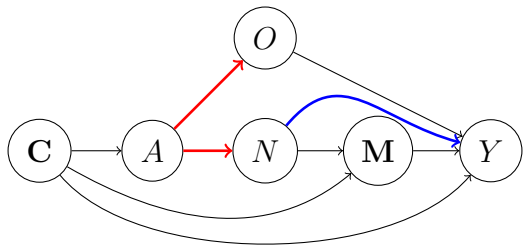
\begin{figure}[h]
    \centering
        \begin{tikzpicture}
        \begin{scope}[on grid,node distance=1.5cm]
        %\draw[help lines](-1,-1) grid(4,4);%\leq -- comment this line to hide the grid
        \node[circle, draw=black] (c) {${\bf C}$};
        \node[circle, draw=black, right= of c] (a) {$A$};
        \node[circle, draw=black, right= of a] (n) {$N$};
        \node[circle, draw=black, above= of n] (o) {$O$};
        \node[circle, draw=black, right= of n] (m) {${\bf M}$};
        \node[circle, draw=black, right= of m] (y) {$Y$};
        \end{scope}
        
        \draw[->] (c) -- (a);
        \draw[->] (n) -- (m);
        \draw[->] (m) -- (y);
        \draw[->, red, line width= 1] (a) -- (o);
        \draw[->, red, line width=1] (a) -- (n);
        \draw[->] (o) -- (y);
        \draw[->, blue, line width= 1] (n) to  [out=50,in=165, looseness=1.25] (y);
        \draw[->] (c) to  [out=330,in=220, looseness=1] (m);
        \draw[->] (c) to  [out=310,in=230, looseness=0.75] (y);
    \end{tikzpicture}
    \caption{The DAG displaying violation of the assumption of no direct effect of $N$ on $Y$. The bold red arrows indicate deterministic relationships. The blue arrow indicates the assumption violation.  }
    \label{fig:dag_sens}
\end{figure}

Given the strength of Assumption \ref{assn:ny}, whose veracity may be questionable in our setting, we now explore sensitivity to its violation. %of one of our key assumptions, Assumption \ref{assn:ny}, which is 
Such a violation is depicted in Figure \ref{fig:dag_sens} by the addition of the blue arrow from $N$ to $Y$. %In the presence of a violation, we will show that the resulting identification will be the original estimand plus a bias term. 
Understanding how large of a deviation from this assumption would need to be present in order to change the direction or significance of our results will provide insight into how sensitive our results are to this strong assumption. We use a similar approach to the sensitivity analysis proposed in \cite{stensrud2021generalized}, though we propose a scalar causally-interpretable sensitivity parameter. Sensitivity analysis for violations of our other key assumption, Assumption \ref{assn:om}, can be found in Appendix \ref{a:sens_om}. However, we argue that the latter assumption seems rather plausible in our setting, since anesthesia seems unlikely to affect any of the variables in ${\bf M}$.

When Assumption \ref{assn:ny} does not hold, we have $\mathrm{Pr}\{Y(n=0, o=1)\leq t\} = \Psi_{01}(P) / \gamma$,
where the sensitivity parameter is defined to be
\begin{equation*}
    \gamma \equiv \frac{\sum_{{\bf m},{\bf c}} \mathrm{Pr}\{Y(n=1,o=1,{\bf m}) \leq t \mid {\bf C} = {\bf c} \} \mathrm{Pr}\{{\bf M}(a=0)={\bf m}, {\bf C}={\bf c}\}}{\sum_{{\bf m},{\bf c}} \mathrm{Pr}\{Y(n=0,o=1,{\bf m}) \leq t \mid {\bf C} = {\bf c}\} \mathrm{Pr}\{{\bf M}(a=0)={\bf m}, {\bf C}={\bf c}\}},
\end{equation*}
\begin{comment}
\begin{equation*}
    \sum_{m,c} \left[\mathrm{Pr}\{Y(n=1,o=1,m) \leq t \mid c \}-\mathrm{Pr}\{Y(n=0,o=1,m) \leq t \mid c\} \right] \mathrm{Pr}\{M(a=0)=m, C=c\},
\end{equation*}
\end{comment}
as we show in Appendix \ref{a:sens_derivation_ny}. 
%To calculate $\psi^{\textrm{anesthesia}}$ on the ratio scale with the violation we can use:
Thus, we have $\psi^{\textrm{anesthesia}} = \gamma^{-1}\Psi_{01}(P)/\Psi_0(P)$. 
%\begin{align}
%    $\psi^{\textrm{anesthesia}} = \{\Psi_{01}(P) - \delta\}/\Psi_0(P)$. 
%\end{align}
The sensitivity parameter $\gamma$ can be interpreted as a type of controlled direct effect \citep{robins1992identifiability} of surgery on $Y$ within strata of ${\bf C}$ that we would see if anesthesia were administered and ${\bf M}$ set to ${\bf m}$, averaged over the joint distribution of the counterfactual ${\bf M}(n=0)$ and ${\bf C}$. Intuitively, this quantifies the direct effect of $N$ on $Y$ with respect to ${\bf M}$ under the distribution of ${\bf M}$ under no surgery. This interpretation is analogous to an alternative interpretation of the identifying functional for the natural direct effect discussed in \cite{van2008direct}. The primary differences are that in our case, we are additionally intervening to set $O=1$ and our contrast is on the relative risk scale.

In practice, the bias term $\gamma$ is unknown; however, it could in principle be identified a randomized trial. For example, this would be the case under Assumption \ref{assn:om} if a two-armed trial was conducted in which anesthesia was administered to all participants and surgery was randomly assigned. %This 4-armed trial is of course unethical and we are not suggesting that it should be conducted. 
%The deconstruction of $\psi^{\textrm{anesthesia}}$ aids in the interpretability of the $\delta$ in the context of the problem.  
Given that we have no such trial (indeed such a trial would be unethical), instead subject-matter experts can speculate about plausible ranges of values for $\gamma$ to adjust the estimate found assuming no violation. %Using a range of potential sensitivity values allows for observation of the behavior of the estimated $\psi^{\textrm{anesthesia}}$.
A stronger interpretation of $\gamma$ is possible under the additional cross-world counterfactual independence assumption that $Y(n=1,o=1,{\bf m})\ci {\bf M}(n=0)\mid {\bf C}$. In particular, we have $\gamma=\mathrm{Pr}[Y\{n=1,o=1,{\bf M}(n=0)\}\leq t]/\mathrm{Pr}[Y\{n=0,o=1,{\bf M}(n=0)\}\leq t]$, which can be interpreted as a modified version of the PNDE of $N$ on $Y$ with respect
to ${\bf M}$ when $O$ has also been intervened on to be set to one, i.e., the
PNDE of surgery on time to DIBD diagnosis with respect to ${\bf M}$ when anesthesia is administered. 
%If our target estimand for the effect of anesthesia was on the difference scale, i.e., $\psi_{01}-\psi_0$, then we could 

We can also rearrange the equation arising when Assumption \ref{assn:ny} is violated to obtain $\Psi_{01}(P)/\Psi_{0}(P) = \gamma \psi_{01} / \psi_0 $, which gives an alternative interpretation of the identifying functional $\Psi_{01}(P)/\Psi_{0}(P)$ as the product of the original effect of interest $\psi_{01}/\psi_0$ capturing the effect of anesthesia and $\gamma$, the direct effect of surgery on DIBD with respect to ${\bf M}$ averaged over the distribution of ${\bf M}(a=0)$ and ${\bf C}$. This product telescopes, reducing to 
\begin{align*}
    \frac{\Psi_{01}(P)}{\Psi_{0}(P)} &= %\frac{\sum_{{\bf m},{\bf c}} \mathrm{Pr}\{Y(n=1,o=1,{\bf m}) \leq t \mid {\bf C} = {\bf c} \} \mathrm{Pr}\{{\bf M}(a=0)={\bf m}, {\bf C}={\bf c}\}}{\mathrm{Pr}\{Y(n=0,o=0)\leq t\}}\\
    %&= 
    \frac{\sum_{{\bf m},{\bf c}} \mathrm{Pr}\{Y(a=1,{\bf m}) \leq t \mid {\bf C} = {\bf c} \} \mathrm{Pr}\{{\bf M}(a=0)={\bf m}, {\bf C}={\bf c}\}}{\sum_{{\bf m},{\bf c}} \mathrm{Pr}\{Y(a=0,{\bf m}) \leq t \mid {\bf C} = {\bf c} \} \mathrm{Pr}\{{\bf M}(a=0)={\bf m}, {\bf C}={\bf c}\}},
\end{align*}
which is precisely a relative risk version of the alternative interpretation of the identifying functional for the natural direct effect discussed in \cite{van2008direct}. Unlike the interpretation for $\gamma$ discussed above, the intervention on $O$ varies between the terms in the contrast rather than being set to one in both as it was above. Furthermore, under Assumption \ref{assn:consistency} (consistency), this quantity is equivalent to the so-called randomized interventional analog to the natural direct effect, which has gained popularity as an alternative to the PNDE when the latter is not identifiable \citep{vanderweele2014effect}. See \cite{miles2023causal} for discussion on the interpretation of this parameter and its limitations as a measure of mediation. Under the additional cross-world counterfactual independence assumption $Y(n=1,o=1,{\bf m})\ci {\bf M}(n=0)\mid {\bf C}$, this further simplifies to $\Psi_{01}(P)/\Psi_{0}(P) = \mathrm{Pr}[Y\{a=1,{\bf M}(a=0)\}\leq t]/\mathrm{Pr}[Y\{a=0,{\bf M}(a=0)\}\leq t\}$. That is, under violations of Assumption \ref{assn:ny}, the identifying functional still can be interpreted as the natural direct effect of $A$ on $Y$ with respect to ${\bf M}$ if the cross-world counterfactual independence holds; there is nothing about this violation that invalidates this identification result from the mediation literature. 

When performing sensitivity analysis, it is typically of interest to find the value of the sensitivity parameter for which the apparent effect is reduced to the null value. In our case, %due to the proportional relationship between $\Psi_{01}(P)/\Psi_{0}(P)$ and $\psi^{\mathrm{anesthesia}}$, 
this value is simply $\gamma=\hat{\Psi}_{01}(P)/\hat{\Psi}_{0}(P)$. Likewise, it is of interest to find the values of the sensitivity parameter for which the upper and lower confidence limits equal the null value, which in our case are simply equal to the upper and lower confidence limits of the point estimate $\hat{\Psi}_{01}(P)/\hat{\Psi}_{0}(P)$, hence solving for these quantities is trivial.
%This is similar in spirit to the generalized front-door estimand when the front-door exclusion restriction is violated discussed in \cite{fulcher2020robust}, which captures a causally interpretable parameter that is distinct from the original estimand of interest. 
%Unfortunately, the identifying functional for our estimand on the ratio scale does not have as clear of an interpretation. To recover a nicer interpretation, we could use an alternative formulation yielding a distinct sensitivity parameter. In particular, when Assumption \ref{assn:ny} is violated, we also have $\Psi_{01}(P)/\Psi_{0}(P) = \gamma\psi^{\mathrm{anesthesia}} $, where the sensitivity parameter is
%\[\gamma \equiv \frac{\sum_{m,c} Pr\{Y(n=1,o=1,m) \leq t \mid c\} Pr\{M(a=0)=m, C=c\}}{\sum_{m,c} Pr\{Y(n=0,o=1,m) \leq t \mid c\} Pr\{M(a=0)=m, C=c\}}.\]
%While $\gamma$ still captures a direct effect of $N$ on $Y$, its interpretation is slightly more complicated due to the counterfactual probabilities being marginalized over before being contrasted. Therefore, we will focus on the sensitivity parameter $\delta$ for the remainder of the article.
% Moving all the extra discussion introducing $L$ to the supplementary materials as I don't think it is very essential

\section{Simulation Study} \label{section:simulation}
%\subsection{Simulated data}
To demonstrate the finite sample performance of our proposed estimation technique and sensitivity analysis, as well as contrast with conclusions one would arrive at by making no attempt to distinguish the effects of surgery and anesthesia, we present results from a simulation study. We first present results under no violations to our exclusion restriction assumptions to demonstrate unbiasedness, then induce a violation of assumption \ref{assn:ny} to explore the degree of sensitivity to such violations. 

We generated four baseline variables ($C_1$, $C_2$, $C_3$, $C_4$), an indicator of surgery with accompanying anesthesia ($A$), two post-exposure variables affected only by surgery ($M_1$, $M_2$), and the time-to-event outcome $Y$. Furthermore, we generated large samples of each counterfactual $Y(n,o)$ for $n=0,1$ representing not receiving vs.~receiving surgery and $o=0,1$ representing not receiving vs.~receiving anesthesia, from which %The cases when $i\not=j$ represent outcomes that will never be observable in our real data. %Therefore, we generate what will happen if the patient received surgery and anesthesia $(Y_{11})$, did not receive either surgery or anesthesia $(Y_{00})$, received only surgery and no anesthesia $(Y_{10})$, and received no surgery but did receive anesthesia $(Y_{01})$. The latter two are not observable in our real setting. 
we computed the true values of $\psi^{\textrm{joint}}$, $\psi^{\textrm{anesthesia}}$ and, $\psi^{\textrm{surgery}}$ via Monte Carlo approximation. Specifically, we used the following data generating process:
\begin{align*}
    (C_1, C_2, C_3, C_4) &\sim \mathcal{N}(0,I), \\
   A &\sim \textrm{Bernoulli}\left\{ \textrm{logit}^{-1}(-2 + C_1 + C_2 + C_3 + C_4)\right\} \\
    N&=A, \; O=A \\
    M_1 &\sim N \times \textrm{Bernoulli}\left\{ \textrm{logit}^{-1}(-1 + C_1 + C_2 + C_3 + C_4 %+\tau O
    )\right\} \\
    M_2 &\sim \textrm{Bernoulli}\left\{ \textrm{logit}^{-1}(-2 + C_1 + C_2 + C_3 + C_4 + N %+\tau O
    )\right\} \\
    Y &= \left\{\frac{-2\log(U)}{\exp(0.25 C_1 + 0.25 C_2 + 0.25C_3 + 0.25 C_4 - 1.5M_1 - 1.5 M_2 + 0.5O + \zeta N)} \right\}^2,
\end{align*}
where $U \sim U(0,1)$ is regarded as unobserved. 
%The $\tau O$ term induces a violation of Assumption \ref{assn:om}. When $\tau = 0$ there is no violation in the assumption.  
The event $M_1=1$ can only occur when $N=1$, mimicking the relationship of some of the elements in ${\bf M}$ with $N$ (or $A$) in our real data scenario. %For some post-exposure variables they are only observed when a mother undergoes surgery, such as post operative pain. 
%The outcome $Y$ was generated by
% \[ 
% \begin{aligned}
% f(y) &= \frac{y^{1/\gamma-1}}{\lambda \gamma}\exp \left( - \frac{y \frac{1}{\gamma}}{\lambda} \right) \\
% \end{aligned}
% \]
% where $\lambda$ and $\gamma$ are user specified values. The model will have $E(Y) = \lambda^{\gamma}\Gamma(\gamma+1) $ and $Var(Y) = (\lambda^2)^{\gamma}(\Gamma(2\gamma +1) -\Gamma^2(v+1))$
% Using the covariates generated we will generate our outcome as:
% \[
% Y = \left(\frac{-\log(U)\lambda}{\exp(0.25 C_1 + 0.25 C_2 + 0.25C_3 + 0.25 C_4 - 1.5M_1 - 1.5 M_2 + 0.5O + \delta N)} \right)^\gamma
% \]
The presence of a nonzero $\zeta$ induces a violation of Assumption \ref{assn:ny}, whereas there is no such violation when $\zeta = 0$. We also introduced uninformative censoring. In our real data scenario, there are two sources that contribute to censoring: random dropout and censoring due to observation time cut-off. We simulated random dropout by $S_1 \sim \textrm{exponential}$ (rate = 0.5) and the second type by $S_2 = 15$, which represents the longest possible follow-up time. Therefore, the observed survival time is $\tilde{Y} = \min(Y, S_1, S_2)$.
% \begin{align*}
%     Y_{\textrm{cens}} &= \min(Y, s_1, s_2),  \\ 
%     \textrm{status} &= I(Y \leq s_1 \ \& \ Y \leq s_2 ).
% \end{align*}
%Currently the number of observed event is around 35\%. 

%When $\delta \not= 0$ then the natural direct effect will be biased. In practice, the true $\delta$ will be unknown. Therefore if a violation in Assumption \ref{assn:ny} is expected then a range of $\delta$ sensitivity values should be used to explore how large of a violation is needed before the results change significantly. When generating data we are able to calculate the true $\delta$ values. Calculations for the bias functions can be found in Appendix \ref{a:data_gen}. 

% description of set up 
We ran 500 simulations with $\zeta=0$ as well as under three violation settings of varying strength: $\zeta \in (0.25, 0.5, 0.75)$. %The extent of the violation introduced for each $\zeta$ is calculated and presented in Appendix \ref{a:math_n_to_y}. 
In the presence of such violations, we present a grid of adjustment values of the sensitivity parameter $\delta$ from $0.9$ to $1.6$. The true population values for the parameters of interest under no assumption violations are $\psi^{\textrm{anesthesia}}=1.28$, $\psi^{\textrm{surgery}}=0.71$, and $\psi^{\textrm{joint}}=0.92$. This specific data generating scenario is one in which the anesthesia causes an increased risk of the outcome compared to that of the joint effect of surgery and anesthesia, which causes a decreased risk of the outcome. While this is unlikely the case in reality, we mean to demonstrate that in general, the joint effect and individual effects need not agree in direction, and that estimating a joint effect and interpreting it only as the effect of one of the exposures could potentially lead to an incorrect qualitative conclusion.

%\subsection{Results}

\begin{figure}[h]
    \centering
    \includegraphics[scale=0.5]{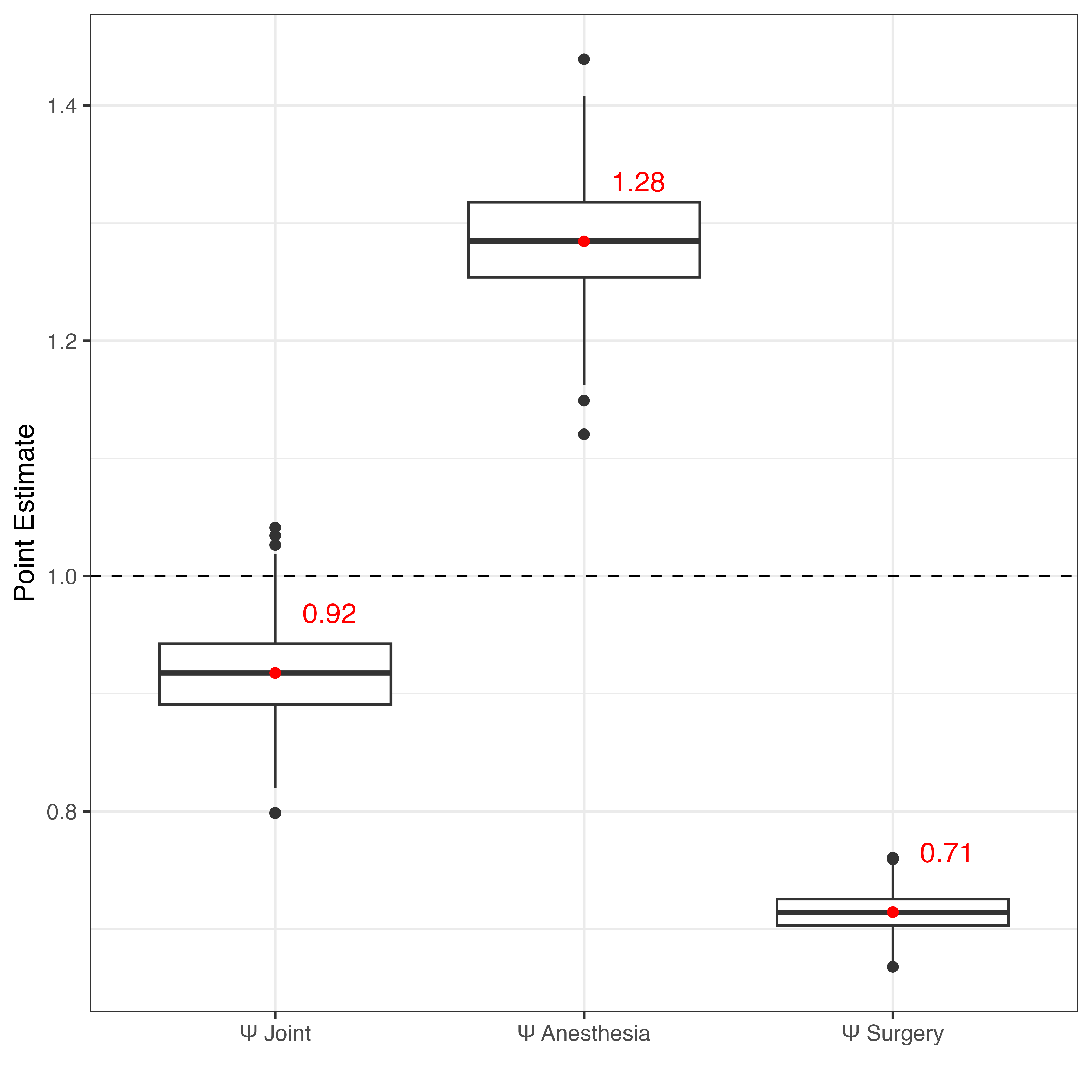} 
    \caption{
    Boxplots of point estimates  from the 500 simulations for the scenario with no assumption violation ($\zeta=0$). The red dots and text label the true parameter values.}% , and the dashed black line indicates the null effect.   } 
    \label{fig:sim_res}
\end{figure}

% need to first describe what happens with zero violations 
We first consider the ideal setting with no assumption violations present ($\zeta=0$). Figure \ref{fig:sim_res} depicts the distributions of the point estimates of the effect of anesthesia ($\psi^{\textrm{anesthesia}}$), surgery ($\psi^{\textrm{surgery}}$), and their joint effect ($\psi^{\textrm{joint}}$). %In red the population true value is plotted. These findings signify that, 
When no assumptions are violated, the estimates approximate their corresponding estimands well. Furthermore, approximate 95\% confidence interval coverage is attained for each estimand, as shown in the right panel of Figure \ref{fig:sim_res}. Given the  lack of alternative methods for handling the extreme positivity violation we are confronting, we instead compare our estimate of $\psi^{\textrm{anesthesia}}$ to the estimate of $\psi^{\textrm{joint}}$, which is what is typically used when attempting to study the effect of anesthesia (though often on different scales). We see that using the $\psi^{\textrm{joint}}$ estimate would typically result in a qualitatively different interpretation than that of the $\psi^{\textrm{anesthesia}}$ estimate.

\begin{figure}[h]
    \centering
    \includegraphics[scale=0.45]{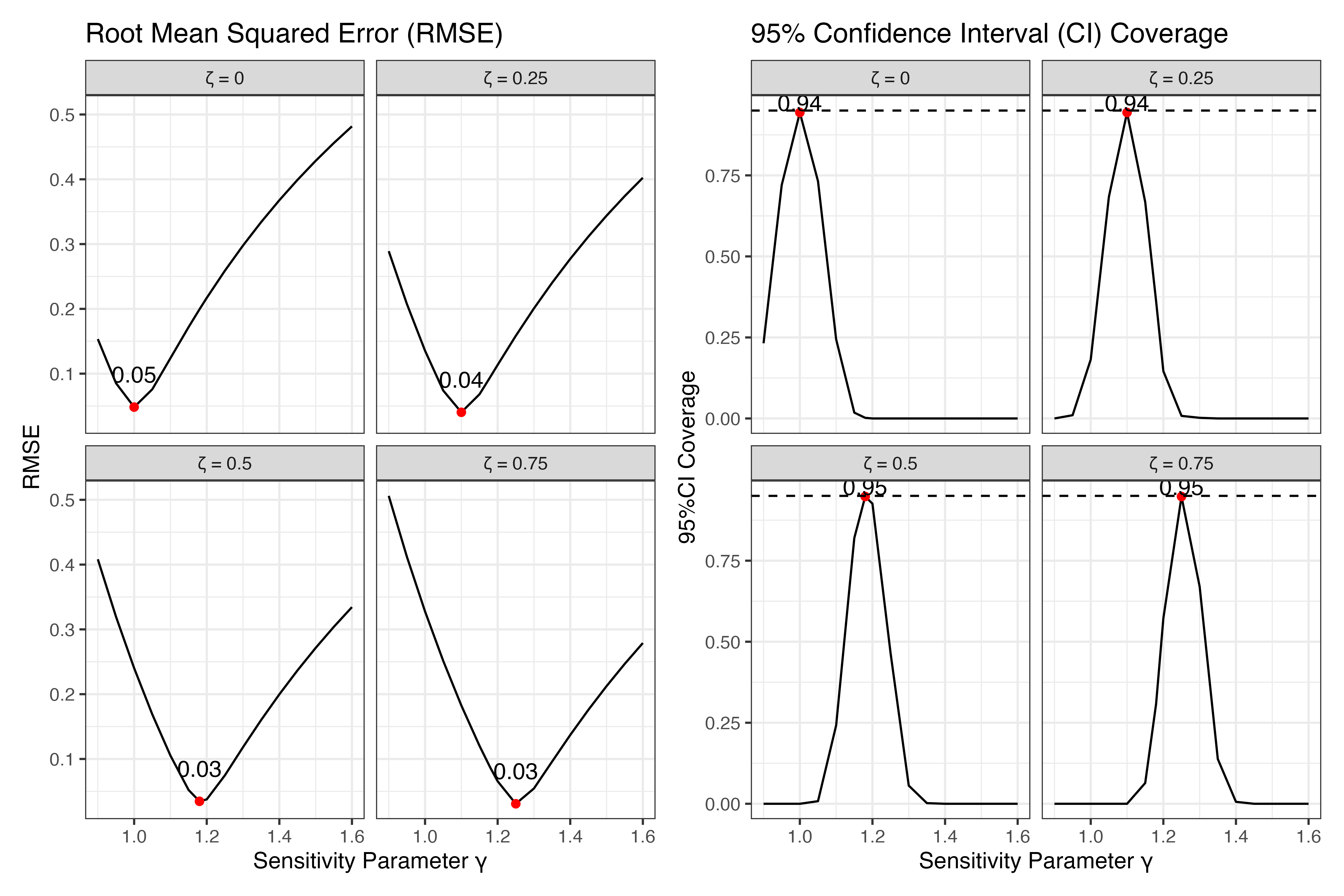}
    \caption{ 
    Results from the sensitivity parameter-adjusted $\psi^{\textrm{anesthesia}}$ estimates from 500 simulations. Each subpanel corresponds to a different $\zeta$ value (0, 0.25, 0.5, 0.75), and the sensitivity parameter $\gamma$ varies along the x-axis. $\gamma = 1$ corresponds to the unadjusted estimate. The left panel depicts the RMSE, and the right panel depicts the 95\% CI coverage probability for each sensitivity parameter value. The dashed lines in the right panel indicate the nominal coverage 0.95. The red points indicate when the sensitivity parameter matches the true underlying $\gamma$.}
    \label{fig:sim_sum_NY}
\end{figure}

Figure \ref{fig:sim_sum_NY} shows the root mean squared error (RMSE) of estimates in the left panel and 95\% confidence interval (CI) coverage in the right panel for $\psi^{\textrm{anesthesia}}$ when a violation of Assumption \ref{assn:ny} is introduced. 
\begin{comment}
\begin{figure}[h]
    \centering
    \includegraphics[scale=0.45]{figures/sim_rmse_coverage_NY_500_additive_20250304.png}
    \caption{ 
    Results from the sensitivity-parameter-adjusted $\psi^{\textrm{anesthesia}}$ estimates from 500 simulations. Each subpanel corresponds to a different $\zeta$ value (0, 0.25, 0.5, 0.75). The sensitivity parameter $\delta$ adjustment value on the $x$-axis. When the sensitivity parameter $\delta = 0$, the estimate corresponds to the unadjusted estimate. The left panel depicts the RMSE, and the right panel depicts the 95\% CI coverage probability for each sensitivity parameter value. The dashed lines in the right panel indicate 0.95. The red points indicate when the sensitivity parameter matches the true parameter.}
    \label{fig:sim_sum_NY}
\end{figure}
\end{comment}
As expected, when the sensitivity parameter $\gamma$ aligns with the true bias $\gamma$ induced by $\zeta$, we observe the lowest RMSE values and the highest 95\% confidence interval coverage in each case. %, indicating that when we correctly adjust for the violation, we closely approximate the truth. 
However, results are sensitive to the violation of Assumption \ref{assn:ny}, and coverage probabilities deteriorate as expected as the sensitivity parameter deviates from the true $\gamma$ value.% This figure also illustrates that failing to adequately adjust for bias, represented by small sensitivity adjustments, or overcompensating with a large sensitivity adjustment, adversely affects the quality of the results. This is evident in the increased RMSE and the absence of 95\% CI coverage. 

\begin{figure}[h]
    \centering
    \includegraphics[scale=0.5]{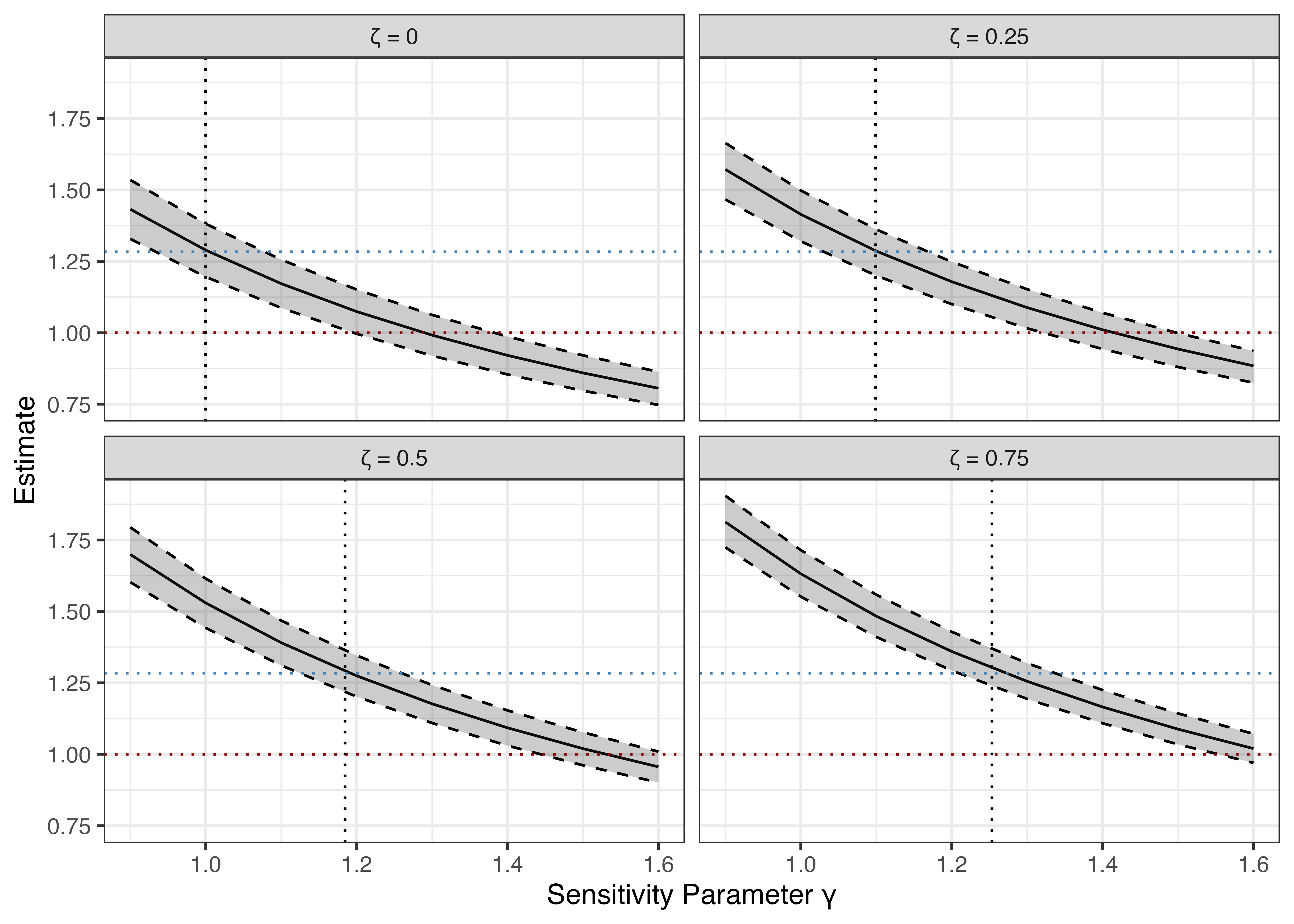} 
    \caption{
    Sensitivity parameter-adjusted $\psi^{\textrm{anesthesia}}$ estimates and 95\% confidence intervals from a single sample.  
    Each subpanel corresponds to a different $\zeta$ value (0, 0.25, 0.5, 0.75). When the sensitivity parameter $\gamma = 0$, the estimate corresponds to the unadjusted estimate. The blue dotted line represents the true parameter value and the red dotted line at one represents the null value. The vertical black dotted line represents the true underlying $\gamma$ induced by $\zeta$. The solid black line indicates the point estimates, and the shaded gray regions are pointwise 95\% confidence intervals across all values of $\gamma$.}
    \label{fig:sim_res_NY}
\end{figure}

Next, we consider results for a single sample under a violation of Assumption \ref{assn:ny}, which are displayed in
Figure \ref{fig:sim_res_NY}.
\begin{comment}
\begin{figure}[h]
    \centering
    \includegraphics[scale=0.5]{figures/sim_res_NY_iter10_additive_20250304.png} 
    \caption{
    Sensitivity-parameter-adjusted $\psi^{\textrm{anesthesia}}$ estimates and 95\% confidence intervals from a single sample.  
    Each subpanel corresponds to a different $\zeta$ value (0, 0.25, 0.5, 0.75). When the sensitivity parameter $\delta = 0$, the estimate corresponds to the unadjusted estimate. The blue dotted line represents the true parameter value and the red dotted line at one represents the null value. The vertical black dotted line represents the true bias induced by $\zeta$. The solid black line indicates the point estimates, and the shaded gray regions are pointwise 95\% confidence intervals across all values of $\delta$.}
    \label{fig:sim_res_NY}
\end{figure} 
\end{comment}
In each panel, the point estimate nearly intersects the point where the blue and black dotted lines cross, indicating that adjusting for the correct value of $\gamma$ nearly recovers the truth. Of course, the true $\gamma$ is not known outside of simulation studies. 
Thus, we now illustrate how to interpret these results using the setting $\zeta=0$ (but ignoring the knowledge of this value). In this case, there is a statistically significant harmful estimated effect of anesthesia. The lower bound of the confidence interval crosses one at about $\gamma=1.18$. Thus, to negate the statistical significance of the apparent harmful effect, the direct effect of surgery on DIBD (averaged over ${\bf M}(n=0)$ and ${\bf C}$) would need to be 1.18 on the relative risk scale.  The effect estimate crosses one at $\gamma=1.28$. Thus, to nullify the estimated effect of anesthesia, the direct effect of surgery on DIBD (averaged over ${\bf M}(n=0)$ and ${\bf C}$) would need to be 1.28 on the relative risk scale. Lastly, the upper limit of the confidence interval crosses one at about $\gamma=1.38$. Thus, to reverse the statistical significance of the apparent harmful effect in the direction of being beneficial, the direct effect of surgery on DIBD (averaged over ${\bf M}(n=0)$ and ${\bf C}$) would need to be 1.38 on the relative risk scale.

\section{MAX Data Application Results} \label{section:real_data}

\subsection{Data description} \label{section:real_data_description}

The study cohort drawn from the Medicaid Analytic eXtract (MAX) data \citep{max} consisted of 16,778,231 mother-child pairs born between 1999 and 2013. We excluded pairs in which the child was censored at birth and in unexposed pairs in which the mother was only eligible for Medicaid in their last trimester of pregnancy. This resulted in 31,494 children who were prenatally exposed to general anesthesia from maternal appendectomy or cholecystectomy, and 14,307,152 unexposed children. Here and henceforth, we use the term ``exposed'' to mean having received both anesthesia and surgery during pregnancy, and ``unexposed'' to mean having received neither. %anesthesia nor surgery during pregnancy.  

The observation window for baseline covariates ${\bf C}$ for exposed mothers is defined as all records up to the date of their procedure. However, the observation window for ${\bf C}$ for unexposed mothers is ambiguous due to their lack of a procedure date. We create approximately the same distribution of observation times among unexposed mothers by assigning unexposed mothers a pseudo-exposure month sampled from the distribution of months at which procedures occurred among the exposed mothers. This is similar to how baseline covariates were defined for unexposed mothers in \cite{ing2024behavioural}, in which unexposed mothers were matched to exposed mothers on the basis of covariates observed within a time window determined by the exposed mothers. We describe the sampling algorithm we used for our procedure in Appendix \ref{a:algorithm_baseline}. Distributions of baseline covariates within the unexposed and exposed groups are comparable, as seen in Table \ref{tab:tab1}, with the majority of covariates having a standardized mean difference below 0.05, and only two covariates having a standardized mean difference larger than 0.2 (pre-pregnancy healthcare utilization of outpatient visits and restricted benefits prior to exposure based on alien status). We included all baseline covariates in ${\bf C}$ and adjusted for them all in our estimation.

%\subsection{Characteristics of the study cohort}

% should I say anything about who had both surgeries? (148 mothers had both)

\subsection{Results}
Provided that Assumptions \ref{assn:consistency}--\ref{assn:censoring} all hold and our statistical models are correctly specified, we have the following estimates and interpretations. The estimated effect of prenatal anesthesia exposure on DIBD diagnosis in children by age $t=10$ is $\hat{\psi}^{\textrm{anesthesia}} = 1.13$ (95\% CI: 1.10, 1.16), indicating a significant increased risk of developing DIBD by age 10 due to anesthesia alone. This estimated effect is stronger than the surgery-specific effect estimate $\hat{\psi}^{\textrm{surgery}} = 1.02$ (95\% CI: 1.005, 1.04), and accounts for most of the estimated joint effect of both surgery and anesthesia $\hat{\psi}^{\textrm{joint}} = 1.16$ (95\% CI: 1.12, 1.19). %Thus, these results suggest that when isolating the effects of anesthesia and surgery on DIBD, most of the observed risk is attributable to exposure to anesthesia. % need another sentence about what this means more generally. 

%$\hat{\psi}^{\textrm{anesthesia}} = 1.133$ (95\% CI: 1.098, 1.162), 
%$\hat{\psi}^{\textrm{surgery}} = 1.021$ (95\% CI: 1.005, 1.039), 
%$\hat{\psi}^{\textrm{joint}} = 1.157$ (95\% CI: 1.124, 1.192). 

\begin{figure}[h]
    \centering
    \includegraphics[scale=0.65]{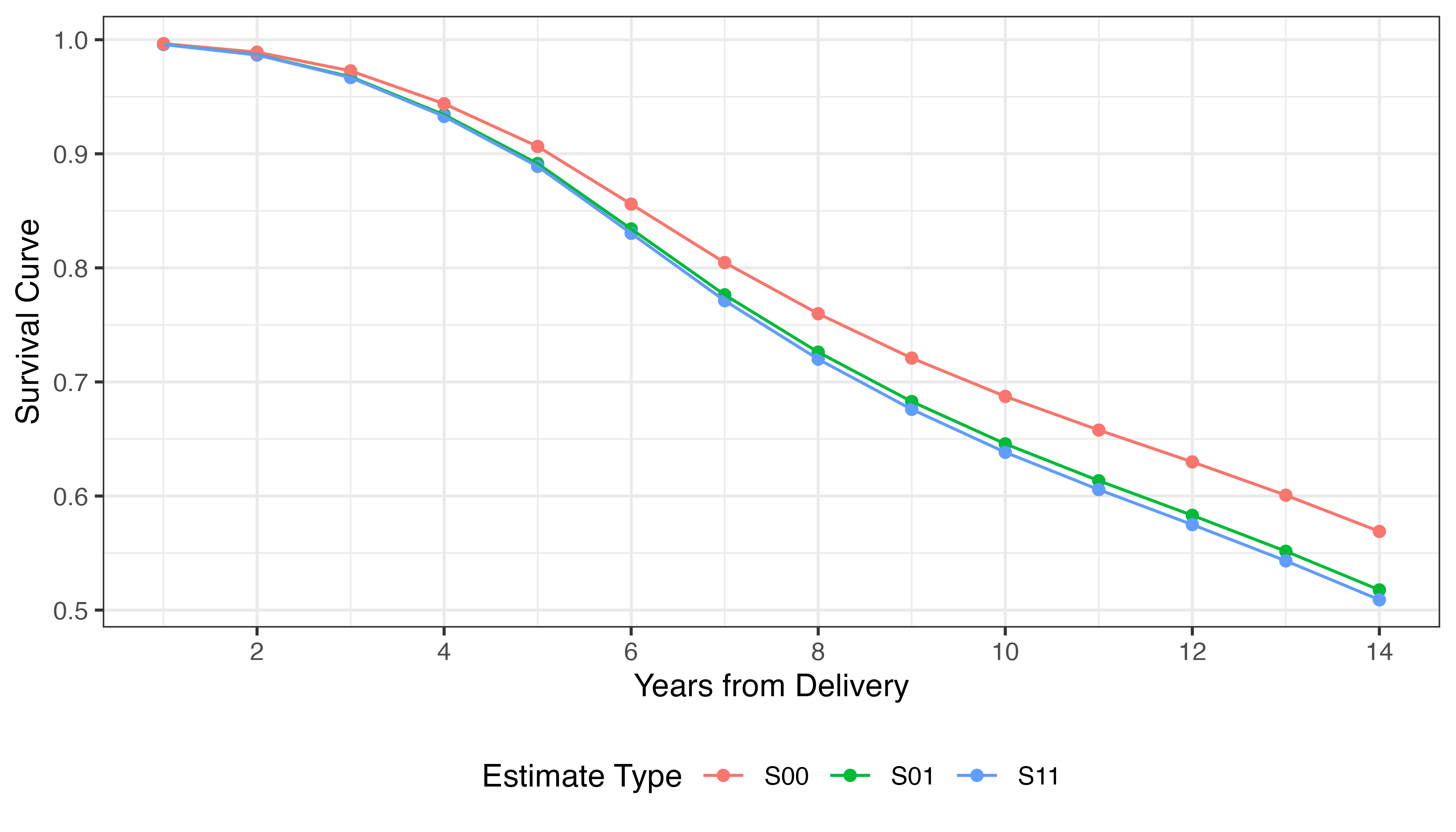}
    \caption{The survival curves for developing a DIBD under various interventions. S00 is the survival probability under no surgery or anesthesia, $\mathrm{Pr}\{Y(n=0,o=0)>t\}$, S11 is the survival probability under both surgery and anesthesia, $\mathrm{Pr}\{Y(n=1,o=1)>t\}$, and S01 is the survival probability under anesthesia but not surgery, $\mathrm{Pr}\{Y(n=0,o=1)>t\}$, from age one to fourteen years old.}
    \label{fig:surv_curv_real_data}
\end{figure}

%When examining how these effects changes across children's age, 
Figure \ref{fig:surv_curv_real_data} depicts the survival curves under each of the interventions setting $(N,O)$ to $(0,0)$, $(0,1)$, and $(1,1)$. The effect of anesthesia on DIBD explains more of the joint effect than does surgery across all ages, though it is possible that this is an artifact of the proportional hazards model assumption. The apparent effect of surgery on the relative risk scale is fairly constant across time, whereas that of anesthesia diminishes over time, as seen in Figure \ref{fig:est_over_time} in Appendix \ref{a:est_over_time}. %the Supplementary Materials.

%as seen by the proximity of the counterfactual survival curves $S_{11}(t)$ (both anesthesia and surgery exposure) and $S_{01}(t)$ (anesthesia only). 
%On the other hand, the curve for $S_{00}(t)$ (no exposure) begins to deviate at ages 4-5 and continues to diverge more sharply through ages 10--14. This divergence may correspond to developmental milestones, such as the start of kindergarten, when access to professionals who can diagnose behavioral symptoms associated with DIBD begins. The deviation in older children (ages 10–14) could reflect both the gradual manifestation of certain disorders and increased access to diagnostic resources in later childhood.

\subsection{Sensitivity analyses} \label{section:sens_analysis_real_data}

\begin{figure}[h]
    \centering
    \includegraphics[scale=0.65]{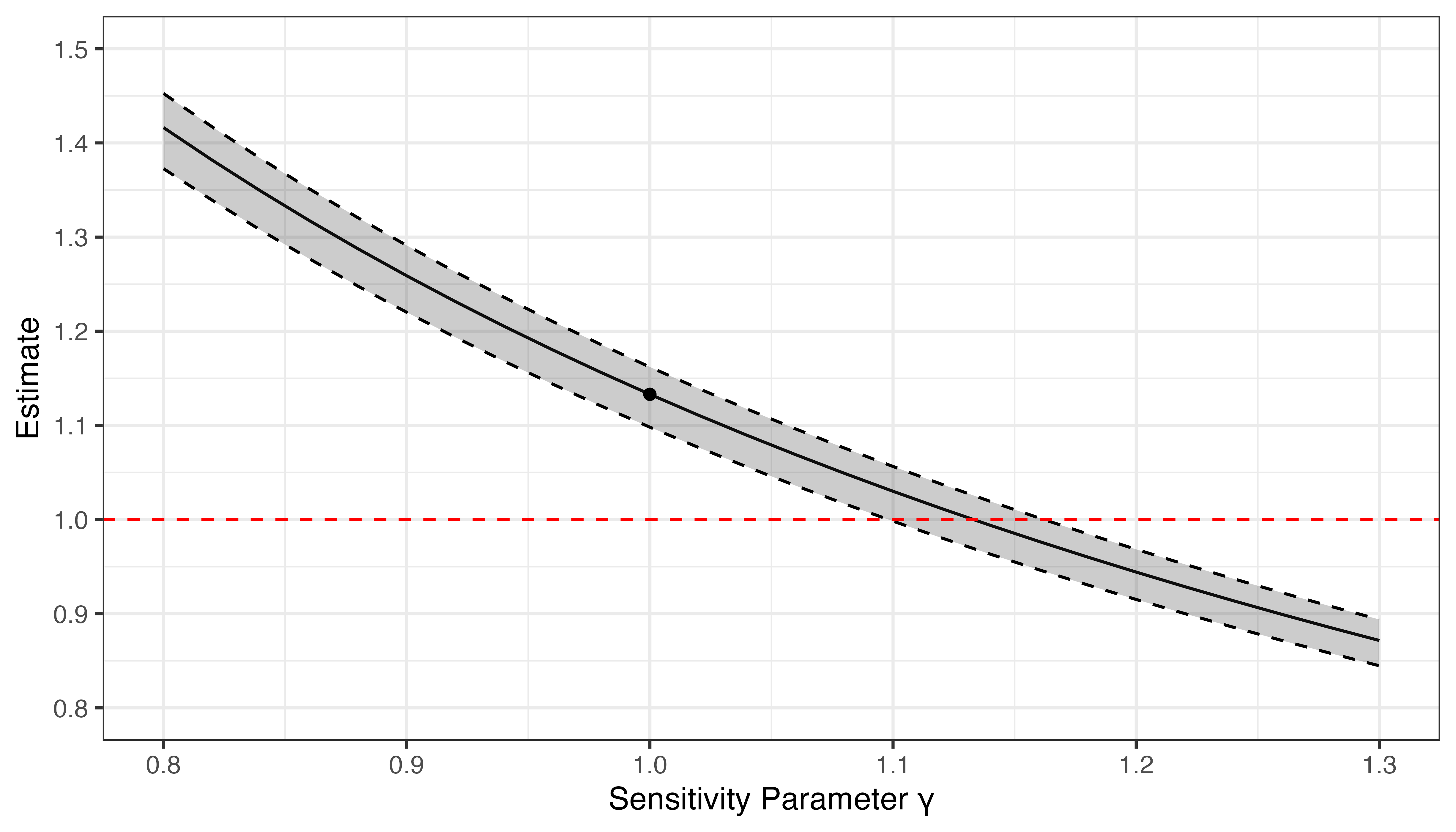}
    \caption{Sensitivity-parameter-adjusted $\psi^{\textrm{anesthesia}}$ estimates and 95\% confidence intervals for $t=10$. When the sensitivity parameter $\gamma=1$, the estimate corresponds to the unadjusted estimate. The red dotted line represents the non-significant result line.}
    \label{fig:sens_res_real_data}
\end{figure}

To understand the sensitivity of our estimate of the effect of anesthesia to violations of Assumption \ref{assn:ny}, we conducted the sensitivity analysis proposed in Section \ref{section:sensitivity} with $\gamma$ ranging from 0.8 to 1.30. Recall that $\gamma$ represents the strength of the direct effect of surgery on DIBD with respect to the ${\bf M}$ variables.  Results are shown in Figure \ref{fig:sens_res_real_data}. 
%The sensitivity parameter-adjusted estimated decreases with $\gamma$. 
At $\gamma = 1.10$, the lower confidence interval crosses the null value one, indicating that if the direct effect of surgery on DIBD with respect to ${\bf M}$ were known to be at least 1.1 on the relative risk scale, there would no longer be a statistically significant harmful effect of anesthesia on DIBD. 
%changing the interpretation of results to no longer significant. 
At about $\gamma = 1.13$, the point estimate crosses the null value of one, indicating that if the direct effect of surgery on DIBD with respect to ${\bf M}$ were known to be at least 1.13 on the relative risk scale, the adjusted estimate of the effect of anesthesia on DIBD would change qualitatively to be null or in the beneficial direction. 
%changing the directions of interpretations of results. 
The upper confidence interval crosses the null value of one at $\gamma = 1.16$, indicating that if the direct effect of surgery on DIBD with respect to ${\bf M}$ were known to be at least 1.16 on the relative risk scale, there would be a statistically significant beneficial effect of anesthesia on DIBD. In our case, a direct effect of $N$ on $Y$ could be through surgery-induced acute inflammation or acetaminophen use. %This is a factor that so far we are assuming to not be influential but through this sensitivity analysis, we can explore its potential influence on our results. 
%\cite{han2021maternal} conducted a systematic review of associations between inflammation during pregnancy and neurodevelopmental disorders. Results concerning acute inflammation were mixed. %The largest studies were by Ginsberg et al. and Werenberg et al., which were European population level studies.  
%\cite{ginsberg2019maternal} found that associations with ADHD were not significant after accounting for social and familial factors. \cite{werenberg2016fever} found little risk of ADHD overall, but increased risk was seen in specific exposure windows. \cite{mann2011maternal} reported an adjusted odds ratio of 1.29 between ADHD and genitourinary infection during pregnancy, though this may have been due to insufficient adjustment for confounding, which was less extensive than that of \cite{ginsberg2019maternal}. 
%Maternal acetaminophen use has been associated with neurodevelopmental outcomes as well, but these results are also mixed. 

\begin{figure}[h]
    \centering
    \includegraphics[scale=0.65]{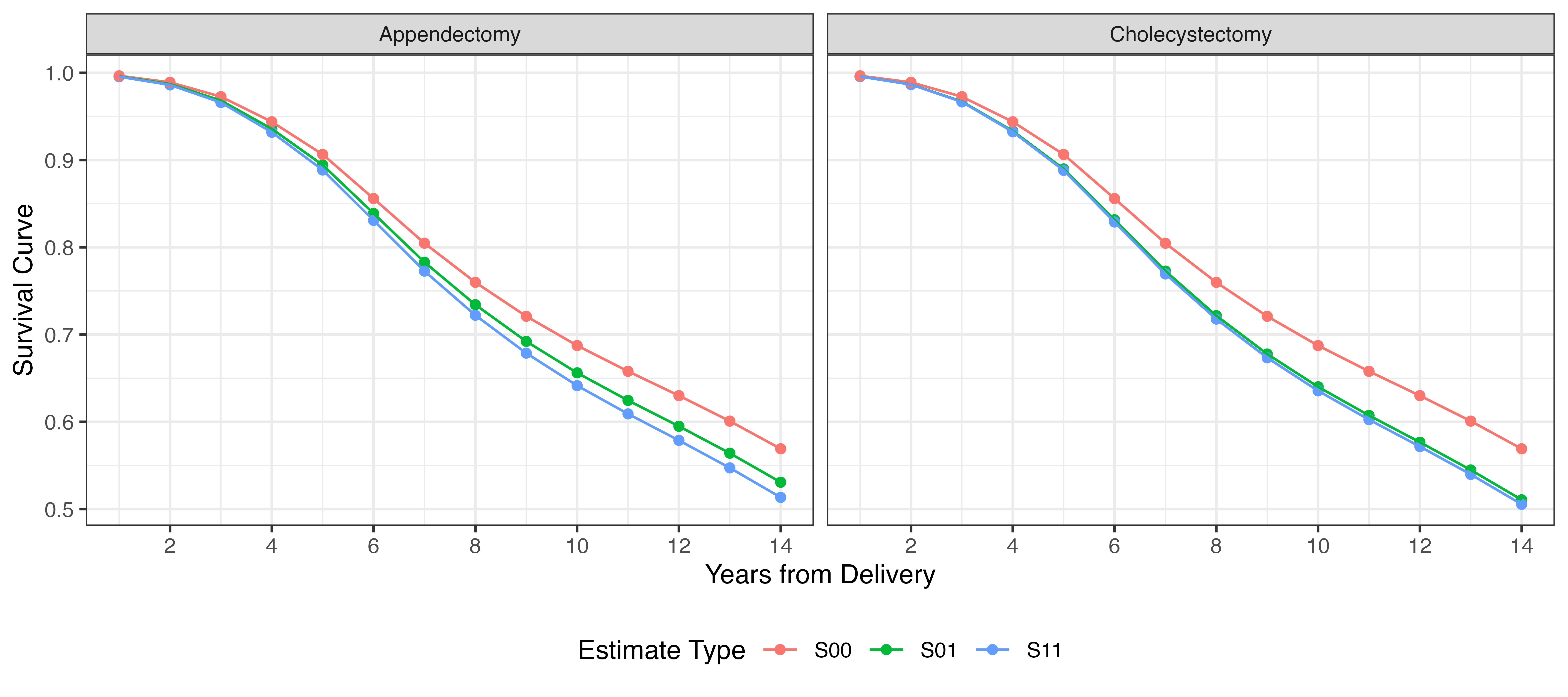}
    \caption{The survival curves for developing a DIBD under various interventions stratified by procedure type. S00 is the survival probability under no surgery or anesthesia, $\mathrm{Pr}\{Y(n=0,o=0)>t\}$, S11 is the survival probability under both surgery and anesthesia, $\mathrm{Pr}\{Y(n=1,o=1)>t\}$, and S01 is the survival probability under anesthesia but not surgery, $\mathrm{Pr}\{Y(n=0,o=1)>t\}$, from age one to fourteen years old.}
    \label{fig:surv_curv_real_data_by_surg}
\end{figure}

We additionally explored potentially differential impacts of the underlying inflammation requiring surgery by conducting separate analyses for mothers who underwent an appendectomy and those who had a cholecystectomy. 
The estimated effect of prenatal anesthesia exposure on DIBD diagnosis in children up to age 10 is 1.10 for the appendectomy group and 1.15 for the cholecystectomy group. The stratified survival curves are shown in Figure \ref{fig:surv_curv_real_data_by_surg}. 
%where the proximity of counterfactual survival curves S11 and S01 for both surgery types is consistent across years with the deviation of S00 starting at year 4. This indicates that the effect of anesthetic exposure on DIBD persists across all ages, more so than the effect of surgery for both surgery types.  
While this stratified analysis does not fully account for the potentially confounding effect of the underlying inflammation, it is notable that the effect estimates are fairly similar to one another despite the differences in the underlying medical condition. We still see that the harmful effect of anesthesia appears to exceed that of surgery among both surgery types across all time points, though the estimated effect of cholecystectomy in the presence of anesthesia is noticeably smaller than that of appendectomy.

\section{Discussion} \label{section:discussion}

This study addresses the complex, yet important challenge of isolating the effect of prenatal exposure to anesthesia from that of surgery using a large Medicaid claims data set. Our findings suggest that anesthesia is responsible for most of the harmful effect of the joint exposure of surgery and anesthesia. If one believes the assumptions required for our estimates to be valid, then this would suggest that resources should be allocated toward finding ways to minimize harmful effects of anesthesia rather than of surgery. 
If delaying treatment until after pregnancy is an option, then this determination should be informed by the joint effect of surgery and anesthesia rather than their individual effects, which in our case was estimated to be harmful. 
We certainly do not mean to suggest that mothers avoid life-saving surgery to prevent fetal anesthesia exposure. However, these findings have important implications for early brain development and can guide decision-making around elective procedures during pregnancy and for young children, particularly for non-essential procedures or the reduction of anesthetic use in young children for diagnostic studies including MRI scans \citep{machado2021predictors,vinson2021trends}. For example, in the case of appendectomies, clinical trials have explored the possibility of treating acute appendicitis with antibiotics as an alternative to surgical removal of the appendix \citep{svensson2015nonoperative, yang2019meta, coda2020randomized}.

There are a number of limitations to our analyses, and the validity of our approach relies on several key assumptions, at least two of which we deem to be quite strong. Perhaps the strongest is the assumption that the variables captured in ${\bf M}$ completely mediate the effect of surgery on DIBD. For this reason, we performed a sensitivity analysis to violations of this assumption, finding a moderate degree of sensitivity. If all other assumptions hold, for there to truly be no effect of anesthesia on DIBD, all of the apparent effect would have to be due to a direct effect of surgery on DIBD with respect to ${\bf M}$. 
%[NOTE: Talk about alternative pathways we were unable to measure and discuss whether we think it would be reasonable for the effect to be this large.] 
% ADDED
Alternative pathways we were unable to measure are inflammation, or maternal acetaminophen use.  
Current research that aims to understand maternal acute inflammation and neurodevelopmental outcomes found mixed results \citep{han2021maternal}. However, \cite{ginsberg2019maternal} looked at maternal infection requiring hospitalization and the development of ADHD in offspring and found that after accounting for social and familial factors that are associated with maternal infection, the association was attenuated.  
Maternal acetaminophen use has been found to be associated with neurodevelopmental outcomes, but these results have also been mixed \citep{ahlqvist2024acetaminophen, baker2025associations}.  
% ADDED
Another strong assumption is that of no unmeasured confounding of the joint effect of surgery and anesthesia on DIBD. Indeed, residual confounding is always a concern in any observational study. In our case, there may be particular concern about the underlying inflammation requiring surgery causing DIBD independently of surgery and anesthesia. We have not addressed this concern in the present article, though we did consider stratified analyses within each surgery type, which did not produce greatly differing results.
Lastly, we relied on an assumption of no informative censoring, which would be violated if patients changed their insurance type away from Medicaid for reasons that also affected their risk of DIBD. This problem is also beyond the scope of this article. %The data set we used is older, covering a period through 2013, and anesthesia techniques may have evolved since then. 
Another potential limitation is that the data we used only covers a period up until 2013. While anesthetic techniques may have evolved since then, general anesthesia remains the most common technique in pregnant women undergoing non-obstetric surgery, particularly for intra-abdominal procedures \citep{brakke2023anaesthesia}.  Pregnancy loss following surgery could have introduced selection bias; however, since only 1\% of fetuses are lost after an appendectomy or cholecystectomy during pregnancy, this is unlikely to meaningfully affect our results \citep{sachs2017risk}. Although we do not claim to have managed to fully address whether anesthesia affects DIBD independently of surgery, we have at least moved in the direction of the desired effect relative to the usual estimation approach that merely targets the joint effect of surgery and anesthesia.

In terms of causal and statistical novelty, we have leveraged existing machinery from the separable effects and mediation literatures to address an extreme form of positivity violation, which is a common challenge encountered not only in causal inference, but also in other domains such as missing data, supervised learning, survey sampling, and data fusion. In doing so, we have inverted the typical application of separable effects, which usually initially focuses on an intermediate event and then comes up with a story about how the exposure can be separated into components to justify the model, to instead begin with separate components that are initially of interest and identifying a set of intermediate variables that justify the model. The former framing has been somewhat controversial. Many find this conceptualization to be illuminating for understanding estimands in mediation and settings with intercurrent events. However, others have criticized this approach, finding it difficult to conceive of real-world examples in which the exposure neatly decomposes into separate components. In our case, the decomposition of the exposure is perfectly natural, and the conceptual burden is instead shifted to identifying a set of intermediate variables that completely mediate the effect of one component while not being affected by the other. While still challenging, we imagine that some critics will find this to be more digestible and transparent than the usual application. Furthermore, the use of multiple intermediate variables makes the exclusion restriction of no direct effect of the component $N$ on $Y$ more plausible than the usual approach in which ${\bf M}$ is univariate and often binary. While this also makes the other exclusion restriction of no effect of the component $O$ on $Y$ less plausible, at least in our case, this assumption seems more reasonable. We have also discussed how this approach might be extended to address milder positivity violations and made connections to front-door and back-door adjustment. Given the strength of our assumptions, we have also developed sensitivity analysis methodology with scalar sensitivity parameters that have a clear causal interpretation. 
%While our approach relies on strong assumptions, 
Lastly, we believe that our approach may be useful in other applications and methodological domains. For example, there are many randomized experiments in which participants in the treated arm receive a package intervention, and there is often interest in understanding the causal effects of the individual components of the intervention. 
Furthermore, extensions or analogues to our approach could potentially address problems of positivity violations that arise in other domains such as those listed above. Such violations may be extreme, as in our case, or more commonplace, as discussed in Section \ref{sec:less-extreme}. To summarize, we have provided a comprehensive illustration of how one can apply a separable effects approach to address an extreme positivity violation using the real-world example of isolating the effects of surgery and anesthesia on developmental and behavioral disorders in children.

\section*{Acknowledgments}
We would like to thank Kara Rudolph for helpful comments and Iv{\'a}n D{\'\i}az for prompting us to explore the connection to front-door adjustment.

\bibliographystyle{apalike}
\bibliography{references}

\begin{thebibliography}{}

\bibitem[Aalen et~al., 2020]{aalen2020time}
Aalen, O.~O., Stensrud, M.~J., Didelez, V., Daniel, R., R{\o}ysland, K., and
  Strohmaier, S. (2020).
\newblock Time-dependent mediators in survival analysis: modeling direct and
  indirect effects with the additive hazards model.
\newblock {\em Biometrical Journal}, 62(3):532--549.

\bibitem[Ahlqvist et~al., 2024]{ahlqvist2024acetaminophen}
Ahlqvist, V.~H., Sj{\"o}qvist, H., Dalman, C., Karlsson, H., Stephansson, O.,
  Johansson, S., Magnusson, C., Gardner, R.~M., and Lee, B.~K. (2024).
\newblock Acetaminophen use during pregnancy and children’s risk of autism,
  adhd, and intellectual disability.
\newblock {\em JAMA}, 331(14):1205--1214.

\bibitem[Baker et~al., 2025]{baker2025associations}
Baker, B.~H., Bammler, T.~K., Barrett, E.~S., Bush, N.~R., Collett, B.~R.,
  Derefinko, K.~J., Enquobahrie, D.~A., Karr, C.~J., LeWinn, K.~Z., Liu, J.,
  et~al. (2025).
\newblock Associations of maternal blood biomarkers of prenatal apap exposure
  with placental gene expression and child attention deficit hyperactivity
  disorder.
\newblock {\em Nature Mental Health}, pages 1--14.

\bibitem[Bartels et~al., 2009]{bartels2009anesthesia}
Bartels, M., Althoff, R.~R., and Boomsma, D.~I. (2009).
\newblock Anesthesia and cognitive performance in children: No evidence for a
  causal relationship.
\newblock {\em Twin Research and Human Genetics}, 12(3):246--253.

\bibitem[Brakke and Sviggum, 2023]{brakke2023anaesthesia}
Brakke, B. and Sviggum, H. (2023).
\newblock Anaesthesia for non-obstetric surgery during pregnancy.
\newblock {\em BJA Education}, 23(3):78--83.

\bibitem[{Centers for Medicare \& Medicaid Services.}, 2023]{max}
{Centers for Medicare \& Medicaid Services.} (2023).
\newblock Medicaid analytic extract ({MAX}) general information.
\newblock
  https://www.cms.gov/data-research/computer-data-systems/medicaid-data-sources-general-information/medicaid-analytic-extract-max-general-information.

\bibitem[Chen et~al., 2020]{chen2020doubly}
Chen, Y., Li, P., and Wu, C. (2020).
\newblock Doubly robust inference with nonprobability survey samples.
\newblock {\em Journal of the American Statistical Association},
  115(532):2011--2021.

\bibitem[Cochran and Rubin, 1973]{cochran1973controlling}
Cochran, W.~G. and Rubin, D.~B. (1973).
\newblock Controlling bias in observational studies: A review.
\newblock {\em Sankhy{\=a}: The Indian Journal of Statistics, Series A}, pages
  417--446.

\bibitem[{Coda Collaborative}, 2020]{coda2020randomized}
{Coda Collaborative} (2020).
\newblock A randomized trial comparing antibiotics with appendectomy for
  appendicitis.
\newblock {\em New England Journal of Medicine}, 383(20):1907--1919.

\bibitem[Crump et~al., 2009]{crump2009dealing}
Crump, R.~K., Hotz, V.~J., Imbens, G.~W., and Mitnik, O.~A. (2009).
\newblock Dealing with limited overlap in estimation of average treatment
  effects.
\newblock {\em Biometrika}, 96(1):187--199.

\bibitem[Di~Maria and Didelez, 2024]{di2024longitudinal}
Di~Maria, C. and Didelez, V. (2024).
\newblock Longitudinal mediation analysis with multilevel and latent growth
  models: A separable effects causal approach.
\newblock {\em BMC Medical Research Methodology}, 24(1):248.

\bibitem[Didelez, 2019]{didelez2019defining}
Didelez, V. (2019).
\newblock Defining causal mediation with a longitudinal mediator and a survival
  outcome.
\newblock {\em Lifetime Data Analysis}, 25:593--610.

\bibitem[Ende et~al., 2024]{ende2024behavioural}
Ende, H.~B., Habib, A.~S., Lim, G., Landau, R., Beilin, Y., and Wong, C.~A.
  (2024).
\newblock Behavioural disorders after prenatal exposure to anaesthesia for
  maternal surgery: Is it the anaesthesia or the surgery? {Comment on Br J
  Anaesth} 2024; 132: 899-910.
\newblock {\em British Journal of Anaesthesia}, pages S0007--0912.

\bibitem[Ginsberg et~al., 2019]{ginsberg2019maternal}
Ginsberg, Y., D'Onofrio, B.~M., Rickert, M.~E., Class, Q.~A., Rosenqvist,
  M.~A., Almqvist, C., Lichtenstein, P., and Larsson, H. (2019).
\newblock Maternal infection requiring hospitalization during pregnancy and
  attention-deficit hyperactivity disorder in offspring: A quasi-experimental
  family-based study.
\newblock {\em Journal of Child Psychology and Psychiatry}, 60(2):160--168.

\bibitem[Han et~al., 2021]{han2021maternal}
Han, V.~X., Patel, S., Jones, H.~F., Nielsen, T.~C., Mohammad, S.~S., Hofer,
  M.~J., Gold, W., Brilot, F., Lain, S.~J., Nassar, N., et~al. (2021).
\newblock Maternal acute and chronic inflammation in pregnancy is associated
  with common neurodevelopmental disorders: a systematic review.
\newblock {\em Translational Psychiatry}, 11(1):71.

\bibitem[Hern{\'a}n, 2018]{hernan2018c}
Hern{\'a}n, M.~A. (2018).
\newblock The c-word: Scientific euphemisms do not improve causal inference
  from observational data.
\newblock {\em American Journal of Public Health}, 108(5):616--619.

\bibitem[Ing et~al., 2012]{ing2012long}
Ing, C., DiMaggio, C., Whitehouse, A., Hegarty, M.~K., Brady, J., von
  Ungern-Sternberg, B.~S., Davidson, A., Wood, A.~J., Li, G., and Sun, L.~S.
  (2012).
\newblock Long-term differences in language and cognitive function after
  childhood exposure to anesthesia.
\newblock {\em Pediatrics}, 130(3):e476--e485.

\bibitem[Ing et~al., 2021]{ing2021prospectively}
Ing, C., Jackson, W.~M., Zaccariello, M.~J., Goldberg, T.~E., McCann, M.-E.,
  Grobler, A., Davidson, A., Sun, L., Li, G., and Warner, D.~O. (2021).
\newblock Prospectively assessed neurodevelopmental outcomes in studies of
  anaesthetic neurotoxicity in children: A systematic review and meta-analysis.
\newblock {\em British Journal of Anaesthesia}, 126(2):433--444.

\bibitem[Ing et~al., 2024]{ing2024behavioural}
Ing, C., Silber, J.~H., Lackraj, D., Olfson, M., Miles, C.~H., Reiter, J.~G.,
  Jain, S., Chihuri, S., Guo, L., Gyamfi-Bannerman, C., et~al. (2024).
\newblock Behavioural disorders after prenatal exposure to anaesthesia for
  maternal surgery.
\newblock {\em British Journal of Anaesthesia}.

\bibitem[Ing et~al., 2022]{ing2022anesthesia}
Ing, C., Warner, D.~O., Sun, L.~S., Flick, R.~P., Davidson, A.~J., Vutskits,
  L., McCann, M.~E., O’Leary, J., Bellinger, D.~C., Rauh, V., et~al. (2022).
\newblock Anesthesia and developing brains: Unanswered questions and proposed
  paths forward.
\newblock {\em Anesthesiology}, 136(3):500--512.

\bibitem[Jevtovic-Todorovic et~al., 2003]{jevtovic2003early}
Jevtovic-Todorovic, V., Hartman, R.~E., Izumi, Y., Benshoff, N.~D., Dikranian,
  K., Zorumski, C.~F., Olney, J.~W., and Wozniak, D.~F. (2003).
\newblock Early exposure to common anesthetic agents causes widespread
  neurodegeneration in the developing rat brain and persistent learning
  deficits.
\newblock {\em Journal of Neuroscience}, 23(3):876--882.

\bibitem[Kennedy, 2019]{kennedy2019nonparametric}
Kennedy, E.~H. (2019).
\newblock Nonparametric causal effects based on incremental propensity score
  interventions.
\newblock {\em Journal of the American Statistical Association},
  114(526):645--656.

\bibitem[Kurth et~al., 2006]{kurth2006results}
Kurth, T., Walker, A.~M., Glynn, R.~J., Chan, K.~A., Gaziano, J.~M., Berger,
  K., and Robins, J.~M. (2006).
\newblock Results of multivariable logistic regression, propensity matching,
  propensity adjustment, and propensity-based weighting under conditions of
  nonuniform effect.
\newblock {\em American Journal of Epidemiology}, 163(3):262--270.

\bibitem[Li et~al., 2018]{li2018balancing}
Li, F., Morgan, K.~L., and Zaslavsky, A.~M. (2018).
\newblock Balancing covariates via propensity score weighting.
\newblock {\em Journal of the American Statistical Association},
  113(521):390--400.

\bibitem[Li and Luedtke, 2023]{li2023efficient}
Li, S. and Luedtke, A. (2023).
\newblock Efficient estimation under data fusion.
\newblock {\em Biometrika}, 110(4):1041--1054.

\bibitem[Loepke and Soriano, 2008]{loepke2008assessment}
Loepke, A.~W. and Soriano, S.~G. (2008).
\newblock An assessment of the effects of general anesthetics on developing
  brain structure and neurocognitive function.
\newblock {\em Anesthesia \& Analgesia}, 106(6):1681--1707.

\bibitem[Machado-Rivas et~al., 2021]{machado2021predictors}
Machado-Rivas, F., Leitman, E., Jaimes, C., Conklin, J., Caruso, P.~A., Liu,
  C.~A., and Gee, M.~S. (2021).
\newblock Predictors of anesthetic exposure in pediatric {MRI}.
\newblock {\em American Journal of Roentgenology}, 216(3):799--805.

\bibitem[Maltzahn et~al., 2024]{maltzahn2024separable}
Maltzahn, N.~N., Mehlum, I.~S., and Gran, J.~M. (2024).
\newblock Separable and controlled direct effects for competing events:
  Estimation of component specific effects on sickness absence.
\newblock {\em Statistics in Medicine}, 43(22):4305--4327.

\bibitem[Miles, 2023]{miles2023causal}
Miles, C.~H. (2023).
\newblock On the causal interpretation of randomised interventional indirect
  effects.
\newblock {\em Journal of the Royal Statistical Society Series B: Statistical
  Methodology}, 85(4):1154--1172.

\bibitem[Neyman, 1923]{neyman1923application}
Neyman, J. (1923).
\newblock On the application of probability theory to agricultural experiments.
  {E}ssay on principles.
\newblock {\em Ann. Agricultural Sciences}, pages 1--51.

\bibitem[Pearl, 1995]{pearl1995causal}
Pearl, J. (1995).
\newblock Causal diagrams for empirical research.
\newblock {\em Biometrika}, 82(4):669--688.

\bibitem[Pearl, 2001]{pearl2001direct}
Pearl, J. (2001).
\newblock Direct and indirect effects.
\newblock In {\em Proceedings of the Seventeenth Conference on Uncertainty in
  Artificial Intelligence}, UAI'01, page 411–420. Morgan Kaufmann Publishers
  Inc.

\bibitem[Pearl, 2009]{pearl_2009}
Pearl, J. (2009).
\newblock {\em Causality}.
\newblock Cambridge University Press.

\bibitem[Reighard et~al., 2022]{reighard2022anesthetic}
Reighard, C., Junaid, S., Jackson, W.~M., Arif, A., Waddington, H., Whitehouse,
  A.~J., and Ing, C. (2022).
\newblock Anesthetic exposure during childhood and neurodevelopmental outcomes:
  A systematic review and meta-analysis.
\newblock {\em JAMA Network Open}, 5(6):e2217427--e2217427.

\bibitem[Robins, 2000]{robins2000marginal}
Robins, J.~M. (2000).
\newblock Marginal structural models versus structural nested models as tools
  for causal inference.
\newblock In {\em Statistical Models in Epidemiology, the Environment, and
  Clinical Trials}, pages 95--133. Springer.

\bibitem[Robins and Greenland, 1992]{robins1992identifiability}
Robins, J.~M. and Greenland, S. (1992).
\newblock Identifiability and exchangeability for direct and indirect effects.
\newblock {\em Epidemiology}, 3(2):143--155.

\bibitem[Robins and Richardson, 2010]{robins_2010}
Robins, J.~M. and Richardson, T.~S. (2010).
\newblock Alternative graphical causal models and the identification of direct
  effects.
\newblock {\em Causality and Psychopathology: Finding the Determinants of
  Disorders and Their Cures}, 84:103--158.

\bibitem[Robins et~al., 2022]{robins2022interventionist}
Robins, J.~M., Richardson, T.~S., and Shpitser, I. (2022).
\newblock An interventionist approach to mediation analysis.
\newblock In {\em Probabilistic and Causal Inference: The Works of Judea
  Pearl}, pages 713--764.

\bibitem[Robins et~al., 1994]{robins1994estimation}
Robins, J.~M., Rotnitzky, A., and Zhao, L.~P. (1994).
\newblock Estimation of regression coefficients when some regressors are not
  always observed.
\newblock {\em Journal of the American Statistical Association},
  89(427):846--866.

\bibitem[Rubin, 1974]{rubin1974estimating}
Rubin, D.~B. (1974).
\newblock Estimating causal effects of treatments in randomized and
  nonrandomized studies.
\newblock {\em Journal of Educational Psychology}, 66(5):688.

\bibitem[Rubin, 1981]{rubin1981bayesian}
Rubin, D.~B. (1981).
\newblock The {B}ayesian bootstrap.
\newblock {\em The Annals of Statistics}, pages 130--134.

\bibitem[Sachs et~al., 2017]{sachs2017risk}
Sachs, A., Guglielminotti, J., Miller, R., Landau, R., Smiley, R., and Li, G.
  (2017).
\newblock Risk factors and risk stratification for adverse obstetrical outcomes
  after appendectomy or cholecystectomy during pregnancy.
\newblock {\em JAMA Surgery}, 152(5):436--441.

\bibitem[Shpitser et~al., 2017]{shpitser2017modeling}
Shpitser, I., Tchetgen~Tchetgen, E.~J., and Andrews, R. (2017).
\newblock Modeling interference via symmetric treatment decomposition.
\newblock {\em arXiv preprint arXiv:1709.01050}.

\bibitem[Singh et~al., 2024]{singh2024anesthetic}
Singh, T., Pitts, A., Miles, C.~H., and Ing, C. (2024).
\newblock Anesthetic exposure during early childhood and neurodevelopmental
  outcomes: Our current understanding.
\newblock {\em Current Anesthesiology Reports}, 14(1):15--24.

\bibitem[Stensrud et~al., 2021]{stensrud2021generalized}
Stensrud, M.~J., Hern{\'a}n, M.~A., Tchetgen~Tchetgen, E.~J., Robins, J.~M.,
  Didelez, V., and Young, J.~G. (2021).
\newblock A generalized theory of separable effects in competing event
  settings.
\newblock {\em Lifetime Data Analysis}, 27(4):588--631.

\bibitem[Stensrud et~al., 2023]{stensrud2023conditional}
Stensrud, M.~J., Robins, J.~M., Sarvet, A., Tchetgen~Tchetgen, E.~J., and
  Young, J.~G. (2023).
\newblock Conditional separable effects.
\newblock {\em Journal of the American Statistical Association},
  118(544):2671--2683.

\bibitem[Stensrud et~al., 2022]{strensrud_2022}
Stensrud, M.~J., Young, J.~G., Didelez, V., Robins, J.~M., and Hern{\'a}n,
  M.~A. (2022).
\newblock Separable effects for causal inference in the presence of competing
  events.
\newblock {\em Journal of the American Statistical Association},
  117(537):175--183.

\bibitem[Svensson et~al., 2015]{svensson2015nonoperative}
Svensson, J.~F., Patkova, B., Almstr{\"o}m, M., Naji, H., Hall, N.~J., Eaton,
  S., Pierro, A., and Wester, T. (2015).
\newblock Nonoperative treatment with antibiotics versus surgery for acute
  nonperforated appendicitis in children: A pilot randomized controlled trial.
\newblock {\em Annals of Surgery}, 261(1):67--71.

\bibitem[{U.S. Food and Drug Administration (FDA)}, 2017]{FDA}
{U.S. Food and Drug Administration (FDA)} (2017).
\newblock {FDA} drug safety communication: {FDA} approves label changes for use
  of general anesthetic and sedation drugs in young children.
\newblock
  https://www.fda.gov/drugs/drug-safety-and-availability/fda-drug-safety-communication-fda-approves-label-changes-use-general-anesthetic-and-sedation-drugs.

\bibitem[van~der Laan and Petersen, 2008]{van2008direct}
van~der Laan, M.~J. and Petersen, M.~L. (2008).
\newblock Direct effect models.
\newblock {\em The International Journal of Biostatistics}, 4(1).

\bibitem[VanderWeele, 2011]{vanderweele2011causal}
VanderWeele, T.~J. (2011).
\newblock Causal mediation analysis with survival data.
\newblock {\em Epidemiology}, 22(4):582--585.

\bibitem[VanderWeele and Hernan, 2013]{vanderweele2013causal}
VanderWeele, T.~J. and Hernan, M.~A. (2013).
\newblock Causal inference under multiple versions of treatment.
\newblock {\em Journal of Causal Inference}, 1(1):1--20.

\bibitem[VanderWeele et~al., 2014]{vanderweele2014effect}
VanderWeele, T.~J., Vansteelandt, S., and Robins, J.~M. (2014).
\newblock Effect decomposition in the presence of an exposure-induced
  mediator-outcome confounder.
\newblock {\em Epidemiology}, 25(2):300--306.

\bibitem[Vinson et~al., 2021]{vinson2021trends}
Vinson, A.~E., Peyton, J., Kordun, A., Staffa, S.~J., and Cravero, J. (2021).
\newblock Trends in pediatric {MRI} sedation/anesthesia at a tertiary medical
  center over time.
\newblock {\em Pediatric Anesthesia}, 31(9):953--961.

\bibitem[Vutskits, 2024]{vutskits2024developmental}
Vutskits, L. (2024).
\newblock Developmental anaesthesia neurotoxicity in humans: Finding the sweet
  spot?
\newblock {\em British Journal of Anaesthesia}.

\bibitem[Vutskits and Xie, 2016]{vutskits2016lasting}
Vutskits, L. and Xie, Z. (2016).
\newblock Lasting impact of general anaesthesia on the brain: Mechanisms and
  relevance.
\newblock {\em Nature Reviews Neuroscience}, 17(11):705--717.

\bibitem[Wanis et~al., 2024]{wanis2024separable}
Wanis, K.~N., Stensrud, M.~J., and Sarvet, A.~L. (2024).
\newblock Separable effects for adherence.
\newblock {\em American Journal of Epidemiology}, page kwae277.

\bibitem[Wilder et~al., 2009]{wilder2009early}
Wilder, R., Flick, R., Sprung, J., Katusic, S., Barbaresi, W., Mickelson, C.,
  Gleich, S., Schroeder, D., Weaver, A., and Warner, D. (2009).
\newblock Early exposure to anesthesia and learning disabilities in a
  population-based birth cohort.
\newblock {\em Anesthesiology}, 110(4):796--804.

\bibitem[Yang et~al., 2019]{yang2019meta}
Yang, Z., Sun, F., Ai, S., Wang, J., Guan, W., and Liu, S. (2019).
\newblock Meta-analysis of studies comparing conservative treatment with
  antibiotics and appendectomy for acute appendicitis in the adult.
\newblock {\em BMC Surgery}, 19:1--10.

\bibitem[Young et~al., 2020]{Young2020causal}
Young, J.~G., Stensrud, M.~J., Tchetgen~Tchetgen, E.~J., and Hern{\'a}n, M.~A.
  (2020).
\newblock A causal framework for classical statistical estimands in
  failure-time settings with competing events.
\newblock {\em Statistics in Medicine}, 39(8):1199--1236.

\bibitem[Zadrozny, 2004]{zadrozny2004learning}
Zadrozny, B. (2004).
\newblock Learning and evaluating classifiers under sample selection bias.
\newblock In {\em Proceedings of the Twenty-First International Conference on
  Machine Learning}, page 114.

\end{thebibliography}

\newpage 
\appendix
%\section{Figures and Tables}

%----------------------------------------
% new data gen 
%---------------------------------------

\section{Data Source Appendix} \label{a:data_source_a}

\subsection{Table of baseline covariates} \label{a:baseline_table}

\begin{longtable}{lrrr} 
\caption{Selected list of covariates and standardized differences from the MAX data from 1999--2013.} 
\label{tab:tab1} \\ \toprule
& \textbf{Unexposed}  & \textbf{Exposed}  & \textbf{SMD}$^1$  \\ 
&  ($n=14,307,152$) &  ($n=31,494$) &  \\ \midrule
\multicolumn{4}{l}{ } \\
\textbf{Child sex} & & & \\
Female                                                       & 7026294 (49.11\%)                                                        & 15538 (49.34\%)                                                    & 0.005                   \\
Male                                                         & 7280858 (50.89\%)                                                        & 15956 (50.66\%)                                                    & -0.005                  \\
\multicolumn{4}{l}{\textbf{Child year of birth}} \\ 
1999                                                         & 591827 (4.14\%)                                                          & 605 (1.92\%)                                                       & -0.130                  \\
2000                                                         & 719627 (5.03\%)                                                          & 1175 (3.73\%)                                                      & -0.064                  \\
2001                                                         & 741219 (5.18\%)                                                          & 1647 (5.23\%)                                                      & 0.002                   \\
2002                                                         & 802861 (5.61\%)                                                          & 1809 (5.74\%)                                                      & 0.006                   \\
2003                                                         & 830586 (5.81\%)                                                          & 1957 (6.21\%)                                                      & 0.017                   \\
2004                                                         & 938407 (6.56\%)                                                          & 2014 (6.39\%)                                                      & -0.007                  \\
2005                                                         & 996852 (6.97\%)                                                          & 2207 (7.01\%)                                                      & 0.002                   \\
2006                                                         & 1037046 (7.25\%)                                                         & 2371 (7.53\%)                                                      & 0.011                   \\
2007                                                         & 1049494 (7.34\%)                                                         & 2250 (7.14\%)                                                      & -0.007                  \\
2008                                                         & 1081052 (7.56\%)                                                         & 2395 (7.6\%)                                                       & 0.002                   \\
2009                                                         & 1095352 (7.66\%)                                                         & 2594 (8.24\%)                                                      & 0.021                   \\
2010                                                         & 1058182 (7.4\%)                                                          & 2499 (7.93\%)                                                      & 0.020                   \\
2011                                                         & 1097355 (7.67\%)                                                         & 2567 (8.15\%)                                                      & 0.018                   \\
2012                                                         & 1156106 (8.08\%)                                                         & 2842 (9.02\%)                                                      & 0.034                   \\
2013                                                         & 1111186 (7.77\%)                                                         & 2562 (8.13\%)                                                      & 0.014                   \\
%\multicolumn{4}{l}{\parbox{15cm}{\textbf{Maternal Age}}}  \\
\multicolumn{4}{l}{\textbf{Maternal Age}} \\  
18 under                                                     & 755834 (5.28\%)                                                          & 1124 (3.57\%)                                                      & -0.083                  \\
18-24                                                        & 7134138 (49.86\%)                                                        & 16780 (53.28\%)                                                    & \textbf{0.068}          \\
25-29                                                        & 3536477 (24.72\%)                                                        & 8334 (26.46\%)                                                     & 0.040                   \\
30-34                                                        & 1864778 (13.03\%)                                                        & 3616 (11.48\%)                                                     & -0.047                  \\
35-39                                                        & 813453 (5.69\%)                                                          & 1336 (4.24\%)                                                      & -0.066                  \\
40-44                                                        & 191536 (1.34\%)                                                          & 292 (0.93\%)                                                       & -0.039                  \\
45 plus                                                      & 10936 (0.08\%)                                                           & 12 (0.04\%)                                                        & -0.016                  \\
\multicolumn{4}{l}{\textbf{Maternal Race}}\\
Hawaiian/Pacific Islander                                    & 86187 (0.6\%)                                                            & 131 (0.42\%)                                                       & -0.026                  \\
American Indian / Alaskan Native                             & 219326 (1.53\%)                                                          & 911 (2.89\%)                                                       & \textbf{0.093}          \\
Asian                                                        & 266677 (1.86\%)                                                          & 238 (0.76\%)                                                       & -0.098                  \\
Aferican American/Black                                      & 2462449 (17.21\%)                                                        & 3156 (10.02\%)                                                     & -0.211                  \\
White                                                        & 4516172 (31.57\%)                                                        & 12751 (40.49\%)                                                    & \textbf{\textit{0.187}}          \\
Unknown                                                      & 6756341 (47.22\%)                                                        & 14307 (45.43\%)                                                    & -0.036                  \\
\multicolumn{4}{l}{  \textbf{Maternal Ethnicity}} \\ 
Hispanic                                                     & 2864336 (20.02\%)                                                        & 6255 (19.86\%)                                                     & -0.004                  \\
Non-Hispanic                                                 & 6701014 (46.84\%)                                                        & 15542 (49.35\%)                                                    & \textbf{0.050}          \\
Unknown                                                      & 4741802 (33.14\%)                                                        & 9697 (30.79\%)                                                     & -0.050                  \\
\multicolumn{4}{l}{  \textbf{Income by zip code} }\\
Quartile 1                                                   & 3350807 (23.42\%)                                                        & 6782 (21.53\%)                                                     & -0.045                  \\
Quartile 2                                                   & 3433539 (24\%)                                                           & 7762 (24.65\%)                                                     & 0.015                   \\
Quartile 3                                                   & 3424101 (23.93\%)                                                        & 8007 (25.42\%)                                                     & 0.035                   \\
Quartile 4                                                   & 3416039 (23.88\%)                                                        & 7630 (24.23\%)                                                     & 0.008                   \\
Unknown                                                      & 682666 (4.77\%)                                                          & 1313 (4.17\%)                                                      & -0.029                  \\
Enrolled in Medicaid for   disability (\%)                   & 201822 (1.41\%)                                                          & 609 (1.93\%)                                                       & 0.041                   \\
%\multicolumn{4}{l}{\parbox{15cm}{\textbf{Filled prescriptions   for psychotropic medications prior to exposure}}}      
\multicolumn{4}{l}{  \textbf{Filled prescriptions for psychotropic medications prior to exposure} }\\
Antidepressants                                              & 443094 (3.1\%)                                                           & 2197 (6.98\%)                                                      & \textbf{\textit{0.178}}         \\
ADHD medication                                              & 37241 (0.26\%)                                                           & 139 (0.44\%)                                                       & 0.031                   \\
Antipsychotics                                               & 64741 (0.45\%)                                                           & 306 (0.97\%)                                                       & \textbf{0.062}          \\
Mood Stabilizers                                             & 77440 (0.54\%)                                                           & 354 (1.12\%)                                                       & \textbf{0.064}          \\
Sedative/anxiolytics                                         & 211586 (1.48\%)                                                          & 1084 (3.44\%)                                                      & \textbf{\textit{0.127}}          \\
\multicolumn{4}{l}{  \textbf{Maternal medical comorbidities} }\\
Asthma                                                       & 259107 (1.81\%)                                                          & 1190 (3.78\%)                                                      & \textbf{\textit{0.120}}          \\
Cardiac valvular disease                                     & 25714 (0.18\%)                                                           & 81 (0.26\%)                                                        & 0.017                   \\
Chronic ischemic heart disease                               & 6581 (0.05\%)                                                            & 27 (0.09\%)                                                        & 0.015                   \\
Chronic liver disease                                        & 19935 (0.14\%)                                                           & 273 (0.87\%)                                                       & \textbf{\textit{0.103}}          \\
Chronic congestive heart   failure                           & 264 (0\%)                                                                & (-) (0\%)                                                          & 0.003                   \\
Congenital heart disease                                     & 26549 (0.19\%)                                                           & 76 (0.24\%)                                                        & 0.012                   \\
Human immunodeficiency virus                                 & 11809 (0.08\%)                                                           & 21 (0.07\%)                                                        & -0.006                  \\
Overweight or obesity                                        & 168385 (1.18\%)                                                          & 940 (2.98\%)                                                       & \textbf{\textit{0.127}}          \\
Pre-existing diabetes                                        & 127642 (0.89\%)                                                          & 476 (1.51\%)                                                       & \textbf{0.057}          \\
Pre-existing hypertension                                    & 157429 (1.1\%)                                                           & 706 (2.24\%)                                                       & \textbf{0.089}          \\
Pulmonary hypertension                                       & 1869 (0.01\%)                                                            & (-) (0.02\%)                                                       & 0.005                   \\
Sickle cell disease                                          & 5487 (0.04\%)                                                            & 27 (0.09\%)                                                        & 0.019                   \\
Smoking/Tobacco Use                                          & 274236 (1.92\%)                                                          & 1384 (4.39\%)                                                      & \textbf{\textit{0.142}}          \\
Underweight                                                  & 2164 (0.02\%)                                                            & (-) (0.01\%)                                                       & -0.005                  \\
Chronic renal disease                                        & 24279 (0.17\%)                                                           & 176 (0.56\%)                                                       & \textbf{0.065}          \\
\multicolumn{4}{l}{  \textbf{Maternal medical comorbidities associated with appendectomy or cholecystectom} }\\
Diverticular disease                                         & 3122 (0.02\%)                                                            & 30 (0.1\%)                                                         & 0.030                   \\
Gout                                                         & 380 (0\%)                                                                & (-) (0.01\%)                                                       & 0.011                   \\
Irritable bowel syndrom                                      & 14993 (0.1\%)                                                            & 117 (0.37\%)                                                       & \textbf{0.055}          \\
Hypercholesterolemia                                         & 19689 (0.14\%)                                                           & 72 (0.23\%)                                                        & 0.021                   \\
Thalassemia                                                  & 1514 (0.01\%)                                                            & (-) (0.01\%)                                                       & 0.002                   \\
Urinary calculus                                             & 40365 (0.28\%)                                                           & 381 (1.21\%)                                                       & \textbf{\textit{0.108}}          \\
\multicolumn{4}{l}{  \textbf{Pre-pregnancy healthcare utalization}}\\
Emergency room visits                                        & 564116 (0.04\%)                                                          & 2934 (0.09\%)                                                      & \textbf{\textit{0.181}}          \\
Outpatient visits                                            & 6929477 (0.48\%)                                                         & 28315 (0.9\%)                                                      & \textbf{\textit{0.301}}          \\
Inpatient visits                                             & 226166 (0.02\%)                                                          & 1230 (0.04\%)                                                      & \textbf{\textit{0.133}}          \\
\multicolumn{4}{l}{  \textbf{Conditions during pregnancy prior to exposure} }\\
Placental abruption                                          & 10934 (0.08\%)                                                           & 40 (0.13\%)                                                        & 0.016                   \\
Placenta accreta                                             & 2298 (0.02\%)                                                            & (-) (0\%)                                                          & -0.013                  \\
Chorioamnionitis                                             & 6489 (0.05\%)                                                            & 18 (0.06\%)                                                        & 0.005                   \\
Gestational diabetes                                         & 80158 (0.56\%)                                                           & 255 (0.81\%)                                                       & 0.030                   \\
Early or threatened labour                                   & 283997 (1.99\%)                                                          & 930 (2.95\%)                                                       & \textbf{0.062}          \\
Feotal heart rate abnormality                                & 77226 (0.54\%)                                                           & 217 (0.69\%)                                                       & 0.019                   \\
Gestational hypertension                                     & 29123 (0.2\%)                                                            & 80 (0.25\%)                                                        & 0.011                   \\
Multiple gestation                                           & 112510 (0.79\%)                                                          & 330 (1.05\%)                                                       & 0.027                   \\
Oligohydramnios                                              & 29533 (0.21\%)                                                           & 51 (0.16\%)                                                        & -0.010                  \\
Preeclampsia - Mild                                          & 18242 (0.13\%)                                                           & 44 (0.14\%)                                                        & 0.003                   \\
Preeclampsia/eclampsia - Severe                              & 6903 (0.05\%)                                                            & (-) (0.03\%)                                                       & -0.010                  \\
Polyhydramnios                                               & 14430 (0.1\%)                                                            & 26 (0.08\%)                                                        & -0.006                  \\
Premature rupture of membranes                               & 6846 (0.05\%)                                                            & (-) (0.02\%)                                                       & -0.014                  \\
Placenta previa                                              & 80017 (0.56\%)                                                           & 292 (0.93\%)                                                       & 0.043                   \\
\multicolumn{4}{l}{  \textbf{Maternal psychiatric comorbidities}}\\
ADHD and other DBD            & 55932 (0.39\%)                                                           & 182 (0.58\%)                                                       & 0.027                   \\
Adjustment disorder                                          & 71312 (0.5\%)                                                            & 332 (1.05\%)                                                       & \textbf{0.063}          \\
Alcohol related disorders                                    & 41105 (0.29\%)                                                           & 183 (0.58\%)                                                       & 0.045                   \\
Anxiety disorder                                             & 239485 (1.67\%)                                                          & 1169 (3.71\%)                                                      & \textbf{\textit{0.126}}          \\
Bipolar disorder                                             & 77134 (0.54\%)                                                           & 408 (1.3\%)                                                        & \textbf{0.079}          \\
Depression                                                   & 311910 (2.18\%)                                                          & 1604 (5.09\%)                                                      & \textbf{\textit{0.156}}          \\
Substance related disorder                                   & 153557 (1.07\%)                                                          & 563 (1.79\%)                                                       & \textbf{0.060}          \\
Eating disorder                                              & 3734 (0.03\%)                                                            & (-) (0.03\%)                                                       & 0.005                   \\
Other mental disorders                                       & 97697 (0.68\%)                                                           & 447 (1.42\%)                                                       & \textbf{0.072}          \\
Schizophrenia                                                & 17835 (0.12\%)                                                           & 64 (0.2\%)                                                         & 0.019                   \\
\multicolumn{4}{l}{  \textbf{Medicaid   eligibility prior to exposure (mean months)} }  \\
Not eligible for Medicaid                                  & 58450447 (4.09\%)                                                        & 100898 (3.2\%)                                                     & -0.298                  \\
Comprehensive   plan only                                    & 11031135 (0.77\%)                                                        & 26693 (0.85\%)                                                     & 0.037                   \\
Dental   plan only                                           & 2235078 (0.16\%)                                                         & 5458 (0.17\%)                                                      & 0.019                   \\
Behavioral   plan only                                       & 2784909 (0.19\%)                                                         & 10403 (0.33\%)                                                     & \textbf{\textit{0.121}}          \\
Primary   care case management plan only                     & 3465580 (0.24\%)                                                         & 11634 (0.37\%)                                                     & \textbf{0.096}          \\
Other   managed care plan only                               & 2348135 (0.16\%)                                                         & 7234 (0.23\%)                                                      & \textbf{0.073}          \\
Comprehensive   and dental plan                              & 3704722 (0.26\%)                                                         & 8903 (0.28\%)                                                      & 0.018                   \\
Comprehensive and behavioral plan                            & 3111115 (0.22\%)                                                         & 9493 (0.3\%)                                                       & \textbf{0.068}          \\
Comprehensive and other managed care plan                  & 1156348 (0.08\%)                                                         & 2848 (0.09\%)                                                      & 0.014                   \\
Comprehensive, dental, and behavioral plan                 & 495525 (0.03\%)                                                          & 2090 (0.07\%)                                                      & \textbf{0.059}          \\
PCCM and dental plan               & 130353 (0.01\%)                                                          & 869 (0.03\%)                                                       & \textbf{0.059}          \\
PCCM and behavioral plan           & 1112359 (0.08\%)                                                         & 3997 (0.13\%)                                                      & \textbf{0.062}          \\
PCCM and other managed care plan   & 916312 (0.06\%)                                                          & 3439 (0.11\%)                                                      & \textbf{0.067}          \\
PCCM, dental, and behavioral plan  & 9871 (0\%)                                                               & 45 (0\%)                                                           & 0.009                   \\
Dental and behavioral plan                                 & 118284 (0.01\%)                                                          & 555 (0.02\%)                                                       & 0.037                   \\
Other combinations                                         & 555900 (0.04\%)                                                          & 2007 (0.06\%)                                                      & 0.047                   \\
Fee   for service                                            & 27541538 (1.93\%)                                                        & 65799 (2.09\%)                                                     & \textbf{0.067}          \\
Managed care plan status is unknown                        & 40936 (0\%)                                                              & 46 (0\%)                                                           & -0.016                  \\
\multicolumn{4}{l}{  \textbf{Private insurance prior to exposure (mean months)} }  \\
%Not eligible for Medicaid or no private insurance coverage 
No Medicaid or private insurance & 115657969 (8.08\%)                                                       & 255647 (8.12\%)                                                    & 0.013                   \\
Private insurance by 3rd party or state        & 3505586 (0.25\%)                                                         & 6691 (0.21\%)                                                      & -0.031                  \\
Invalid or missing data                                    & 43764 (0\%)                                                              & 73 (0\%)                                                           & -0.007                  \\
\multicolumn{4}{l}{  \textbf{Restricted benefits prior to exposure (mean months)} } \\
Entitled to full scope of Medicaid benefits                & 58481325 (4.09\%)                                                        & 100934 (3.2\%)                                                     & -0.299                  \\
Restricted benefits on alien status                  & 45095576 (3.15\%)                                                        & 128197 (4.07\%)                                                    & \textbf{\textit{0.296}}          \\
Restricted benefits on Medicaid dual eligibility     & 5778858 (0.4\%)                                                          & 10765 (0.34\%)                                                     & -0.044                  \\
Restricted benefits pregnancy-related services         & 22038 (0\%)                                                              & 18 (0\%)                                                           & -0.013                  \\
Restricted benefits pregnancy-related services         & 6678445 (0.47\%)                                                         & 15706 (0.5\%)                                                      & 0.024                   \\
Restricted benefits other reasons                      & 1221002 (0.09\%)                                                         & 2528 (0.08\%)                                                      & -0.008                  \\
Only family planning services                              & 1854997 (0.13\%)                                                         & 3949 (0.13\%)                                                      & -0.006                  \\
Benefit restrictions are unknown                           & 28987 (0\%)                                                              & 44 (0\%)                                                           & -0.010 \\ \bottomrule
\multicolumn{4}{l}{\textbf{Abbreviations}: SMD: standardized mean differences,}\\ 
\multicolumn{4}{l}{PCCM: primary care case management,} \\
\multicolumn{4}{l}{DBD: disruptive behavioral disorders. } \\
\multicolumn{4}{l}{$^1$Bold text identifies cells where SMD $>0.05$ or $<0.1$;}\\ 
\multicolumn{4}{l}{bold \& italic text identifies cells where SMD$>0.1$;}\\
\multicolumn{4}{l}{(-) denotes a number $<11$ that cannot be displayed.}\\
\end{longtable}

\subsection{Algorithm for establishing baseline variables}
\label{a:algorithm_baseline}

\begin{figure}[H]
    \centering
    \includegraphics[scale=0.45]{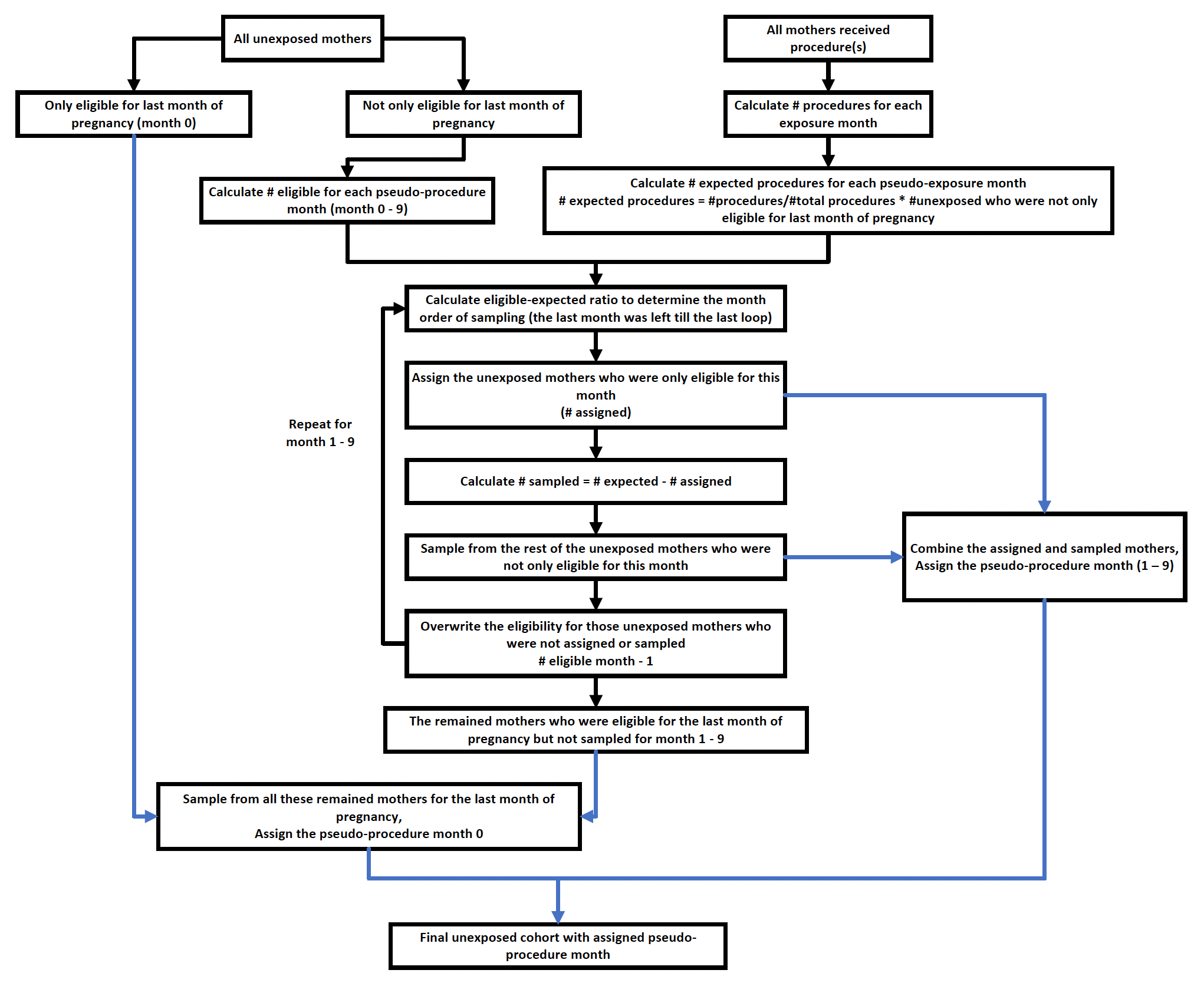}
    \caption{Algorithm to establish pseudo-procedure months for the unexposed group, which were used to determine the observation windows for the baseline covariates $\bf C$. }
    \label{fig:flow_chart}
\end{figure}

We used a previously constructed nationwide cohort of Medicaid insured mothers with appendectomy or cholecystectomy during pregnancy between 1999 and 2013. These mothers were eligible for Medicaid at time of delivery, and an estimated last menstrual period (LMP) was calculated for all the mothers. The observation window for the mothers was set from 90 days prior to the estimated LMP or first day of enrollment, whichever came last, to the date of delivery. Thus, the mothers were observed up to 335 days for those who had preterm deliveries, and up to 360 days for non-preterm deliveries. All mothers' claims billed during this observation window were evaluated. %To make the unexposed and exposed groups as comparable as possible, the observation period for baseline variables needs to be established for both groups. 
For the exposed group, baseline variables were collected from 90 days prior to the estimated LMP up until the time of procedure. To achieve a similar distribution of observation periods for the unexposed group, we used the algorithm in Figure \ref{fig:flow_chart} to produce a pseudo-procedure date for the unexposed mothers as opposed to using the full duration of their pregnancy as their observation window. %as the baseline observation window for the unexposed group. Creating the pseudo-procedure date for the unexposed group ensures that unexposed mothers did not have a longer period to observe baseline variables than that of exposed mothers. 

%The algorithm, displayed in Figure \ref{fig:flow_chart}, shows the decision-making steps. The main goal is to assign the unexposed mothers a pseudo-procedure month to help establish the baseline variable observation period. Ideally, the distribution of pseudo-procedures per month resembles the distribution of observed procedures per month. 
The maximum observation period was 360 days, and observation months were labeled in reverse order from month zero being the month of delivery to month nine being nine months prior to the delivery, estimated month of LMP. We first determined the proportion of total procedures that were observed during each month of pregnancy, then calculated the number of unexposed mothers per month that would yield the same distribution among the unexposed mothers by multiplying this proportion by the total number of unexposed who were eligible for at least one month prior to the month of delivery.  We excluded mothers who were only eligible during the last month of pregnancy in this step due to an imbalanced number of mothers who fell into this category in the unexposed group compared to the exposed group. These mothers were handled in the final step. We refer to these as the numbers of expected pseudo-procedure dates per month.

%Using these numbers, we then assigned unexposed mothers a pseudo-procedure month in an algorithmic fashion. 
Saving month zero for last, we selected the month with the smallest ratio of unexposed mothers who were eligible for Medicaid during that month. % and several expected pseudo-procedures. 
We then assigned the mothers who were only eligible for Medicaid during that month to this pseudo-procedure month, and then randomly selected other eligible mothers until achieving the desired number of expected pseudo-procedures. This process was then repeated until all months except month zero were completed.

After assigning unexposed mothers to pseudo-procedure months in months one to nine, we combined the remaining unexposed mothers with the unexposed mothers who were only eligible for the month of deliveries. Thus, for month zero, the majority of the mothers to be sampled enrolled in Medicaid right before deliveries, which meant that we may not have obtained enough Medicaid claims to evaluate baseline characteristics. To adjust for unmeasured confounding due to such limited healthcare utilization information, the remaining mothers not assigned pseudo-procedures month were assigned month zero. The mothers who were only eligible for month zero were randomly selected until we achieved the total expected pseudo-procedure for month zero. After this sampling process, 737,141 mothers who were only eligible for the month of delivery were excluded from the final analysis, accounting for 4.9\% of all unexposed mothers. %There were 737,141 left and they were excluded from the analysis. % need to write why we did this.

%In the end, we constructed the sub-cohort for the prenatal anesthetic exposure for the mothers who received procedures and those who did not. The total number of unexposed mothers were those who were not only eligible for the month of delivery and the distribution of pseudo-procedure month is the same as the exposed mothers.

\subsection{ ${\bf M}$ variable ICD-9 procedure, ICD-9 diagnosis and CPT codes}

\begin{table}[H] \footnotesize
    \centering
    \begin{tabular}{p{0.44\linewidth}|p{0.54\linewidth}} \toprule 
        \textbf{$\bf M$ Variable Label}  & \textbf{Codes} \\ \hline 
         Peri-operative complications: & \\
         \quad Post-operative infection  &  \textit{ICD-9 Diagnosis codes:} 038.0--038.12, 038.19, 038.2, 038.3, 038.40--038.44, 038.49, 038.8, 038.9, 567.22, 569.5, 785.52, 785.59, 995.91, 995.92, 998.0\\
        \quad Hemorrhage & \textit{ICD-9 Diagnosis codes:} 998.11--998.12
         \\
        \quad & \textit{ICD-9 Procedure codes:}  18.09, 28.7, 38.80--38.89, 39.41, 39.98, 49.95, 54.0, 54.12, 57.93, 59.19, 60.94, 61.0, 69.98, 70.14, 71.09, 75.91, 75.92, 86.04 \\ 
         \quad Wound dehiscence &  \textit{ICD-9 Procedure codes:} 54.61 \\
        \quad Abdominopelvic accidental puncture/laceration &  \textit{ICD-9 Diagnosis codes:} %9982, E8700-E8709.\\
        998.2,E870.0--E870.9 \\
         \quad Blood transfusion &  \textit{ICD-9 Diagnosis codes:} %9996, 9997, E8760, \\
         999.6, 999.7, E876.0\\
         & \textit{ICD-9 Procedure codes:} 99.04, 99.05\\%9904, 9905. \\
         \quad Presence of a foreign body &  \textit{ICD-9 Diagnosis code:} 998.4, 998.7, E871.0--E871.9 \\ %9984, 9987, E8710--E8719.\\
         \quad Intra-operative or peri-operative aspiration &  \textit{ICD-9 Diagnosis codes:} 507.0 \\ %5070. \\
         \quad Post-operative intubation &  \textit{ICD-9 Procedure codes:} 96.04 \\ \hline % 9604.\\ \hline 
     Need for re-operation within 30 days &  \textit{ICD-9 Procedure codes:} 51.37, 51.79, 51.84, 51.85, 51.87, 51.88, 51.98, 52.94, 54.11, 54.19, 54.21, 54.51, 54.59, 87.53\\ 
     %5185, 5187, 5188, 5421, 5451, 5184, 5419, 5179, 8753, 5137, 5198, 5294, 5411, 5459\\ 
     & \textit{CPT codes:} 43260--43262, 43264, 43265, 43268, 43269, 43271, 44180, 47420, 47510, 47511, 47555, 47579, 47900, 49000, 49002, 49320, 49322, 49323, 49329, 58660 \\ \hline %43262, 43264, 43268, 43260, 49320, 43261, 43265, 43269, 43271, 44180, 47420, 47510, 47511, 47555, 47579, 47900, 49000, 49002, 49322, 49323, 49329, 58660 \\ \hline 
     %
    %Pain (defined as opioid prescriptions $> 7$ days) plus refill & \\ \hline 
    %
    %Peri-operative length of stay $> 7$ days & \\ \hline 
    %
     Cesarean delivery &  \textit{ICD-9 Procedure codes:}% 740, 741, 742, 744, 7499\\
     74.0--74.2, 74.4, 74.99 \\ 
     &  \textit{CPT codes:} 59510, 59514, 59515\\ 
    \hline 
     Pre-term birth and low birth weight &  \textit{ICD-9 Diagnosis codes:} 644.00, 644.20, 765.00--765.29%64400, 64420, 76500--76529 
     \\ \hline 
     Thromboembolism &  \textit{ICD-9 Diagnosis codes:} 
     415.11, 415.19, 451.11, 451.19, 451.2, 451.81, 451.9, 453.40--453.42, 453.9 \\ %41511, 41519, 45111, 45119, 4512, 45181, 4519, 45340--45342, 4539\\ %\hline 
     %ER visits unrelated to prenatal care & \\ 
     \bottomrule 
    \end{tabular}
    \caption{ICD-9 procedure, ICD-9 diagnosis and CPT codes used to create the post-exposure variables related to surgical procedure.}
    \label{tab:m_icd_9_codes}
\end{table}

\section{Identification of $\psi_{01}$} \label{a:identification}
Here, we reproduce a version of the identification result from \cite{robins_2010} adapted to our setting based on assumptions given in Section \ref{section:assumptions}. Observe that no part of this derivation depends on positivity of $O$ with respect to $N$ or vice versa.
\begin{align*}
    \psi_{01} &= Pr\{Y(n=0, o=1)\leq t\}\\
    &= \sum_{m,c} \int^t_{-\infty} f^{int}_{n=0,o=1}(y,m,c) dy \\
    &= \sum_{m,c} \int^t_{-\infty} f(y\mid O=1, m,c)f(m\mid N=0,c)f(c) dy &&\text{\small(by the truncated g-formula)} \\
    %&= \sum_{m,c} \int^t_{-\infty} f(y\mid O=1, N=1, m,c) f(m\mid N=0, O=0,c)f(c) dy &&\text{\small(since $O=N$)}  \\
    &=\sum_{m,c} \int^t_{-\infty} f(y\mid A=1, m, c) f(m\mid A=0, c)f(c)  dy &&\text{\small\text{\small(since $O=N=A$)}}\\
    &= \sum_{m,c} Pr(Y\leq t \mid A=1, M=m,C=c )f(m\mid A=0,c)f(c),
\end{align*}
where $f^{int}_{n=0,o=1}(y,m,c)$ is the joint distribution of $Y$, $M$, and $C$ under the intervention setting $N=0$ and $O=1$.

\section{Connection to Front-Door Adjustment}
\label{a:fdc}
% \subsection{Proof for equation $Pr\{Y(n=0) \leq t\}$}
% \label{a:fdc1}
% {\footnotesize
% \begin{align*}
%     \sum_{o=0}^1 \mathbb{E}\{Y(n,o)\}Pr(O=o) &=  \sum_{a=0}^1 \sum_m \mathbb{E}(Y\mid O=a, N=a, M=m) Pr(M=m \mid N=0, O=0) Pr(A=a) \\
%     &\overset{A7}{=} \sum_m Pr(M=m \mid N=0) \sum_{a=0}^1 \mathbb{E}[Y\mid O=a, N=a, M=m] Pr(A=a) \\
%     &\overset{A5}{=} \sum_m Pr(M=m \mid N=0) \sum_{n=0}^1 \mathbb{E}[Y\mid N=n, M=m]  Pr(N=n)\\
%     &= \mathbb{E}[Y(n=0)].
% \end{align*}}

% \subsection{Proof for equation \ref{eq:fdc_secondproof}}
% \label{a:fdc2}
% {\footnotesize
% \begin{align*}
%     \mathbb{E}[Y(n=0,o=1)] &= \mathbb{E}[Y(n=0,o=1)] + \mathbb{E}[Y(a=0)] \frac{Pr(A=0)}{Pr(A=1)} - \mathbb{E}[Y(a=0)] \frac{Pr(A=0)}{Pr(A=1)} \\
%     &= \left\{ \mathbb{E}[Y(n=0,o=1)] + \mathbb{E}[Y(a=0)]Pr(A=0) - \mathbb{E}[Y(a=0)]Pr(A=0)\right\}  / Pr(A=1) \\
%     &= \left\{ \sum_O \mathbb{E}[Y(n,o)]P(O=o) - \mathbb{E}[Y(a=0)]Pr(A=0)\right\}  / Pr(A=1) \\
%     &= \left\{ \mathbb{E}[Y(n=0)]- \mathbb{E}[Y(a=0)]Pr(A=0)\right\}  / Pr(A=1). 
% \end{align*}
% }

%\subsection{Connection of identification to Section \ref{section:identification_est}}
%\label{a:fdc3}
\begin{align*}
    Pr&\{Y(n=0,o=1)\leq t\} \\
    &= \left[Pr\{Y(n=0)\leq t \}- Pr\{Y(a=0)\leq t \}Pr(O=0)\right] / Pr(O=1) \\
    &= \sum_m Pr(M=m \mid N=0) \sum_{n} Pr(Y \leq t \mid N=n, M=m ) Pr(N=n)/ Pr(O=1) \\
    &\quad - Pr\{Y(a=0)\leq t \}Pr(O=0)/ Pr(O=1) \\
    &=  \sum_m Pr(M=m \mid N=0) Pr(Y \leq t \mid N=0, M=m ) Pr(N=0)/ Pr(O=1) \\
    &\quad +\sum_m Pr(M=m \mid N=0) Pr(Y \leq t \mid N=1, M=m ) Pr(N=1) / Pr(O=1)  \\
    &\quad - Pr\{Y(a=0)\leq t \}Pr(O=0) / Pr(O=1) \\
    &=  \sum_m Pr(M=m \mid A=0) Pr(Y \leq t \mid A=0, M=m ) Pr(A=0)/ Pr(A=1) \\
    &\quad +\sum_m Pr(M=m \mid A=0) Pr(Y \leq t \mid A=1, M=m ) Pr(A=1) / Pr(A=1)  \\
    &\quad - Pr\{Y(a=0)\leq t \}Pr(A=0) / Pr(A=1) \\
    &=  Pr(Y \leq t \mid A=0) Pr(A=0)/ Pr(A=1) \\
    &\quad +\sum_m Pr(M=m \mid A=0) Pr(Y \leq t \mid A=1, M=m )   \\
    &\quad - Pr(Y\leq t \mid A=0)Pr(A=0) / Pr(A=1) \\
    &= \sum_m Pr(Y \leq t \mid A=1, M=m ) Pr(M=m \mid A=0).
\end{align*}
This aligns with the identification formula for $\Psi_{01}(P)$ when $C$ is the empty set, and can easily be extended to cases with nonempty $C$ by conditioning on $C$ throughout.

\section{Sensitivity Analysis Results}

\subsection{Sensitivity to violations of assumption \ref{assn:ny}} \label{a:sens_derivation_ny}

Here we consider the sensitivity of our results to violations of Assumption \ref{assn:ny}. First, we have
\begin{align*}
    f^{int}_{n=0,o=1}(y,m,c) &= f_{Y(n=0,o=1)\mid M(n=0,o=1),C}(y\mid m,c)f_{M(n=0,o=1)\mid C}(m\mid c)f_C(c) \\
    &= f_{Y(n=0,o=1)\mid M(n=0,o=1),C}(y\mid m,c)f_{M(n=0)\mid C}(m\mid c)f_C(c),
\end{align*}
where for the second factor, we have
\begin{align*}
    f_{M(n=0,o=1)\mid C}(m\mid c) &= Pr\{M(n=0)=m\mid C=c\}\\
    &= Pr\{M(n=0)=m\mid N=0, C=c\}\\
    &= Pr(M=m\mid A=0, C=c).
\end{align*}
Thus,
\begin{align*}
    \psi_{01} &= \sum_{m,c}\int_{-\infty}^t f^{int}_{n=0,o=1}(y,m,c)\; dy \\
    &= \sum_{m,c} Pr\{Y(n=0,o=1)\leq t\mid M(n=0,o=1)=m,C=c\}Pr(M=m\mid A=0, C=c)f_C(c) \\
    %&\quad \times \frac{\sum_{m,c} Pr\{Y(n=1,o=1)\leq t\mid M(n=1,o=1)=m,C=c\}Pr(M=m\mid A=0, C=c)f_C(c)}{\sum_{m,c} Pr\{Y(n=1,o=1)\leq t\mid M(n=1,o=1)=m,C=c\}Pr(M=m\mid A=0, C=c)f_C(c)} \\
    &= \sum_{m,c} Pr\{Y(n=1,o=1)\leq t\mid M(n=1,o=1)=m,C=c\}Pr(M=m\mid A=0, C=c)f_C(c) \\
    &\quad \times \frac{\sum_{m,c} Pr\{Y(n=0,o=1)\leq t\mid M(n=0,o=1)=m,C=c\}Pr(M=m\mid A=0, C=c)f_C(c)}{\sum_{m,c} Pr\{Y(n=1,o=1)\leq t\mid M(n=1,o=1)=m,C=c\}Pr(M=m\mid A=0, C=c)f_C(c)} \\
    &= \sum_{m,c} Pr\{Y(n=1,o=1)\leq t\mid M(n=1,o=1)=m, N=1, O=1, C=c\}Pr(M=m\mid A=0, c)f_C(c) \\
    &\quad \times \frac{\sum_{m,c} Pr\{Y(n=0,o=1)\leq t\mid M(n=0,o=1)=m,C=c\}Pr\{M(n=0)=m\mid C=c\}f_C(c)}{\sum_{m,c} Pr\{Y(n=1,o=1)\leq t\mid M(n=1,o=1)=m,C=c\}Pr\{M(n=0)=m\mid C=c\}f_C(c)}\\
    &= \sum_{m,c} Pr(Y\leq t\mid M=m, A=1, C=c)Pr(M=m\mid A=0, C=c)f_C(c) \\
    &\quad \times \frac{\sum_{m,c} Pr\{Y(n=0,o=1,m)\leq t\mid C=c\}Pr\{M(n=0)=m\mid C=c\}f_C(c)}{\sum_{m,c} Pr\{Y(n=1,o=1,m)\leq t\mid C=c\}Pr\{M(n=0)=m\mid C=c\}f_C(c)} \\
    &= \Psi_{01}(P) \times \frac{1}{\gamma}.
\end{align*}
Under the additional cross-world counterfactual independence assumption $Y(n=1,o=1,m)\ci M(n=0)\mid C$, we have
\begin{align*}
    \gamma =&\ \frac{\sum_{m,c} Pr\{Y(n=1,o=1,m)\leq t\mid C=c\}Pr(M(n=0)=m\mid C=c)f_C(c)}{\sum_{m,c} Pr\{Y(n=0,o=1,m)\leq t\mid C=c\}Pr(M(n=0)=m\mid C=c)f_C(c)} \\
    =& \frac{\sum_{m,c} Pr\{Y(n=1,o=1,m)\leq t\mid M(n=0)=m, C=c\}Pr(M(n=0)=m\mid C=c)f_C(c)}{\sum_{m,c} Pr\{Y(n=0,o=1,m)\leq t\mid M(n=0)=m, C=c\}Pr(M(n=0)=m\mid C=c)f_C(c)} \\ 
    =& \frac{\sum_{m,c} Pr\{Y(n=1,o=1,M(n=0))\leq t\mid M(n=0)=m, C=c\}Pr(M(n=0)=m\mid C=c)f_C(c)}{\sum_{m,c} Pr\{Y(n=0,o=1,M(n=0))\leq t\mid M(n=0)=m, C=c\}Pr(M(n=0)=m\mid C=c)f_C(c)} \\
    =& \frac{Pr\{Y(n=1,o=1,M(n=0))\leq t\}}{ Pr\{Y(n=0,o=1,M(n=0))\leq t\}}.
\end{align*}
Irrespective of which interpretation of $\delta$ one uses, altogether, we have 
\[\psi^{\textrm{anesthesia}} = \frac{\psi_{01}}{\psi_0} = \frac{\Psi_{01}(P)}{\Psi_0(P)}\times \frac{1}{\gamma}.\]

Additionally, under $Y(n=1,o=1,m)\ci M(n=0)\mid C$,
\begin{align*}
    \Psi_{01}(P)/\Psi_0(P) &= (\psi_{01}/\psi_0)\gamma \\
    &= \frac{Pr\{Y(n=0,o=1)\leq t\}}{Pr\{Y(n=0,o=0)\leq t\}}\times\frac{Pr[Y\{n=1,o=1,M(n=0)\}\leq t]}{Pr[Y\{n=0,o=1,M(n=0)\}\leq t]} \\
    &= \frac{Pr[Y\{n=1,o=1,M(n=0)\}\leq t]}{Pr\{Y(n=0,o=0)\leq t\}} \\
    &= \frac{Pr[Y\{a=1,M(a=0)\}\leq t]}{Pr[Y\{a=0,M(a=0)\}\leq t]} \\
    &= PNDE
\end{align*}

\subsection{Alternative DAG structure for violations of A\ref{assn:ny}}
\label{a:alternative_dag}
\begin{figure}[h]
\begin{tabular}{cc}
    
    \begin{tikzpicture}
        \begin{scope}[on grid,node distance=1.75cm]
        %\draw[help lines](-1,-1) grid(4,4);%\leq -- comment this line to hide the grid
        \node[circle, draw=black] (c) {$C$};
        \node[circle, draw=black, right= of c] (a) {$A$};
        \node[circle, draw=black, right= of a] (n) {$N$};
        \node[circle, draw=black, above= of n] (o) {$O$};
        \node[circle, draw=black, right= of n] (m) {$M$};
        \node[circle, draw=black, above= of m] (l) {$L$};
        \node[circle, draw=black, right= of m] (y) {$Y$};
        \end{scope}
        
        \draw[->] (c) -- (a);
        \draw[->] (n) -- (m);
        \draw[->] (m) -- (y);
        \draw[->, red, line width= 1] (a) -- (o);
        \draw[->, red, line width=1] (a) -- (n);
        \draw[->] (o) to  [out=40,in=100, looseness=1.2] (y);
        \draw[->] (n) -- (l);
        \draw[->] (l) -- (m);
        \draw[->]  (l) -- (y);
        \draw[->] (c) to  [out=330,in=220, looseness=1] (m);
        \draw[->] (c) to  [out=310,in=230, looseness=0.75] (y);
        \draw[->] (c) to  [out=60,in=135, looseness=1] (l);
    \end{tikzpicture}  & 
    \begin{tikzpicture}
        \begin{scope}[on grid,node distance=1.75cm]
        %\draw[help lines](-1,-1) grid(4,4);%<-- comment this line to hide the grid
        \node[circle, draw=black] (c) {$C$};
        \node[circle, draw=black, right= of c] (a) {$A$};
        \node[circle, draw=black, right= of a] (n) {$N$};
        \node[circle, draw=black, above= of n] (o) {$O$};
        \node[circle, draw=black, right= of n] (m) {$M$};
        \node[circle, draw=black, above= of m] (l) {$L$};
        \node[circle, draw=black, right= of m] (y) {$Y$};
        \end{scope}
        
        \draw[->] (c) -- (a);
        \draw[->] (n) -- (m);
        \draw[->] (m) -- (y);
        \draw[->, red, line width= 1] (a) -- (o);
        \draw[->, red, line width=1] (a) -- (n);
        \draw[->] (o) to  [out=40,in=100, looseness=1.2] (y);
        \draw[->] (o) -- (l);
        \draw[->] (l) -- (m);
        \draw[->] (n) -- (l);
        \draw[->] (l) -- (y);
        \draw[->] (c) to  [out=330,in=220, looseness=1] (m);
        \draw[->] (c) to  [out=310,in=230, looseness=0.75] (y);
        \draw[->] (c) to  [out=60,in=135, looseness=1] (l);
    \end{tikzpicture} \\
    \textbf{(a)} &  \textbf{(b)}
\end{tabular}
    \centering
    \caption{DAG contains an observed common cause $L$ of the mediator $M$ and outcome $Y$ that is also caused by (a) $N$ and (b) both $N$ and $O$. Let the bold red arrows denote deterministic functions.   }
    \label{fig:dag_sens2.5}
\end{figure}
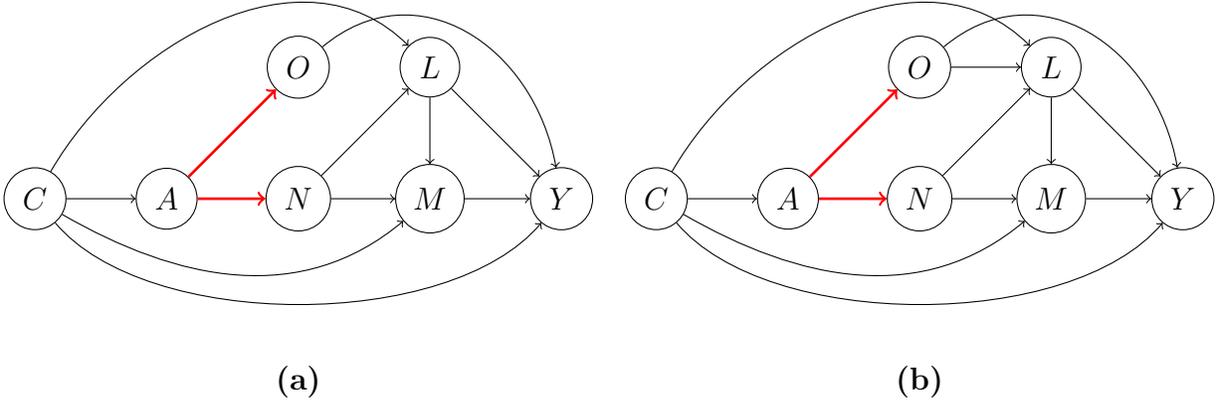

%Given the challenge of interpretability with real data, 
We now introduce an alternative DAG structure originally considered in \cite{robins2022interventionist} (and generalized in \cite{stensrud2021generalized}) that may arise when the assumption of no direct effect of $N$ on $Y$ is violated. Suppose that there is an additional observed variable $L$ that is a common cause of $M$ and $Y$, and is also affected by $A$, %However, we have decomposed $A$ into two separate components: anesthesia ($O$) and surgery ($N$). 
but only through $N$, i.e., $N$ is the only component of $A$ that affects $L$, as illustrated in Figure \ref{fig:dag_sens2.5}.a. Here, there is an indirect effect (with respect to $M$) of $N$ on $Y$ through $L$. Despite this, \cite{robins2022interventionist} showed that we can still identify $\psi^{\textrm{anesthesia}}$ :
%{\small 
\begin{align}
    Pr\{Y(n=0, o=1)\leq t\} = \sum_{m,l,c} Pr(Y\leq t \mid A=1, m, l, c) f(m \mid l, A=0, c) f(l \mid A = 0, c) f(c).
\end{align}
%}
In fact, in our case, this is no different from simply including $L$ in $M$, which is what one would naturally do in our setting. In fact, this is essentially what we have done when performing the exercise of determining which variables to include in $M$. On the other hand, 
%Suppose that in a certain data application, assuming there is no direct effect of $N$ on $Y$ is not plausible. Identifying an intermediate variable(s) ($L$) such that all the effect of $N$ on $Y$ is completely mediated through $L$, can facilitate the identification of $Pr\{Y(n=0, o=1)\leq t\}$.  This idea has been generalized by \citep{stensrud2021generalized}.  I
if the underlying DAG structure is that in Figure \ref{fig:dag_sens2.5}.b, then $Pr\{Y(n=0, o=1)\leq t\}$ is not identifiable.

\begin{comment}
Given the DAG structure in \ref{fig:dag_sens2.5}.a.
\begin{align*}
    f^{int}_{n=0, o=1}(y,m,c,l) &= f(y \mid o=1, m, l, c) f(m \mid l, n=0, c) f(l \mid o = 1) f(c) I(n=0, o=1) \\ 
    &= f(y \mid o=1, m, l, c) f(m \mid l, n=0, c) f(l \mid n=0) f(c) I(n=0, o=1) \\ 
    &= f(y \mid A=1, m, l, c) f(m \mid l, A=0, c) f(l \mid A = 0) f(c).
\end{align*}
Thus making: 
\begin{align*}
    Pr\{Y(n=0, o=1)\leq t\} &= \sum_{m,l,c} \sum_y^t f^{int}_{n=0, o=1}(y,m,c,l) \\
    &= \sum_{m,l,c} \sum_y^t f(y \mid A=1, m, l, c) f(m \mid l, A=0, c) f(l \mid A = 0) f(c)  \\
    &= \sum_{m,l,c} Pr\{Y\leq t \mid A=1, m, l, c\} f(m \mid l, A=0, c) f(l \mid A = 0) f(c).
\end{align*}
\end{comment}

\subsection{Sensitivity to violations of assumption \ref{assn:om}} \label{a:sens_om}
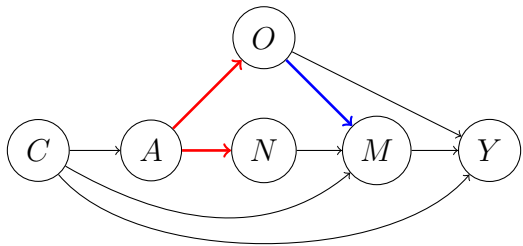
\begin{figure}[H]
    \centering
   % \begin{tabular}{cc}
    \begin{tikzpicture}
        \begin{scope}[on grid,node distance=1.5cm]
        %\draw[help lines](-1,-1) grid(4,4);%\leq -- comment this line to hide the grid
        \node[circle, draw=black] (c) {$C$};
        \node[circle, draw=black, right= of c] (a) {$A$};
        \node[circle, draw=black, right= of a] (n) {$N$};
        \node[circle, draw=black, above= of n] (o) {$O$};
        \node[circle, draw=black, right= of n] (m) {$M$};
        \node[circle, draw=black, right= of m] (y) {$Y$};
        \end{scope}
        
        \draw[->] (c) -- (a);
        \draw[->] (n) -- (m);
        \draw[->] (m) -- (y);
        \draw[->, red, line width= 1] (a) -- (o);
        \draw[->, red, line width=1] (a) -- (n);
        \draw[->] (o) -- (y);
        \draw[->, blue, line width= 1] (o) -- (m);
        \draw[->] (c) to  [out=330,in=220, looseness=1] (m);
        \draw[->] (c) to  [out=310,in=230, looseness=0.75] (y);
    \end{tikzpicture} 
    \caption{Let the bold red arrows denote deterministic functions. Let the Blue line denote the violation of interest. The DAG displays the violation in the direct effect of $O$ on $M$. }
    \label{fig:dag_sens_om}
\end{figure}

%We also explore the sensitivity to violations of Assumption \ref{assn:om}. This violation can be visualized in 
Figure \ref{fig:dag_sens_om} depicts a violation of Assumption \ref{assn:om} with the addition of the blue arrow from $O$ to $M$. We now develop sensitivity analysis for this violation. %In the presence of a violation, we will show that the resulting identification will be the original estimand plug a bias term. Understanding how big of a bias term needs to be present to change the direction and interpretation of results can give insight into how sensitive our results are to this assumption.
%Figure \ref{fig:dag_sens}.b depicts the violation of the assumption by the inclusion of the blue arrow from $O$ to $M$ eliminating the conditional independence between $O$ and $M$. 

%---------------------
When Assumption \ref{assn:om} does not hold, we have
%\begin{equation}
    $Pr\{Y(n=0, o=1)\leq t\} = \Psi_{01}(P) / \eta$,
%\end{equation}
where the sensitivity parameter is defined to be
\begin{equation*}
    \eta \equiv \frac{\sum_{m,c} Pr\{Y(o=1,m) \leq t \mid C = c \} Pr\{M(n=0, o=0)=m, C=c\}}{\sum_{m,c} Pr\{Y(o=1,m) \leq t \mid C = c\} Pr\{M(n=0, o=1)=m, C=c\}},
\end{equation*}
\begin{comment}
    \begin{equation}
    \xi = \sum_{c} \left\{\mathbb{E}[g(M,c) \mid n=0, o=0, c]-\mathbb{E}[g(M,c) \mid n=0,o=1,c] \right\} f(c)
\end{equation}
and $g(m,c) = Pr\{Y \leq t \mid A=1, m,c\}$.
\end{comment}
as we show in the following subsection. %appendix \ref{a:sens_derivation_om}. 
Thus, we have $\psi^{\textrm{anesthesia}} = \eta^{-1}\Psi_{01}(P)/\Psi_0(P)$. 

$\eta$ has a causal interpretation as a contrast of $Pr\{Y(o=1,m) \leq t \mid c\}$ averaged over $m$ according to two different distributions. The first is the distribution of $M$ under no exposure to surgery or anesthesia and baseline covariates, and the second is that of $M$ under no exposure to surgery but exposure to anesthesia and baseline covariates. This interpretation is
analogous to an alternative interpretation of the randomized interventional analog of the natural indirect effect
discussed in \cite{miles2023causal}. The primary differences are that in our case, we
are additionally intervening to set $N = 0$ and our contrast is on the relative risk scale. %Although the interpretation is not straightforward the goal of this quantity is to 
This characterizes the impact that the effect of $O$ on $M$ has on $Y$. If there is either no effect of $O$ on $M$ or no effect of $M$ on $Y$, then $\eta=0$. Conversely, if $\eta$ is substantial, it indicates the presence of an effect of $O$ on $M$ and of $M$ on $Y$, and that at least one of these is large. 

In practice, the bias term $\eta$ is unknown; however, it could in principle be identified in certain randomized trials. For example, this would be the case under Assumption \ref{assn:ny} if a two-armed trial was conducted in which surgery was withheld from all participants and anesthesia was randomly assigned. As before, there exists no such trial; instead subject-matter experts can speculate about plausible ranges of values for $\eta$ to adjust the estimate found assuming no violation. A stronger interpretation of $\eta$ is possible under the additional cross-world counterfactual independence assumption $Y(o=1,m)\ci M(n=0,o=0)\mid C$. In particular, we have
\begin{align*}
    \eta &= \frac{\sum_{m,c} Pr\{Y(o=1,m) \leq t \mid C = c \} Pr\{M(n=0, o=0)=m, C=c\}}{\sum_{m,c} Pr\{Y(o=1,m) \leq t \mid C = c\} Pr\{M(n=0, o=1)=m, C=c\}}\\
    &= \frac{\sum_{m,c} Pr\{Y(o=1,m) \leq t \mid M(n=0, o=0)=m, C = c \} Pr\{M(n=0, o=0)=m, C=c\}}{\sum_{m,c} Pr\{Y(o=1,m) \leq t \mid M(n=0, o=1)=m, C = c\} Pr\{M(n=0, o=1)=m, C=c\}}\\
    &= \frac{\sum_{m,c} Pr[Y\{o=1,M(n=0, o=0)\} \leq t \mid M(n=0, o=0)=m, C = c ] Pr\{M(n=0, o=0)=m, C=c\}}{\sum_{m,c} Pr[Y\{o=1,M(n=0, o=1)\} \leq t \mid M(n=0, o=1)=m, C = c] Pr\{M(n=0, o=1)=m, C=c\}}\\
    &= \frac{Pr[Y\{o=1,M(n=0, o=0)\} \leq t]}{Pr[Y\{o=1,M(n=0, o=1)\} \leq t ]},
\end{align*}
which can be interpreted as a modified version of the total natural indirect effect (TNIE) of $O$ on $Y$ with respect to $M$ when $N$ has also been intervened on to be set to zero, i.e., the TNIE of anesthesia on time to DIBD diagnosis with respect to $M$ when surgery is not administered.

As with the violation of Assumption \ref{assn:ny}, we can also rearrange the equation arising when Assumption \ref{assn:om} is violated to obtain $\Psi_{01}(P)/\Psi_{0}(P) = \eta \psi_{01} / \psi_0 $, which gives an alternative interpretation of the identifying functional $\Psi_{01}(P)/\Psi_{0}(P)$ as the product of the original effect of interest $\psi_{01}/\psi_0$ capturing the effect of anesthesia and $\eta$. %the direct effect of surgery on DIBD with respect to ${\bf M}$ averaged over the distribution of ${\bf M}(a=0)$ and ${\bf C}$. 
As before, this product telescopes, again reducing to 
\begin{align*}
    \frac{\Psi_{01}(P)}{\Psi_{0}(P)} &= %\frac{\sum_{{\bf m},{\bf c}} \mathrm{Pr}\{Y(n=1,o=1,{\bf m}) \leq t \mid {\bf C} = {\bf c} \} \mathrm{Pr}\{{\bf M}(a=0)={\bf m}, {\bf C}={\bf c}\}}{\mathrm{Pr}\{Y(n=0,o=0)\leq t\}}\\
    %&= 
    \frac{\sum_{{\bf m},{\bf c}} \mathrm{Pr}\{Y(a=1,{\bf m}) \leq t \mid {\bf C} = {\bf c} \} \mathrm{Pr}\{{\bf M}(a=0)={\bf m}, {\bf C}={\bf c}\}}{\sum_{{\bf m},{\bf c}} \mathrm{Pr}\{Y(a=0,{\bf m}) \leq t \mid {\bf C} = {\bf c} \} \mathrm{Pr}\{{\bf M}(a=0)={\bf m}, {\bf C}={\bf c}\}},
\end{align*}
i.e., the same quantity we arrived at under the violation of Assumption \ref{assn:ny}. Thus, this has the same interpretation as that given in Section \ref{section:sensitivity}, and is also equivalent to the randomized interventional analog to the natural direct effect. 
Furthermore, under the cross-world counterfactual independence assumption $Y(o=1,m)\ci M(n=0,o=0)\mid C$, we again have $\Psi_{01}(P)/\Psi_{0}(P) = \mathrm{Pr}[Y\{a=1,M(a=0)\}\leq t]/\mathrm{Pr}[Y\{a=0,M(a=0)\}\leq t\}$. That is, under violations of Assumption \ref{assn:om}, the identifying functional still can be interpreted as the natural direct effect of $A$ on $Y$ with respect to $M$ if the cross-world counterfactual independence holds. As in the other exclusion restriction violation case, there is nothing about the extended DAG that invalidates this identification result from the mediation literature. 

%---------------------

\begin{comment}
In the presence of this violation the new identification becomes 
\begin{align}
    \Psi^{\textrm{anesthesia}} &= %\frac{\Psi_{01} +  \xi }{Pr\{Y(N=0, O=0) \leq t\}]} = 
    \frac{\Psi_{01} - \xi }{\Psi_0} .
\end{align}
The unknown bias quantity is
\begin{equation}
    \xi = \sum_{c} \left\{\mathbb{E}[g(M,c) \mid n=0, o=0, c]-\mathbb{E}[g(M,c) \mid n=0,o=1,c] \right\} f(c)
\end{equation}
and $g(m,c) = Pr\{Y \leq t \mid A=1, m,c\}$.

Interpreting this unknown quantity $\xi$ in terms of our application is not as straightforward compared to the bias term found before. The $\xi$ can be thought of as the contrast of $g(m,c)$ averages over $m$ according to two different distributions. The first is the distribution of $M$ given no exposure to surgery ($N=0$) but exposure to anesthesia ($O=1$) and $C$, the second distributing is of $M$ given no exposure to either surgery or anesthesia ($N=0, 0=0$) and $C$. Although the interpretation is not straightforward the goal of this quantity is to characterize the effect of $O$ on $M$. So, when $\xi=0$, it means there is no influence of $O$ on $M$. Conversely, if $\xi$ is substantial, it indicates a considerable impact of $O$ on $M$ and thus a larger violation. Using the same strategy, subject material knowledge could help propose a range of sensitivity parameters that can be used to adjust our estimate.   
\end{comment}

\subsubsection{Derivation of sensitivity parameter for violation in \ref{assn:om}} \label{a:sens_derivation_om} 

%We will now explore the sensitivity of our results when there is a violation in Assumption \ref{assn:om}. 
First, we have
\begin{align*}
    f^{int}_{n=0,o=1}(y,m,c) &= f_{Y(n=0,o=1)\mid M(n=0,o=1),C}(y\mid m,c)f_{M(n=0,o=1)\mid C}(m\mid c)f_C(c) 
\end{align*}
where for the first factor, we have
\begin{align*}
    f_{Y(n=0,o=1)\mid M(n=0,o=1),C} &= Pr\{Y(n=0,o=1) \leq t \mid M(n=0,o=1)=m,C\}\\
    &= Pr\{Y(n=0,o=1,m) \leq t \mid M(n=0,o=1)=m,C\}\\
    &= Pr\{Y(n=0,o=1,m) \leq t \mid C\}\\
    &= Pr\{Y(n=1,o=1,m) \leq t \mid C\}\\
    &= Pr\{Y(n=1,o=1,m) \leq t \mid N=1, O=1, M(n=1,o=1)=m,C\}\\
    &= Pr(Y\leq t\mid A=1, M=m, C=c).
\end{align*}
Thus,
\begin{align*}
    \psi_{01} &= \sum_{m,c}\int_{-\infty}^t f^{int}_{n=0,o=1}(y,m,c)\; dy \\
    &= \sum_{m,c} Pr(Y\leq t\mid A = 1,M=m,C=c)Pr\{M(n=0, o=1)=m\mid C=c\}f_C(c) \\
    %&\quad \times \frac{\sum_{m,c} Pr\{Y\leq t\mid A=1, M=m,C=c\}Pr(M(n=0,o=0)=m\mid C=c)f_C(c)}{\sum_{m,c} Pr\{Y\leq t\mid A=1, M=m,C=c\}Pr(M(n=0, o=0)=m\mid C=c)f_C(c)} \\
    &= \sum_{m,c} Pr(Y\leq t\mid A = 1,M=m,C=c)Pr\{M(n=0, o=0)=m\mid C=c\}f_C(c) \\
    &\quad \times \frac{\sum_{m,c} Pr(Y\leq t\mid A=1, M=m,C=c)Pr\{M(n=0,o=1)=m\mid C=c\}f_C(c)}{\sum_{m,c} Pr(Y\leq t\mid A=1, M=m,C=c)Pr\{M(n=0, o=0)=m\mid C=c\}f_C(c)} \\
    &= \sum_{m,c} Pr(Y\leq t\mid A = 1,M=m,C=c)Pr\{M(n=0, o=0)=m\mid N=0, O=0, C=c\}f_C(c) \\
    &\quad \times \frac{\sum_{m,c} Pr(Y\leq t\mid A=1, M=m,C=c)Pr\{M(n=0,o=1)=m\mid C=c\}f_C(c)}{\sum_{m,c} Pr(Y\leq t\mid A=1, M=m,C=c)Pr\{M(n=0, o=0)=m\mid C=c\}f_C(c)} \\
    &= \sum_{m,c} Pr(Y\leq t\mid A = 1,M=m,C=c)Pr(M=m\mid A=0, C=c)f_C(c) \\
    &\quad \times \frac{\sum_{m,c} Pr(Y\leq t\mid A=1, M=m,C=c)Pr\{M(n=0,o=1)=m, C=c\}}{\sum_{m,c} Pr(Y\leq t\mid A=1, M=m,C=c)Pr\{M(n=0, o=0)=m, C=c\}} \\
    &= \Psi_{01}(P) \times \frac{1}{\eta}.
\end{align*}
\begin{comment}
Under additional assumptions
\begin{align*}
    \eta =&\frac{\sum_{m,c} Pr\{Y\leq t\mid A=1, M=m,C=c\}Pr(M(n=0,o=0)=m\mid C=c)f_C(c)}{\sum_{m,c} Pr\{Y\leq t\mid A=1, M=m,C=c\}Pr(M(n=0, o=1)=m\mid C=c)f_C(c)} \\
    =&\frac{\sum_{m,c} Pr\{Y\leq t\mid A=1, M=m,C=c\}Pr(M(n=0,o=0)=m\mid n=0, o=0, C=c)f_C(c)}{\sum_{m,c} Pr\{Y\leq t\mid A=1, M=m,C=c\}Pr(M(n=0, o=1)=m\mid n=0, o=1, C=c)f_C(c)} \\ 
    =&\frac{\sum_{c} E[g(M,c) \mid n=0, o=0, C=c]f_C(c)}{\sum_{c} E[g(M,c) \mid n=0, o=1, C=c]f_C(c)} 
\end{align*}
where $g(M,c) = Pr\{Y\leq t\mid A=1, M=m,C=c\}$. 
Irrespective of which interpretation of $\eta$ one uses, 
\end{comment}
Altogether, we have 
\[\psi^{\textrm{anesthesia}} = \frac{\psi_{01}}{\psi_0} = \frac{\Psi_{01}(P)}{\Psi_0(P)}\times \frac{1}{\eta}.\]

\subsubsection{Alternative DAG structure for violation A\ref{assn:om} result}
\label{a:alternative_dag_om}

\begin{figure}[H]
\begin{tabular}{cc}
    
    \begin{tikzpicture}
        \begin{scope}[on grid,node distance=1.75cm]
        %\draw[help lines](-1,-1) grid(4,4);%\leq -- comment this line to hide the grid
        \node[circle, draw=black] (c) {$C$};
        \node[circle, draw=black, right= of c] (a) {$A$};
        \node[circle, draw=black, right= of a] (n) {$N$};
        \node[circle, draw=black, above= of n] (o) {$O$};
        \node[circle, draw=black, right= of n] (m) {$M$};
        \node[circle, draw=black, above= of m] (l) {$L$};
        \node[circle, draw=black, right= of m] (y) {$Y$};
        \draw[->] (c) to  [out=60,in=135, looseness=1] (l);
        \end{scope}
        
        \draw[->] (c) -- (a);
        \draw[->] (n) -- (m);
        \draw[->] (m) -- (y);
        \draw[->, red, line width= 1] (a) -- (o);
        \draw[->, red, line width=1] (a) -- (n);
        \draw[->] (o) to  [out=40,in=100, looseness=1.2] (y);
        \draw[->] (o) -- (l);
        \draw[->] (l) -- (m);
        \draw[->]  (l) -- (y);
        \draw[->] (c) to  [out=330,in=220, looseness=1] (m);
        \draw[->] (c) to  [out=310,in=230, looseness=0.75] (y);
    \end{tikzpicture}  & 
    \begin{tikzpicture}
        \begin{scope}[on grid,node distance=1.75cm]
        %\draw[help lines](-1,-1) grid(4,4);%<-- comment this line to hide the grid
        \node[circle, draw=black] (c) {$C$};
        \node[circle, draw=black, right= of c] (a) {$A$};
        \node[circle, draw=black, right= of a] (n) {$N$};
        \node[circle, draw=black, above= of n] (o) {$O$};
        \node[circle, draw=black, right= of n] (m) {$M$};
        \node[circle, draw=black, above= of m] (l) {$L$};
        \node[circle, draw=black, right= of m] (y) {$Y$};
        \end{scope}
        
        \draw[->] (c) -- (a);
        \draw[->] (n) -- (m);
        \draw[->] (m) -- (y);
        \draw[->, red, line width= 1] (a) -- (o);
        \draw[->, red, line width=1] (a) -- (n);
        \draw[->] (o) to  [out=40,in=100, looseness=1.2] (y);
        \draw[->] (o) -- (l);
        \draw[->] (l) -- (m);
        \draw[->] (n) -- (l);
        \draw[->] (l) -- (y);
        \draw[->] (c) to  [out=330,in=220, looseness=1] (m);
        \draw[->] (c) to  [out=310,in=230, looseness=0.75] (y);
        \draw[->] (c) to  [out=60,in=135, looseness=1] (l);
    \end{tikzpicture} \\
    \textbf{(a)} &  \textbf{(b)}
\end{tabular}
    \centering
    \caption{DAG contains an observed common cause $L$ of the mediator $M$ and outcome $Y$ that is also caused by (a) $O$ and (b) both $O$ and $N$. Let the bold red arrows denote deterministic functions.   }
    \label{fig:dag_sens2.5_om}
\end{figure}
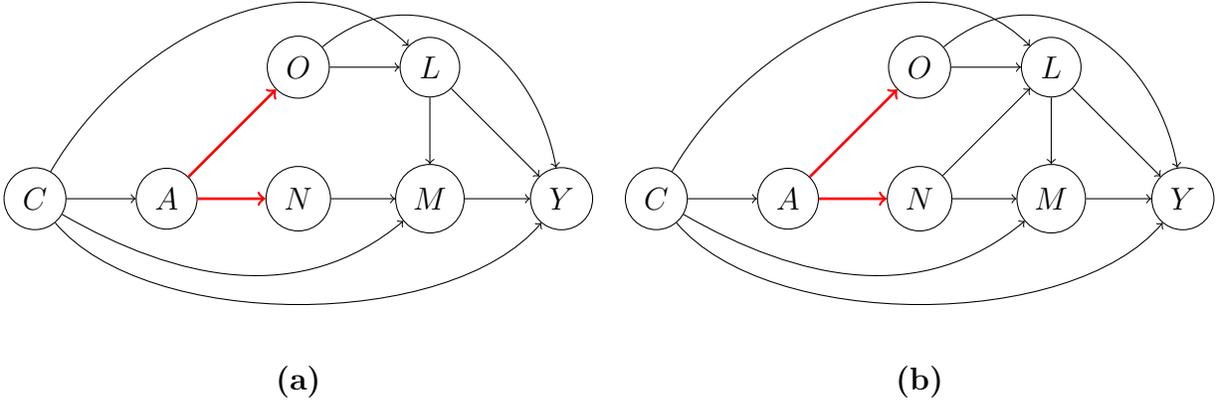

%Given the challenge of interpretability with real data, we propose an alternative DAG structure and narrative to help avoid violations when the assumption of no effect of $O$ on $M$ is not plausible. 
We now introduce an alternative DAG structure originally considered in \cite{robins2022interventionist} (and generalized in \cite{stensrud2021generalized}) that may arise when the assumption of no effect of $O$ on $M$ is violated. 
Suppose that there is an additional observed variable $L$ that is a common cause of the mediator $M$ and outcome $Y$, and is also influenced by $A$, but only through $O$, i.e., $O$ is the only component of $A$ that affects $L$, as illustrated in Figure \ref{fig:dag_sens2.5_om}.a. %Here, there is an indirect effect of $O$ on $M$ through $L$. 
Despite this, \cite{robins2022interventionist} showed that we can still identify $\psi^{\textrm{anesthesia}}$: %\citep{robins2022interventionist}. The identification result is 
{\small 
\begin{align}
    Pr\{Y(n=0, o=1)\leq t\} = \sum_{m,l,c} Pr(Y\leq t \mid A=1, m, l, c) f(m \mid l, A=0, c) f(l \mid A = 1, c) f(c).
\end{align}
}
The derivation of this result is given below. Suppose that in a certain data application, assuming there is no effect of $O$ on $M$ is not plausible. Then identification of an intermediate variable(s) ($L$) such that all the effect of $O$ on $M$ is completely mediated through $L$ facilitates the identification of $Pr\{Y(n=0, o=1)\leq t\}$.  %This idea has been generalized by \citep{stensrud2021generalized}.  
On the other hand, if the underlying DAG structure is that of Figure \ref{fig:dag_sens2.5_om}.b, then $Pr\{Y(n=0, o=1)\leq t\}$ is not identifiable. In our case, we feel that this assumption is reasonable, and so we do not speculate about such potential variables belonging in $L$.

Given the DAG structure in \ref{fig:dag_sens2.5_om}.a. 
\begin{align*}
    f^{int}_{n=0, o=1}(y,m,c,l) &= f(y \mid O=1, m, l, c) f(m \mid l, N=0, c) f(l \mid O = 1,c) f(c)\\ 
    &= f(y \mid A=1, m, l, c) f(m \mid l, A=0, c) f(l \mid A = 1,c) f(c).
\end{align*}
Thus making: 
\begin{align*}
    Pr\{Y(n=0, o=1)\leq t\} &= \sum_{m,l,c} \int_{-\infty}^t f^{int}_{n=0, o=1}(y,m,c,l) \; dy \\
    &= \sum_{m,l,c} \int_{-\infty}^t f(y \mid A=1, m, l, c) \; dy \; f(m \mid l, A=0, c) f(l \mid A = 1, c) f(c)  \\
    &= \sum_{m,l,c} Pr(Y\leq t \mid A=1, m, l, c) f(m \mid l, A=0, c) f(l \mid A = 1, c) f(c).
\end{align*}

\section{Simulation Study for Violation of $O$ to $M$}

\subsection{Data generation}
To generate a violation of Assumption \ref{assn:om}, we modified the generation of the post-exposure variables:
\begin{align*}
    M_1 &\sim N \times \textrm{Bernoulli}\left( p = \textrm{logit}^{-1}[-2 + C_1 + C_2 + C_3 + C_4 + N +\xi O
    ]\right) \\
    M_2 &\sim \textrm{Bernoulli}\left( p = \textrm{logit}^{-1}[-2 + C_1 + C_2 + C_3 + C_4 + N +\tau O
    ]\right). 
\end{align*}
$\xi \neq  O$ induces a violation of Assumption \ref{assn:om}. %When $\xi = 0$ there is no violation in the assumption.  The rest of the data generating process will stay consistent. 

\subsection{Simulation results}

\begin{comment}
\begin{figure}[H]
    \centering
    \includegraphics[scale=0.45]{figures/sim_rmse_coverage_OM_500_additive_20250304.png}
    \caption{ 
    Summary results from the sensitivity-parameter-adjusted $\psi^{\textrm{anesthesia}}$ estimates from 500 simulations. Each subpanel corresponds to a different $\xi$ values (0, 0.25, 0.5, 0.75).The sensitivity parameter $\tau$ adjustment value on the $x$-axis. When the sensitivity parameter $\tau = 0$, the estimate corresponds to the unadjusted estimate. The left panel depicts the RMSE, and the right panel depicts the 95\% CI coverage probability for each sensitivity parameter value. The dashed lines in the right panel indicate 0.95. The red points indicate when the sensitivity parameter matches the true parameter. }
    \label{fig:sim_sum_OM}
\end{figure}
\end{comment}

\begin{figure}[H]
    \centering
    \includegraphics[scale=0.45]{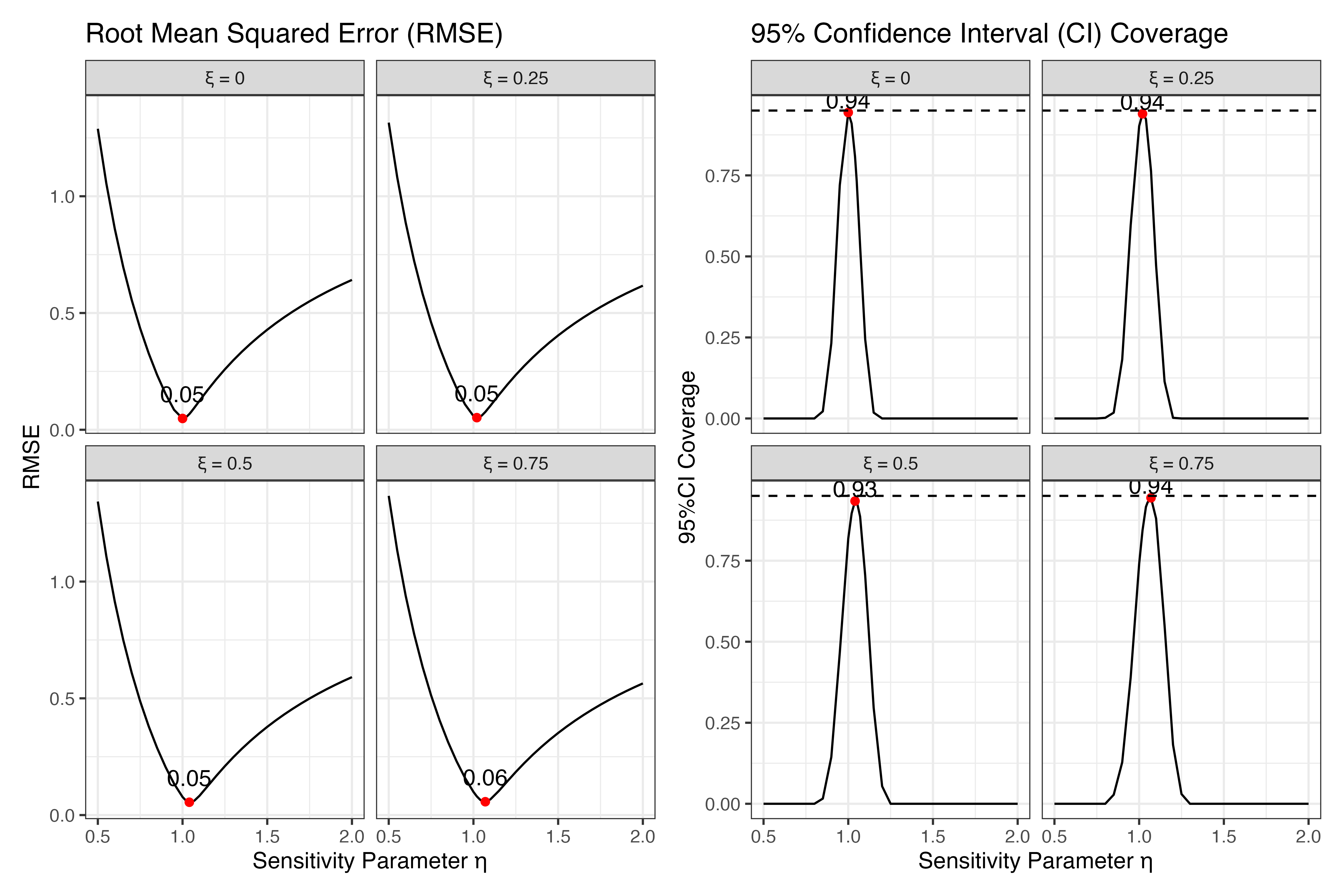}
    \caption{ 
    Summary results from the sensitivity-parameter-adjusted $\psi^{\textrm{anesthesia}}$ estimates from 500 simulations. Each subpanel corresponds to a different value of $\xi$ (0, 0.25, 0.5, 0.75). The $x$-axis is the sensitivity parameter $\eta$ adjustment value. When the sensitivity parameter $\eta = 0$, the estimate corresponds to the unadjusted estimate. The left column depicts the RMSE, and the right column depicts the 95\% CI coverage probability for each sensitivity parameter value. The dashed lines in the right panel indicate the nominal coverage probability of 0.95. The red points indicate when the sensitivity parameter matches the truth. }
    \label{fig:sim_sum_OM}
\end{figure}

\begin{comment}
\begin{figure}[H]
    \centering
    \includegraphics[scale=0.5]{figures/sim_res_OM_iter10_additive_20250304.png} 
    \caption{
    Sensitivity-parameter-adjusted $\psi^{\textrm{anesthesia}}$ estimates and 95\% confidence intervals from a single sample. 
    Each subpanel corresponds to a different $\xi$ values (0, 0.25, 0.5, 0.75). When sensitivity parameter $\tau$ = 0, the estimate corresponds to the unadjusted estimate. The blue dotted line represents the true parameter value and the red dotted line at one represents the null value. The vertical black dotted line represents the true bais induced by $\xi$. The solid black line indicates the point estimates, and the shaded gray regions are pointwise 95\% confidence intervals.}
    \label{fig:sim_res_OM}
\end{figure}
\end{comment}
\begin{figure}[H]
    \centering
    \includegraphics[scale=0.5]{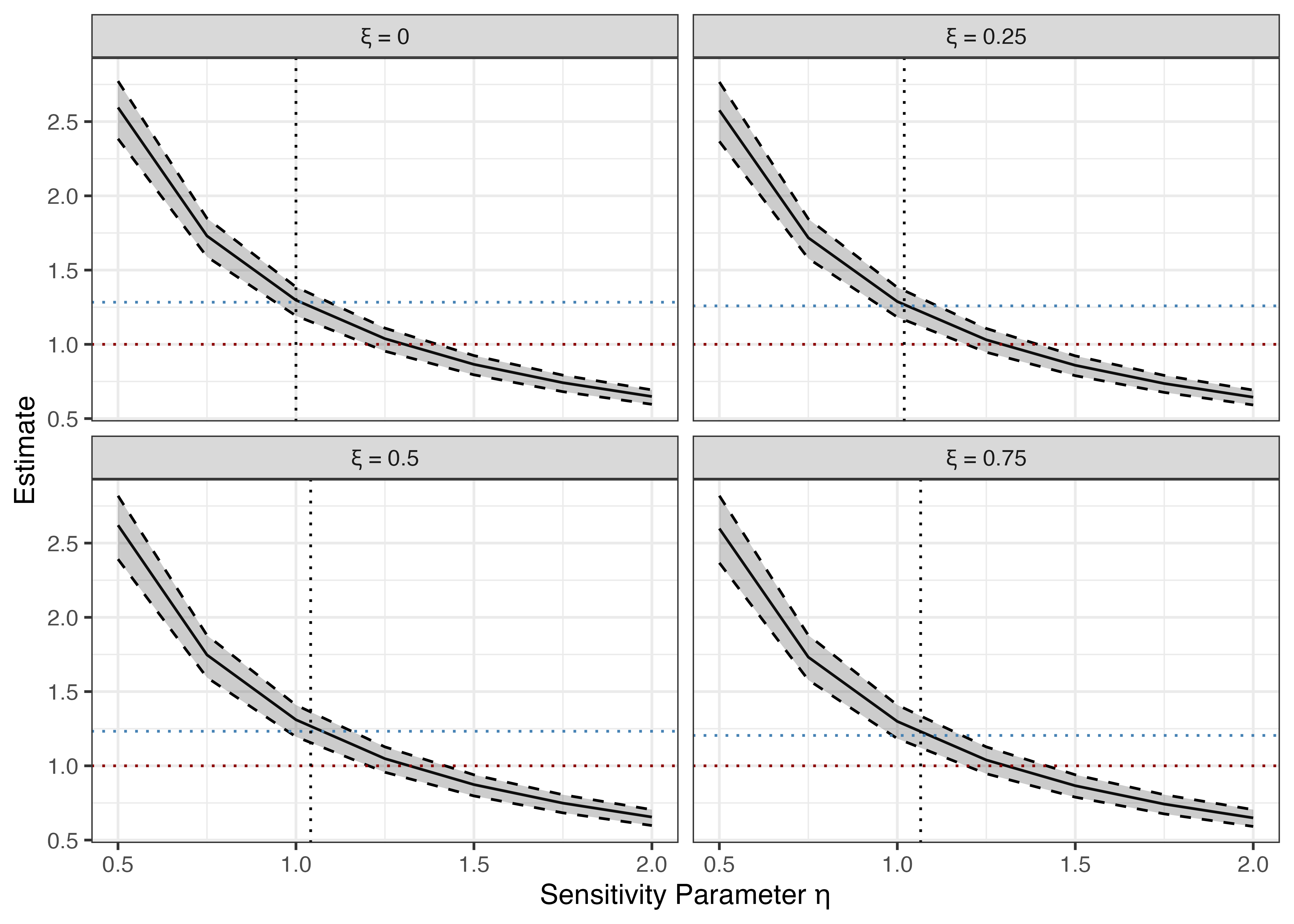} 
    \caption{
    Sensitivity-parameter-adjusted $\psi^{\textrm{anesthesia}}$ estimates and 95\% confidence intervals from a single sample. 
    Each subpanel corresponds to a different value of $\xi$ (0, 0.25, 0.5, 0.75). When $\eta$ = 0, the estimate corresponds to the unadjusted estimate. The blue dotted line represents the true parameter value and the red dotted line at one represents the null value. The vertical black dotted line represents the true bias induced by $\xi$. The solid black line indicates the point estimates, and the shaded gray regions are pointwise 95\% confidence intervals.}
    \label{fig:sim_res_OM}
\end{figure}

\section{Application Results: Estimates Over Time}
\label{a:est_over_time}

\begin{figure}[h!]
    \centering
    \includegraphics[width=0.75\linewidth]{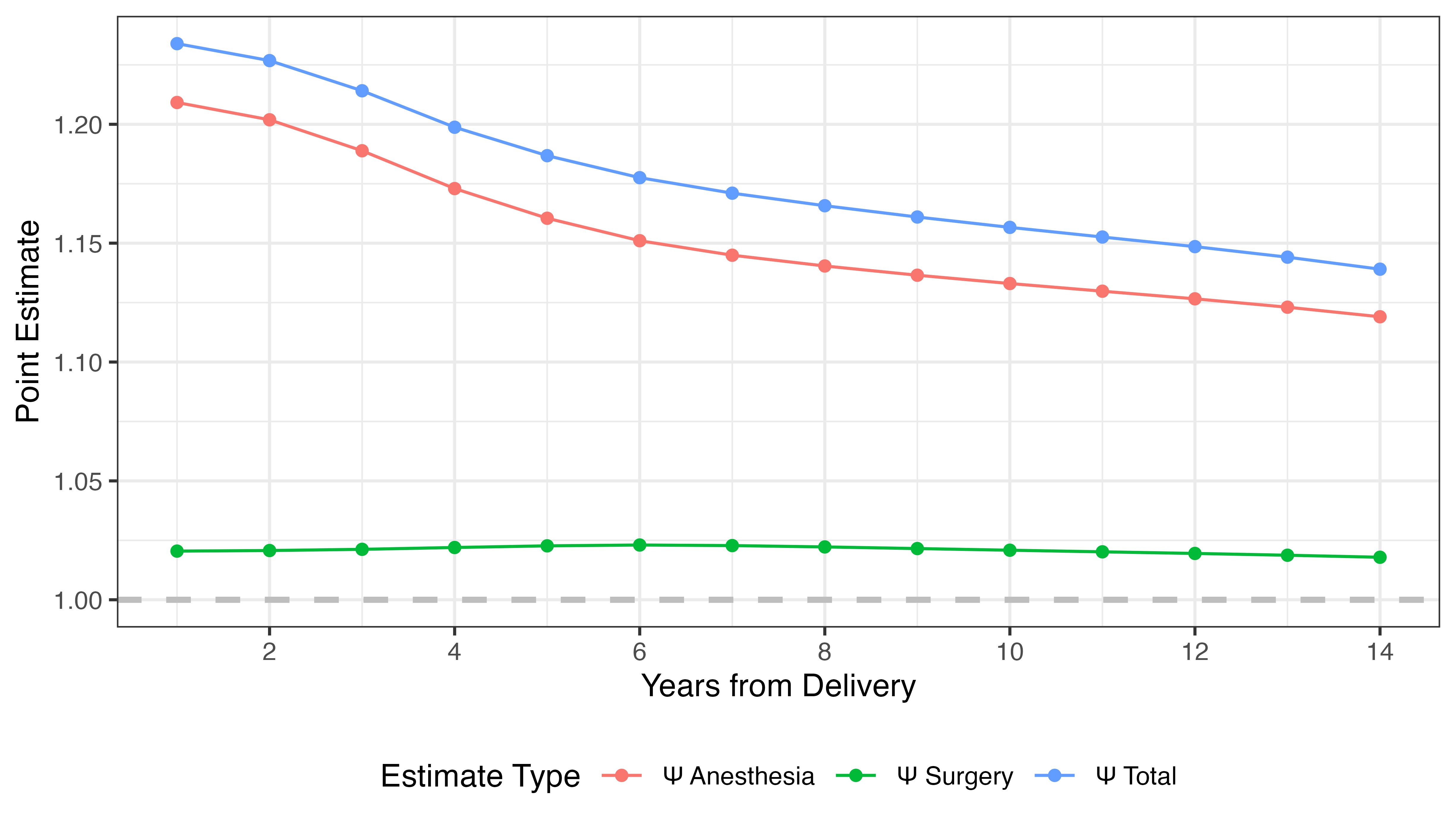}
    \caption{Point estimates of the relative risks of developing a DIBD attributed to anesthesia $\psi^{\textrm{anesthesia}}$, surgery $\psi^{\textrm{surgery}}$, and their joint effect $\psi^{\textrm{total}}$, from age one to fourteen years old. }
    \label{fig:est_over_time}
\end{figure}

\begin{figure}[h!]
    \centering
    \includegraphics[width=0.75\linewidth]{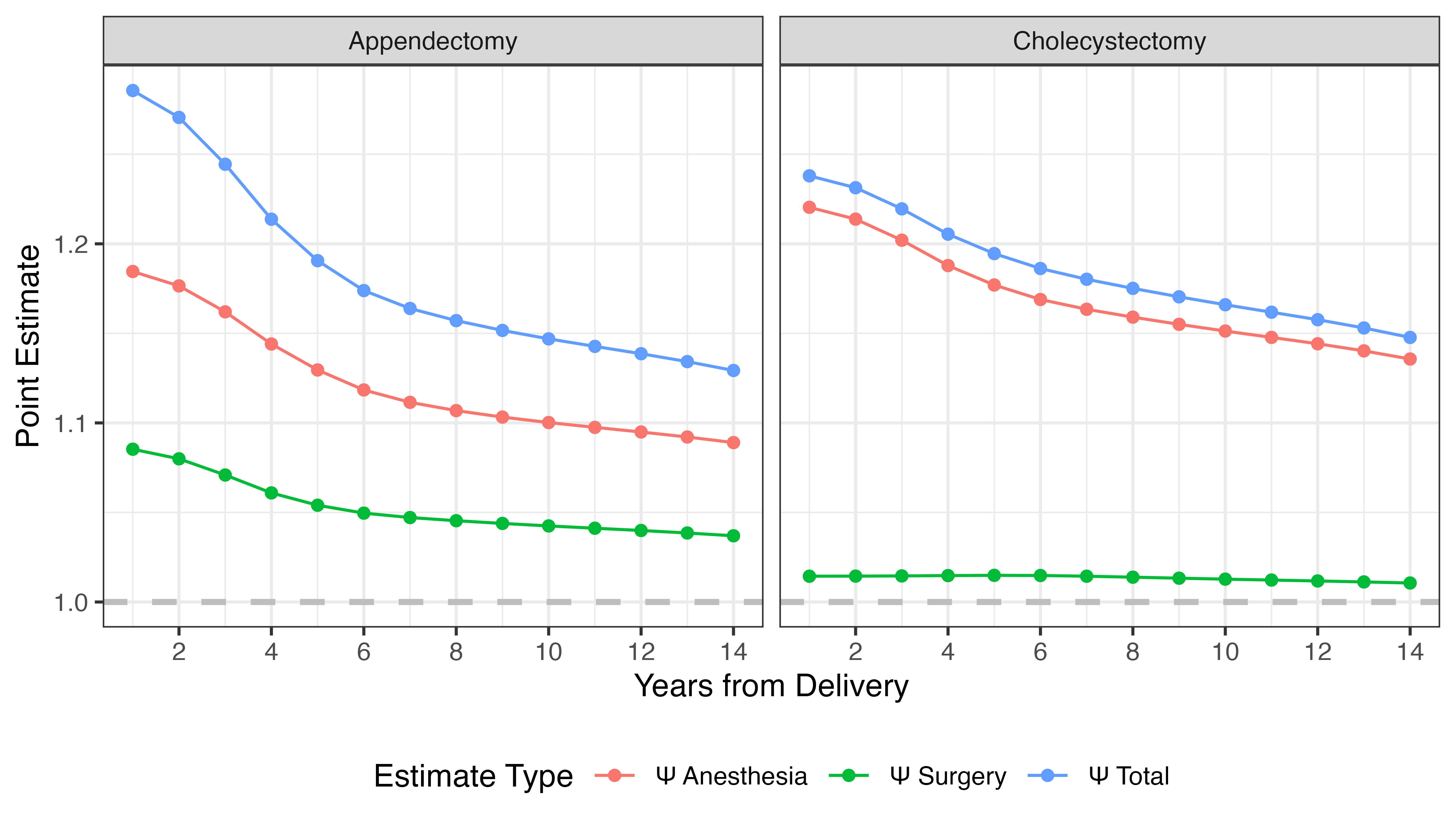}
    \caption{Surgery-specific point estimates of the relative risks of developing a DIBD attributed to anesthesia $\psi^{\textrm{anesthesia}}$, surgery $\psi^{\textrm{surgery}}$, and their joint effect $\psi^{\textrm{total}}$, from age one to fourteen years old. The left panel shows effect estimates comparing appendectomy and accompanying anesthesia to no surgery or anesthesia; the right panel shows effect estimates comparing cholecystectomy and accompanying anesthesia to no surgery or anesthesia. }
    \label{fig:est_over_time_by_surg}
\end{figure}

\end{document}